%% file: main.tex
\documentclass[acmtog,nonacm]{acmart}
\graphicspath{{figures/}}
\input{header.tex}

\begin{document}

\title{Digital 3D Smocking Design}  
\author{Jing Ren}
\affiliation{\institution{ETH Zurich}
\country{Switzerland}
}
\email{jing.ren@inf.ethz.ch}

\author{Aviv Segall}
\affiliation{\institution{ETH Zurich}
\country{Switzerland}
}
\email{aviv.segall@inf.ethz.ch}

\author{Olga Sorkine-Hornung}
\affiliation{\institution{ETH Zurich}
\country{Switzerland}
}
\email{sorkine@inf.ethz.ch}



\begin{abstract}
% Intro sentence.
We develop an optimization-based method to model \emph{smocking}, a surface embroidery technique that provides decorative geometric texturing while maintaining stretch properties of the fabric. 
% Define smocking
During smocking, multiple pairs of points on the fabric are stitched together, creating non-manifold geometric features and visually pleasing textures.
% What is the challenge of smocking?
% First, the conceptual: the result is unpredictable, you often don't know what you will get until you actually make it. This motivates a digital simulation tool that would visualize the smocking result you would get, without having to make it physically by hand.
Designing smocking patterns is challenging, because the outcome of stitching is unpredictable: the final texture is often revealed only when the whole smocking process is completed, necessitating painstaking physical fabrication and time consuming trial-and-error experimentation. This motivates us to seek a digital smocking design method.
% Second, the digital simulation challenge: traditional surface deformation and cloth simulation methods fail
Straightforward attempts to compute smocked fabric geometry using surface deformation or cloth simulation methods fail to produce realistic results, likely due to the intricate structure of the designs, the large number of contacts and high-curvature folds.
%The main challenge of the smocking design problem is that the \emph{sparse} stitching constraints are not sufficient to solve or simulate the deformation of the fabric as a shape deformation or cloth dynamics problem.
%
% How we solve the challenge
We instead formulate smocking as a graph embedding and shape deformation problem. We extract a coarse graph representing the fabric and the stitching constraints, and then derive the graph structure of the smocked result. We solve for the 3D embedding of this graph, which in turn reliably guides the deformation of the high-resolution fabric mesh. 
%from a coarse representation of the fabric to handle the input stitching constraints, which also provides information on how to embed the graph in 3D. We then deform the fabric, represented as a finer grid, guided by the solved embedding of the coarser representation. 
%
% Bragging about our method and results
Our optimization based method is simple, efficient, and flexible, which allows us to build an interactive system for smocking pattern exploration. 
To demonstrate the accuracy of our method, we compare our results to real fabrications on a large set of smocking patterns.
\end{abstract}

%     \centering
%     \vspace{-3pt}
%     \includegraphics[width=1\linewidth]{figures/teaser.png}
%     \vspace{-22pt}
%     \caption{Our method can simulate the 3D textures from smocking, a decorative embroidery technique of stitching points from a fabric to create pleats. Here we show some examples of smocked sleeves using our method, where volumetric smocking textures can create natural folds.}
%     \label{fig:teaser:sleeves}
\pagestyle{plain}
\thispagestyle{empty} 
\maketitle

\begin{figure}[!t]
    \centering
    \begin{overpic}[trim=0cm 0cm 0cm 0cm,clip,width=1\linewidth,grid=false]{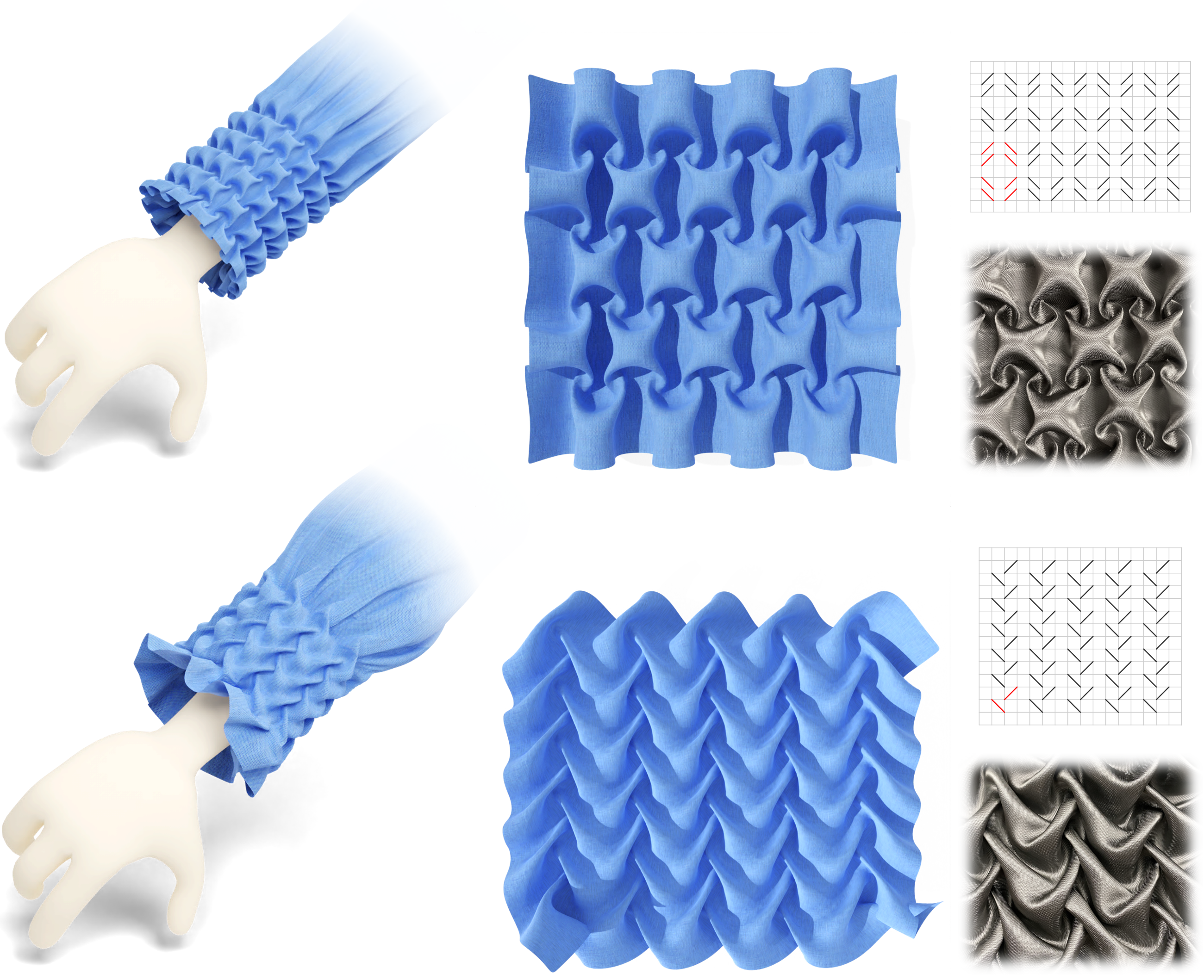}\vspace{-9pt}
    \put(79.5,77){\footnotesize\itshape smocking pattern}
    \put(83,61){\footnotesize\itshape fabrication}
    \put(50,77){\footnotesize\itshape simulated result}
    
    \put(79.5,37){\footnotesize\itshape smocking pattern}
    \put(83,18.5){\footnotesize\itshape fabrication}
    \put(50,33){\footnotesize\itshape simulated result}
    \end{overpic}\vspace{-9pt}
    \caption{We model \emph{smocking}, a decorative geometric cloth texturing technique, where pairs of  points are stitched together to form a pleated pattern. Here we show examples of smocked sleeves produced using our method, where the volumetric smocked textures create natural folds. The geometry of the smocked fabric computed with our method based on the input pattern closely matches the physically fabricated counterpart (photos in grey).}\label{fig:teaser:sleeves}\vspace{-12pt}
\end{figure}
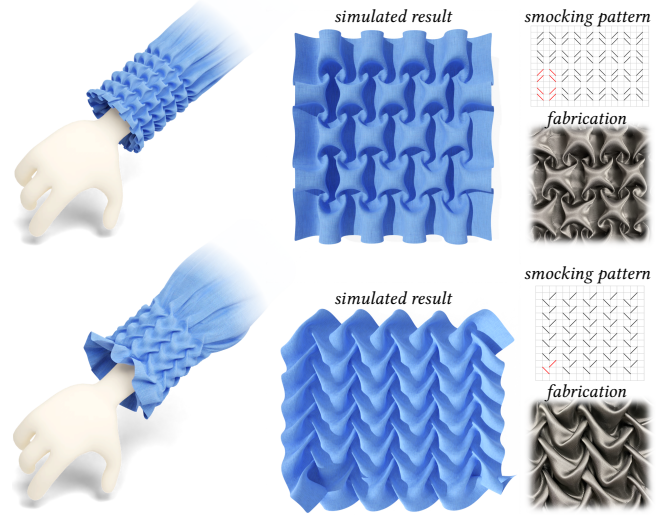

\input{sections/intro.tex}
\input{sections/relatedwork.tex}
\input{sections/background.tex}

\input{sections/method.tex}
\input{sections/results.tex}
\input{sections/conclusion.tex}

\bibliographystyle{ACM-Reference-Format}
\bibliography{bibliography}

\appendix
\input{appendix/appendix_metric.tex}

\input{appendix/appendix_ui.tex}
\input{appendix/appendix_formulation.tex}

\input{appendix/appendix_additional_res.tex}

\end{document}

%% file: header.tex
\usepackage{booktabs} % For formal tables
\usepackage{units} % for nice formatting of values with units

\usepackage[ruled]{algorithm2e} % For algorithms

\SetAlFnt{\small}
\SetAlCapFnt{\small}
\SetAlCapNameFnt{\small}
\SetAlCapHSkip{0pt}
\usepackage{hhline}
\usepackage{overpic}
\usepackage{bbold} % all ten digits with \mathbb
\usepackage{amsfonts} 
\usepackage{amsmath} 
\usepackage{pgfplots}
\usepackage{pgf,tikz}
\usepackage{tkz-euclide}
\usepackage{bm}
\usepackage{wrapfig}
\usepackage{multirow}
\usepackage{mathtools}
\usepackage{enumerate}
\usepackage{enumitem}
\usepackage{verbatim}

\usepackage{dsfont}
\newcommand{\R}{\mathds{R}} 
\renewcommand{\S}{\mathcal{S}}
\newcommand{\M}{\mathcal{M}}
\newcommand{\G}{\mathcal{G}}
\newcommand{\V}{\mathcal{V}}
\newcommand{\E}{\mathcal{E}}

\renewcommand{\L}{\mathcal{L}}
\renewcommand{\P}{\mathcal{P}}
\newcommand{\N}{\mathcal{N}}
\newcommand{\X}{\mathbf{X}}
\newcommand{\x}{\mathbf{x}}
\newcommand{\Su}{\S_u}
\newcommand{\Sp}{\S_p}

\usepackage{subcaption}

\newcommand{\etal}{et al.}

\usepackage{soul}

\usepackage{listings}
\usepackage{color, colortbl} %red, green, blue, yellow, cyan, magenta, black, white
\definecolor{mygreen}{RGB}{28,172,0} % color values Red, Green, Blue
\definecolor{mylilas}{RGB}{170,55,241}
\definecolor{ffzzcc}{rgb}{1,0.6,0.8}
\definecolor{mytbcol}{RGB}{175,227,246}
\definecolor{mypink}{RGB}{254, 197, 187}
\definecolor{tabyellow}{HTML}{7AAAF2} %yellow: ffd60a

\theoremstyle{definition}

\newtheorem{defi}{Definition}[section]

\usepackage{lipsum}

\newcommand{\myspace}{\mkern9mu}

\DeclareFontFamily{U}{mathx}{\hyphenchar\font45}
\DeclareFontShape{U}{mathx}{m}{n}{<-> mathx10}{}
\DeclareSymbolFont{mathx}{U}{mathx}{m}{n}
\usepackage{tipa}
\UndeclareTextCommand{\!}{T3}
\DeclareTextCommand{\tipaEXCLAM}{T3}{}
\DeclareRobustCommand{\!}{%
  \ifmmode\mskip-\thinmuskip\else\expandafter\tipaEXCLAM\fi
}

\usepackage{booktabs}       % toprules
\usepackage{mathrsfs}
\usetikzlibrary{arrows}
\usetikzlibrary{matrix}
\usetikzlibrary{positioning,calc,fadings}
\usetikzlibrary{backgrounds, fit}
\pgfplotsset{compat=1.14}
\usepackage{pgfplotstable}

\definecolor{myblue}{RGB}{23, 195, 178}
\definecolor{myyellow}{RGB}{255, 186, 8}
\definecolor{mygray}{rgb}{0.5,0.5,0.5}
\definecolor{myred}{RGB}{254, 109, 115}

\definecolor{colblue}{HTML}{7fc8f8}
\definecolor{colyellow}{HTML}{ffe45e}

\definecolor{colvu}{HTML}{99d98c}
\definecolor{colvp}{HTML}{7fc8f8}
\definecolor{coleu}{HTML}{ff8fab}
\definecolor{colep}{HTML}{ffd000} %ffee99}
\definecolor{coled}{HTML}{99d98c}

\newcommand{\arap}{{\textsc{arap}}}
\newcommand{\cipc}{{\textsc{C-IPC}}}
\newcommand{\arcsim}{{\textsc{ArcSim}}}

\newif\ifdraft

\draftfalse

\ifdraft
\newcommand{\JR}[1]{{\color{orange}[\textbf{JR:} #1]}}
\newcommand{\AS}[1]{{\color{blue}[\textbf{AS:} #1]}}
\newcommand{\OSH}[1]{{\color{magenta}[\textbf{OSH:} #1]}}

\newcommand{\todo}[1]{{\color{red}[\textbf{TODO:} #1]}}

\newcommand{\jr}[1]{{\color{orange}#1}}

\newcommand{\osh}[1]{{\color{orange}#1}}

\else
\newcommand{\JR}[1]{}
\newcommand{\AS}[1]{}
\newcommand{\OSH}[1]{}

\newcommand{\todo}[1]{}

\newcommand{\jr}[1]{{\color{black}#1}}

\newcommand{\osh}[1]{{\color{black}#1}}

\fi

\newcommand{\secref}[1]{Sec.~\ref{#1}}
\newcommand{\secreflist}[2]{Sections~\ref{#1} and~\ref{#2}}

\newcommand{\figref}[1]{Fig.~\ref{#1}}

\newcommand{\figreflist}[2]{Figures~\ref{#1} and~\ref{#2}}

\newcommand{\tabref}[1]{Table~\ref{#1}}

\newcommand{\eqnref}[1]{Eq.~(\ref{#1})}

\usepackage{pifont}% http://ctan.org/pkg/pifont
\newcommand{\cmark}{\ding{52}}%
\newcommand{\xmark}{{\ding{56}}}%
\usepackage{accsupp}
\newcommand{\checked}{%
  \BeginAccSupp{method=hex,unicode}%
  \text{\rlap{\hskip 2pt \raisebox{0.5pt}{\normalsize{$\backprime$}}}\cmark}%
  \EndAccSupp{}%
}

%% file: sections/intro.tex
\section{Introduction}\label{sec:intro}

Smocking is a surface embroidery technique used in textile design that serves two main purposes: 
it is highly decorative and provides ornamentation,
and it also has the practical benefit of allowing a close fit of the garment while maintaining a certain degree of stretch. Consequently, smocking is an artistic means of controlling a garment’s fullness, thereby creating more shape for the wearer~\cite{durand1979smocking,banner2022make}.
Moreover, smocking can act as reinforcement and insulation, padding over areas such as shoulders and chest to add durability to the garment~\cite{toplis2021hidden}.
Garments made with this technique are called \emph{smock-frocks} or \emph{smocks}.

\setlength{\columnsep}{3pt}%
\setlength{\intextsep}{0pt}%
\begin{wrapfigure}{r}{0.49\linewidth}
\centering
\begin{overpic}[trim=0cm 0cm 0cm 0cm,clip,width=1\linewidth,grid=false]{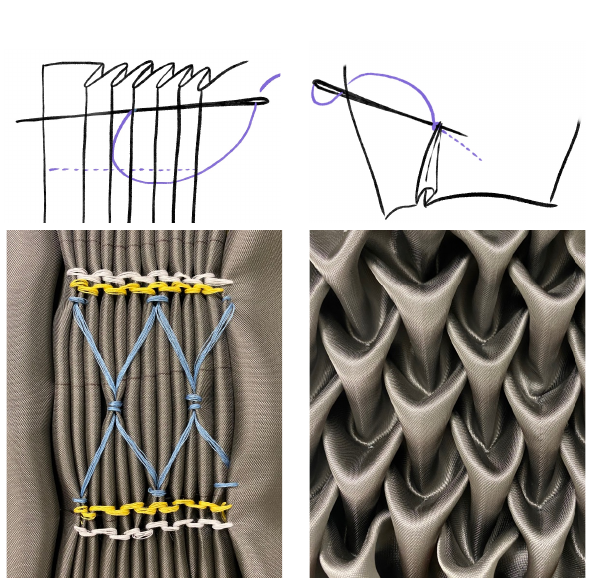}
\put(3,92.5){\footnotesize\emph{English smocking}}
\put(54,92.5){\footnotesize\emph{Canadian smocking}}
\end{overpic}
%\caption{\textbf{English v.s. Canadian Smocking}}\label{fig:intro:smocking_type}
\end{wrapfigure}
Smocking can be roughly categorized into two styles according to the embroidering process: \emph{English smocking}, where the pleating and stitching are done sequentially, and \emph{Canadian smocking}, where the stitching generates the pleating simultaneously. In traditional English smocking, the fabric is first folded into close and \emph{uniform} pleats,
and then the gathered threads are used as a guide to embroider rows of stitches through the pleats. The stitches remain visible and play the main decorative role, akin to standard 2D embroidery, whereas the pleating serves mainly to create a thicker base medium and generate folds when transitioning between smocked and non-smocked parts of the fabric.
%we first gather the fabric into close and \emph{uniform} pleats and then use the gathered threads as a guide to embroider rows of stitches through the pleats. 
The final appearance of an English smocking pattern is predictable, since the pleats are pre-folded and the embroidery patterns are determined by the alignment of stitches.
In Canadian smocking, the fabric is pleated by stitching it locally, connecting or ``pinching'' pairs of points in a special pattern. The stitches are invisible in the final result, and the decorative, \emph{geometric} texture is formed by the pleats themselves.
The final appearance of a Canadian smocking pattern is much harder to predict based on the given stitch pattern; the geometric texture is often revealed only
when the whole smocking process is completed, making its design a challenging trial-and-error process~\cite{efrat2016hybrid}.

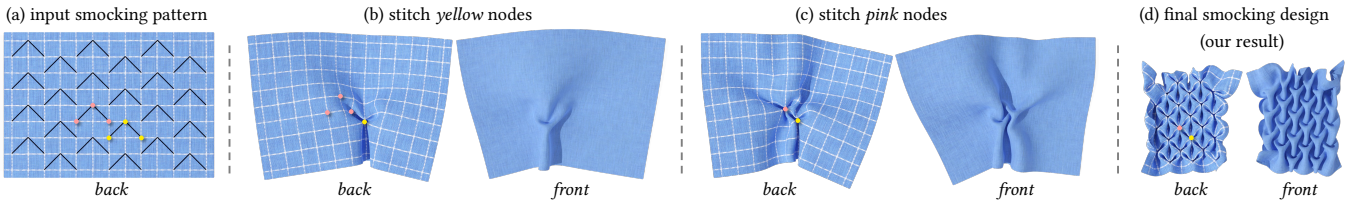
\begin{figure*}
    \begin{overpic}[trim=0cm 0cm 0cm -1.5cm,clip,width=1\linewidth,grid=false]{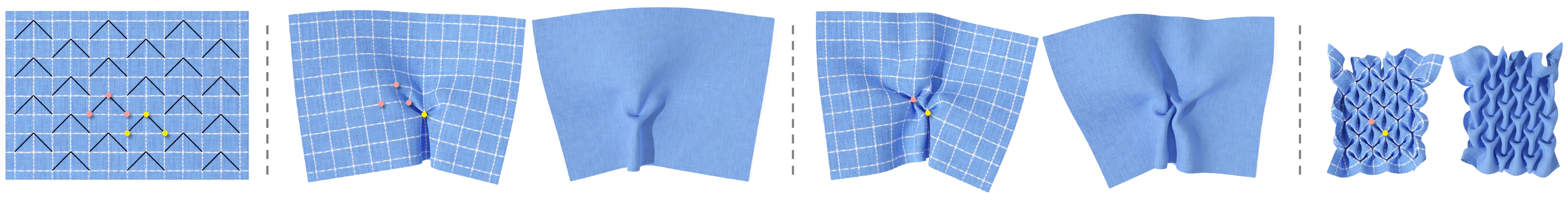}
    \put(0.5,13){\footnotesize (a) input smocking pattern}
    \put(27,13){\footnotesize (b) stitch \emph{yellow} nodes}
    \put(59,13){\footnotesize (c) stitch \emph{pink} nodes}
    \put(84.5,13){\footnotesize (d) final smocking design}
    \put(89,11){\footnotesize (our result)}
    \put(7,0){\footnotesize \itshape back}
    \put(25,0){\footnotesize\itshape back}
    \put(41,0){\footnotesize\itshape front}
    \put(56.5,0){\footnotesize\itshape back}
    \put(74,0){\footnotesize\itshape front}
    \put(87,0){\footnotesize\itshape back}
    \put(95,0){\footnotesize\itshape front}
    \end{overpic}\vspace{-6pt}
    \caption{\textbf{The smocking process.} A smocking pattern (a) consists of stitching lines (the polylines in black). Each stitching line needs to be contracted into a point by gathering and sewing together its nodes (b, c). Sewing all stitching lines reveals the final geometric texture (d). All pieces of fabric are shown in scale; note that smocking shrinks the starting piece of cloth significantly, since multiple points are pinched together.} \label{fig:intro:eg_smocking}
\end{figure*}

In this work, we therefore focus on \emph{Canadian smocking}, with the goal of creating a digital framework for design and preview, where users can explore and experiment with various stitching patterns and  visualize the smocking results without having to sew them physically (\figref{fig:teaser:sleeves}). 
%where the volumetric textures formed during pleating and stitching create \emph{non-manifold} geometric features, which makes the fabric itself more fascinating and the underlying mathematical problem more challenging and interesting. 
% At the same time, the unpredictability of smocking by its nature gives artists a lot of freedom and fun for smocking pattern exploration. At the same time, the unpredictability can be problematic since the exploration can take long time due to trial and error.
%
We investigate a mathematical formalization of the smocking problem and design an automatic and efficient algorithm to compute smocked fabric geometry based on input patterns. 
Surprisingly, approaching smocking modeling in a straightforward way as a constrained surface deformation or cloth simulation problem generally fails to satisfy all point-to-point stitching constraints and deliver faithful results, likely due to the intricate, essentially non-manifold structure of the design, the very large number of contacts and the high-curvature folds.
We instead formulate smocking as a graph embedding optimization problem that guides the cloth  deformation. We extract a coarse graph representing the fabric and the stitching constraints, and then derive the graph structure of the smocked result. We solve for the 3D embedding of this graph, which in turn reliably guides the deformation of the high-resolution fabric mesh.
%
%We instead formulate the smocking design as a graph embedding and shape deformation problem, where we solve for the 3D positions of each vertex on the fabric after some vertices have been stitched together. The final design is expected to have \emph{visually pleasant} pleats or 3D textures, that are \emph{realistic} enough.
%
To demonstrate the
accuracy of our method, we compare our results to real fabrications on a
variety of smocking patterns.

\textit{Contributions.} In this work we propose
(1) the {first} formalization of smocking design as a graph embedding and shape deformation problem,
(2) an efficient algorithm to compute the smocked fabric geometry from a given  pattern, enabling 
(3) an interactive tool for designing smocking patterns.

%% file: sections/relatedwork.tex
\section{Related Work}\label{sec:related}
\textit{Smocking.}
The word \emph{smock} comes from the Anglo-Saxon word \emph{smocc}, the name of an outer sack-like garment, later called \emph{smock frock}, which was worn over a farmer's other clothes to protect them from getting soiled~\cite{durand1979smocking,toplis2021hidden,spufford2017clothing}.
%
%A \emph{smock} generally refers to women's garments, while a \emph{frock} refers to men's clothing in the seventeenth century~\cite{spufford2017clothing}.
We refer the interested readers to a recent book \textit{``The hidden history of the smock frock''}~\cite{toplis2021hidden} for details.

Existing research on smocking is related to adult education \cite{bauer1992smocking}, 
psychology \cite{elbyaly2022investigating},
or bedroom decorations marketing~\cite{joseph2011lattice}.
\citet{efrat2016hybrid} propose a digital design tool for smocking, where users can tile predefined unit smocking patterns and then print them on fabric. Their system does not  visualize the result: the smocking itself needs to be completed manually by sewing the physical fabric. The authors note that the complexity of  smocking  makes it difficult to automate, and that predicting the smocking result of a given pattern is challenging.

\citet{lind2019smocked} explores the design of colorful jacquard woven patterns that serve as templates for the smocking stitches, such that the woven pattern shapes the fabric.
The jacquard patterns are customized for different smocking patterns. During the design and experimentation, the patterned fabrics have to be produced on a jacquard machine and then smocked manually in each design iteration.
\citet{kim2020study} emulates smocking in a step-by-step manner in a commercial virtual clothing software~\cite{CLO3D}, manually creating each stitch by simulating a tacking and folding step.
{Online creators  post similar manual techniques to model smocking details in digital garments~\cite{youtube2021smocking}}.
In contrast, our method is fully automatic and efficiently simulates the entire smocked shape.
%, which can be beneficial for parametric design~\cite{efrat2016hybrid} or textile design~\cite{lind2019smocked}.

\textit{Physically based cloth simulation.}
Following the pioneering work of \citet{terzopoulos1987elastically}, different elastic models have been studied for representing cloth dynamics, including 
finite element methods \cite{baraff1998large,narain2012adaptive}, mass-spring systems \cite{choi2002cloth,Liu2013FSM}, 
and yarn-level cloth simulation~\cite{kaldor2010efficient,cirio2014yarn}.
To accurately model folding or wrinkling of cloth, different collision handling techniques have been proposed~\cite{bridson2005simulation,tang2018cloth,wang2021gpu,Li2021CIPC}.
\citet{chen2021fine} propose a new model based on thin shells to model fine-scale wrinkling.
However, general-purpose cloth simulators struggle with the smocking task, because the pleats are mainly formed by stitches, whose pinching effect is challenging to capture by simulated wrinkles from cloth dynamics alone (see \figref{fig:background:blender} for an example).
FoldSketch~\cite{Li:2018:FoldSketch} is a dedicated inverse modeling system for folds and pleats, where the user sketches the desired folds on the draped 3D garment, and the algorithm adjusts the sewing pattern in order to reproduce them. While very effective for pleats and gathers that extend along \emph{one-dimensional} curved paths, this system is not suitable for sketching smocked pleats, which are arranged in a \emph{two-dimensional} pattern with many occlusions and overlaps.
The smocked appearance is not entirely independent of the fabric type, %i.e., the smocked results can vary somewhat for different types of fabric, 
but the dominant factor that governs the geometry and the regularity of the pleats is the structural stitching pattern, as opposed to the cloth parameters. 
In this work, we focus on the geometric formulation of smocking and assume the fabric to be roughly inextensible~\cite{Goldenthal:ChebyshevCloth:2007}.

\textit{Shape deformation.}
Instead of dynamically simulating cloth, smocking can be seen as an end state of a draping process that can be computed via surface deformation with positional constraints~\cite{DeformationTutorial:2009}.
%~\cite{sederberg86free,floater2003mean,ju2005mean,huang2006subspace,lipman2007gpu,ben2009variational,wang2015linear}. 
%See the surveys~\cite{botsch2007linear,nieto2013cage} for more detailed discussions of general shape deformation approaches.
Among the many surface deformation methods, as-rigid-as-possible deformation ({\arap})~\cite{arap} models least-squares isometric deformations, which can be used as a  stand-in for inextensible cloth. 
Deformation methods generally do not consider self collisions and contacts and do not do well on the smocking task when applied directly (\figref{fig:background:arap}). In our approach, we encapsulate the contacts, i.e., the sewing constraints, in the smocked graph structure, which guides the subsequent fine-grained deformation to a feasible and faithful configuration.

\textit{Digital design.} 
The subject of our work fits into digital design -- algorithms and systems that assist users in creating digital artifacts before physically fabricating them. Recent examples in this space include  origami~\cite{dudte2016programming}, kirigami~\cite{castle2014making,castle2016additive,jiang2020freeform}, %zippable~\cite{Schueller2018Zippables}, 
knittable stitch meshes~\cite{Wu2019knittable}, 
%auxetic structures~\cite{chen2021bistable}, 
3D weaving~\cite{ren20213d} and quilting~\cite{Leake2021,carlson2015algorithmic,igarashi2015patchy}, among many others. Here we focus on kirigami and quilting, which are more closely related to smocking.

Kirigami is a generalized origami technique where cutting out holes is allowed. It is often employed for regular tessellation patterns, similar to Canadian smocking, which are visually appealing and/or achieve particular mechanical behaviors~\cite{wang2017patterning,an2020programmable}.
\citet{castle2014making,castle2016additive} explore rules for cutting and folding kirigami, while \citet{jiang2020freeform} investigate the inverse problem of designing a kirigami pattern such that the deployed result is similar to a given 3D shape. 
The main difference between kirigami and smocking is the material: kirigami uses paper, which can neither stretch nor shear. 
In contrast, smocking is intended for woven fabric,
%, where warp and weft thread or yarn are weaved together. 
where a certain degree of shearing is possible even if the warp and weft yarns are inextensible. 
Fabric has a much richer set of degrees of freedom when deforming, so that smocking geometry is smoother and generally more varied compared to kirigami.

\citet{Leake2021} formalize the foundation paper piecing process, which is popular for constructing textile patchwork quilts based on printed  patterns.
This work encodes the pattern geometry via a dual hypergraph and investigates whether a given pattern is valid, i.e., pieceable.
The challenge is to solve for the order of placing the fabric pieces to meet the constraints posed by \emph{known} geometry.
In contrast, the challenge of formulating smocking is that the final 3D geometry is \emph{unknown} before the fabrication process is completed. We therefore need to build a graph that can capture the unknown structure information.

%
%The pioneers in this field were \citet{sederberg86free} with Free-Form deformation.
%To alleviate its assumption on regular lattices for spatial deformation specification, many cage-based deformation methods have been developed to deform the interior based on the cage~\cite{floater2003mean,ju2005mean,huang2006subspace,lipman2007gpu,ben2009variational,wang2015linear,botsch2007linear}.
%See a recent survey~\cite{nieto2013cage} for more detailed discussions.
%The linear approaches are robust with relatively low complexity, but the linearization can cause artifacts when large deformation occurs~\cite{botsch2007linear}. 
%This inspires the development of non-linear techniques~\cite{botsch2006primo,arap,botsch2007adaptive,chao2010simple}, which achieve high quality even for large deformations at a cost of efficiency.
%
%The smocking process can potentially be formulated as a low-dimensional control of deforming the input fabric mesh by pinching vertex pairs as annotated in the pattern.
%However, such local deformations are low-dimensional without taking the correlations among pinched vertex pairs into consideration, which makes it hard to obtain a regular deformed shape (see Fig.~\ref{fig:background:arap} for an example). 
%In this work we use a graph to bridge the pinched vertex pairs and to guide the deformation of fabric. 

%% file: sections/background.tex
\section{Preliminaries}\label{sec:background}

Canadian smocking consists of the following steps: (1) preparing a smocking pattern by drawing a grid and designing stitching lines on a piece of fabric; (2) \jr{gathering all grid vertices of one stitching line and sewing them together.}
%which contracts the sewing line into a point. 
The sewing is repeated for all stitching lines. Optionally, one can (3) fold the pleats formed during the stitching in a nicer way and iron the smocked pattern if necessary. See also \figref{fig:intro:eg_smocking} and the accompanying video.
In \secref{sec:background:formulation} we formalize each step and in 
Sections \ref{sec:background:arap} and \ref{sec:background:simulation} we discuss conceivable straightforward approaches.
% methods to solve the digital smocking problem.

\subsection{Notation and problem formulation}
\label{sec:background:formulation}

A classic smocking pattern consists of a piece of fabric with a 2D grid drawn on top of it, and a set of \emph{stitching lines}, each containing a list of grid nodes. A \emph{pleat} is formed when the nodes of one stitching line are gathered and stitched together into a single point. In practice, a stitching line is annotated by a set of connected line segments to visually separate different stitching lines from each other. The overview of the smocking process is illustrated in \figref{fig:intro:eg_smocking}.
%, which would make the fabrication process much easier than annotating each stitching line by a group of grid nodes. %See Fig.~\ref{fig:intro:eg_smocking}~(a) for an example of smocking pattern.
%
\begin{defi}
A \textbf{smocking pattern} $\P = \left(\G, \L\right)$ is a piece of fabric, represented by a graph $\G = \left(\V, \E\right)$ \jr{consisting of vertices $\V$ and edges $\E$}, annotated with a set of stitching lines $\L = \left\{\ell_i\right\}$.
A \textbf{stitching line} $\ell$ is a subset of vertices in $\V$ that are to be stitched together.
\end{defi}

\setlength{\columnsep}{5pt}%
\setlength{\intextsep}{1pt}%
\begin{wrapfigure}{r}{0.48\linewidth}
\centering
\includegraphics[width=1\linewidth]{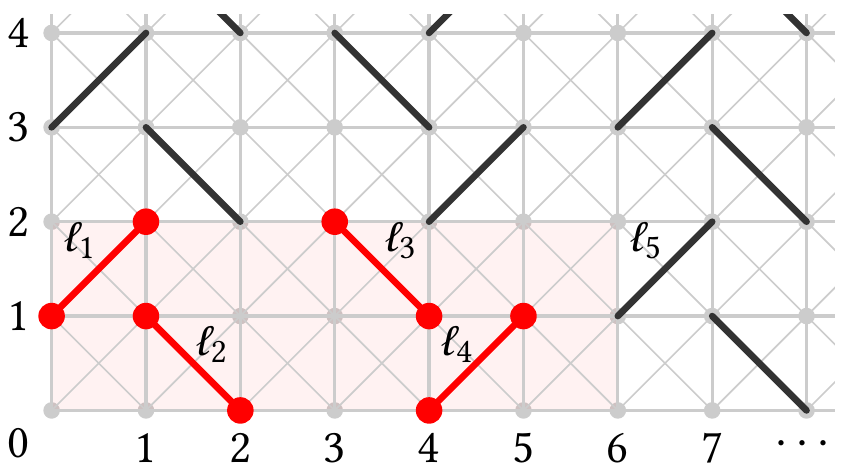}\vspace{-10pt}
\caption{A smocking pattern $\P$.}\label{fig:background:notation}
\end{wrapfigure}
\figref{fig:background:notation} shows a simple smocking pattern, represented by \jr{a} grid, where we denote the vertices as:
$\V = \big\{v_{0,0},\dots, v_{i,j},\dots, v_{n,m}\big\}.$
Then we can read the annotated stitching lines $\L~=~\big\{ \ell_i\big\}$ as:
$\ell_1~=~(v_{0,1} \, , v_{1,2})$, 
$\ell_2~=~(v_{1,1} \, ,v_{2,0})$, 
$\ell_3~= (v_{3,2} \, , v_{4,1})$,
$\ell_4~= (v_{4,0} \, , v_{5,1})$, 
$\ell_5~= (v_{6,1} \, , v_{7,2}), \dots$
% \begin{equation}\begin{split}
%     \ell_1 & = (v_{0,1} \, , v_{1,2}), \quad \ell_2  = (v_{1,1} \, , v_{2,0}), \quad \ell_3 = (v_{3,2} \, , v_{4,1}), \\
%     \ell_4  & = (v_{4,0} \, , v_{5,1}), \quad \ell_5  = (v_{6,1} \, , v_{7,2}), \quad \dots\dots
% \end{split}\end{equation}
%
In practice, a smocking pattern is obtained by tiling a \emph{unit} smocking pattern regularly on the fabric. We delineate the unit smocking pattern by a pink rectangle, and the stitching lines of the unit pattern are marked in red. {Note that we include the diagonals of the grid quads \jr{into the graph edges $\E$}, since they play a role in the subsequent graph embedding. }

During the smocking process, the vertices belonging to the same stitching line (e.g., $v_{0,1}$ and $v_{1,2}$) are gathered and stitched together.
A stitching line can consist of multiple line segments, in which case more than two points need to be stitched at the same time (see \figref{fig:intro:eg_smocking}~(a) for such an example). Our goal is to compute the 
\setlength{\columnsep}{5pt}%
\setlength{\intextsep}{0pt}%
\begin{wrapfigure}{r}{0.46\linewidth}
\centering
\includegraphics[width=1\linewidth]{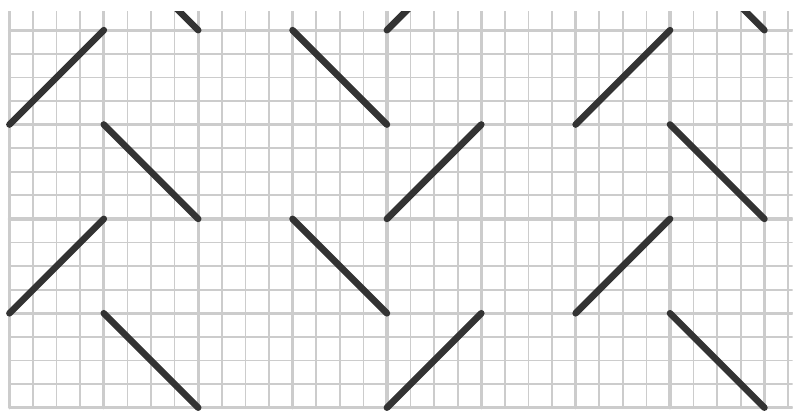}\vspace{-9pt}
\caption{Finer discretization $\widetilde{\P}$.}\label{fig:background:finer_grid}
\end{wrapfigure}
\emph{smocking design}, i.e., the 3D geometric 
texture shape resulting from any given smocking pattern. For this purpose, we use a higher-resolution representation of the fabric, $\widetilde{\P} = (\widetilde{\G}, \L)$, where $\widetilde{\G} = (\widetilde{\V}, \widetilde{\E})$ and $\V \subset \widetilde{\V}$, see~\figref{fig:background:finer_grid}.

\begin{defi}\label{def:smocking design}
The \textbf{smocking design} from a pattern $\P$ is a mesh $\widetilde{\M} = (\widetilde{\X}, \widetilde{\G})$ embedded in 3D, where $\widetilde{\X}\in\R^{\vert \widetilde{\V} \vert \times 3}$ stores the 3D positions $\x_p$ of all nodes $v_p\in\widetilde{\V}$ and satisfies $\x_p = \x_q,\  \forall v_p, v_q\in \ell_i,\ \forall \ell_i\in\L$. We can extract a non-manifold mesh representation $\M'$ from $\widetilde{\M}$ by removing the duplicated vertices and updating the topology of $\widetilde{\G}$ accordingly.
\end{defi}
%

%\OSH{Here, $\X$ and $\V$ have the same number of vertices, but $\V$ is depicted as very coarse. I think we need to add a comment that the grid is densified, and the stitching lines can go across many squares of the refined grid -- make a small inset figure? This would solve the issue of the subsequent two subsections, which show deformation and simulation on a fine grid, but we never showed or mentioned such a fine grid.}

The above definition  seems to imply that the smocking design can be computed using shape deformation or cloth simulation, but these approaches fall short.

%---------------------------------------------------------------------------------------

\subsection{Shape deformation using \arap}\label{sec:background:arap}

We can cast the smocking design computation as a shape deformation problem and easily adapt as-rigid-as-possible deformation (\arap)~\cite{arap} to obtain $\widetilde{\X}$:
%
%Specifically, we would like to update the 3D location $\x_p$ of each vertex $v_p\in\V$ by deforming the initial fabric in an as-rigid-as-possible way and at the same time to make sure that any pair of vertices belong to the same stitching line have the same location:
%
%
\begin{equation}\begin{split}\label{eq:background:arap}
    \min_{\widetilde{\X}\in\R^{\vert \widetilde{\V} \vert \times 3}} \ \ & \sum_{i} \min_{R_i \in SO(3)} \sum_{j\in\N(i)} w_{ij}\left\Vert \left(\x_i - \x_j\right) - R_i \left( \bar{\x}_i - \bar{\x}_j\right) \right\Vert_2^2, \\
    \text{s.t.}\quad & \left\Vert\, \x_p - \x_q\,\right\Vert_2 = \epsilon, \quad \forall (v_p,v_q)\in\ell_k,\ \forall \ell_k\in\L,
\end{split}\end{equation}
where $\bar{\x}_i$ denotes the known starting position of vertex $\x_i$ in the flat fabric piece, $\N(i)$ is the one-ring neighborhood of the $i$-th node in $\widetilde{\G}$, and $w_{ij}$ are the cotangent weights~\cite{meyer2003discrete}. 
%between the $i$-th and $j$-th node \OSH{only needed if the grid is triangulated somehow; if you only use grid neighbors then the weights are just 1}. 
%
The {\arap} energy encourages the edges in $\widetilde{\G}$ to stay rigid and maintain their length.
%after the deformation compared to the initial status of the fabric.
The deformation occurs due to stitching, which is modeled via the constraints. 
For two nodes in the same stitching line, we allow $\epsilon$ distance in the deformed state; $\epsilon$ can either be set to the thickness of the fabric or zero for simplicity.

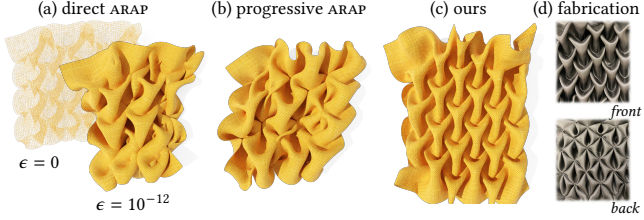
\begin{figure}[t]
    \centering
    \begin{overpic}[trim=0cm 0cm 0cm -1cm,clip,width=1\linewidth,grid=false]{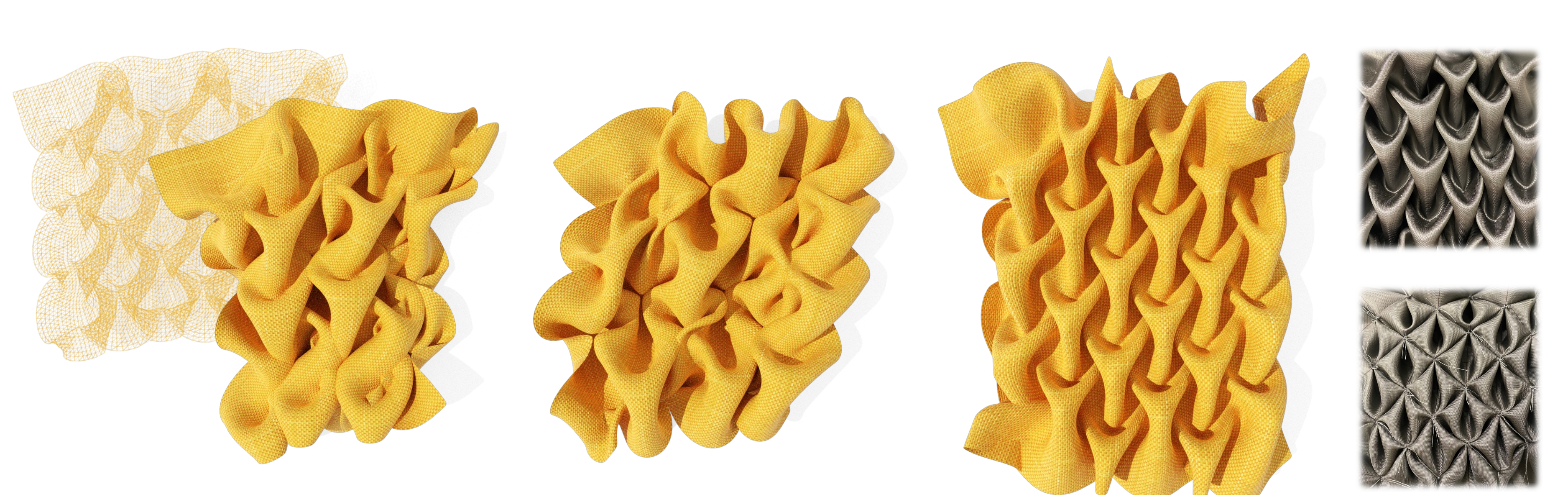}
    \put(6,30){\footnotesize(a) direct {\arap}}
    \put(3,6){\footnotesize $\epsilon = 0$}
    \put(15,-0.5){\footnotesize $\epsilon = 10^{-12}$}
    \put(33,30){\footnotesize (b) progressive {\arap}}
    \put(67,30){\footnotesize (c) ours}
    \put(83,30){\footnotesize (d) fabrication}
    \put(95,14.1){\scriptsize\itshape front}
    \put(95,-1){\scriptsize\itshape back}
    \end{overpic}\vspace{-3pt}
    \caption{Straightforward application of {\arap} deformation~\cite{arap} to the smocking pattern shown in \figref{fig:intro:eg_smocking} fails to recover the expected smocked geometry {(a)}. The value of $\epsilon$ stands for the maximum allowed distance between stitched nodes; when $\epsilon=0$, the deformed mesh stays planar. In (b) we show the result of gradually decreasing $\epsilon$ from \jr{half of the initial length to $0$}. Our method successfully computes the smocking design (c), closely matching its physical fabrication (d).}\label{fig:background:arap}
\end{figure}

\jr{We note that the constraints in \eqnref{eq:background:arap} are nonlinear and non-convex for $\epsilon \neq 0$, so we simplify the constraints by linearization.}
In \figref{fig:background:arap}, we show the smocking design results of {\arap} obtained using three different settings for the same pattern illustrated in \figref{fig:intro:eg_smocking}~(a). 
%Specifically, we solve \eqnref{eq:background:arap} with  \emph{(i)} $\epsilon = 0$, \emph{(ii)} $\epsilon = 10^{-12}$, and \emph{(iii)} gradually decreasing the value of $\epsilon$ \jr{until reaching $0$},
% \jr{using Lagrange-multiplier method with linearized constraints.}
%
\emph{(i)} When $\epsilon = 0$, \jr{we get planar positional constraints $\x_p = \x_q$. Since the initial mesh is planar,} the {\arap} deformation does not manage to get out of the planar configuration and produces a planar self-intersecting mesh (\figref{fig:background:arap}(a), left). 
%This happens because the initial locations for all the vertices are their 2D positions in the fabric, and the closest solution to \eqnref{eq:background:arap} is 2D rotations.
\emph{(ii)} Softening the stitching constraints by setting $\epsilon$ to a small non-zero value allows the optimization to find a non-planar local minimum, but the result is irregularly wrinkled (\figref{fig:background:arap}(a), right), likely because the stitching constraints overpower the optimization, not letting the surface relax. 
\emph{(iii)} In \figref{fig:background:arap}(b) we attempt a progressive strategy, where we iteratively reduce $\epsilon$ \jr{from half of the initial length of the stitching lines to $0$}. The result is better, but still not sufficiently regular.

\subsection{Cloth simulation} 
\label{sec:background:simulation}
\begin{figure}[!t]
    \centering
    \begin{overpic}[trim=0cm 0cm 0cm -2.26cm,clip,width=1\linewidth,grid=false]{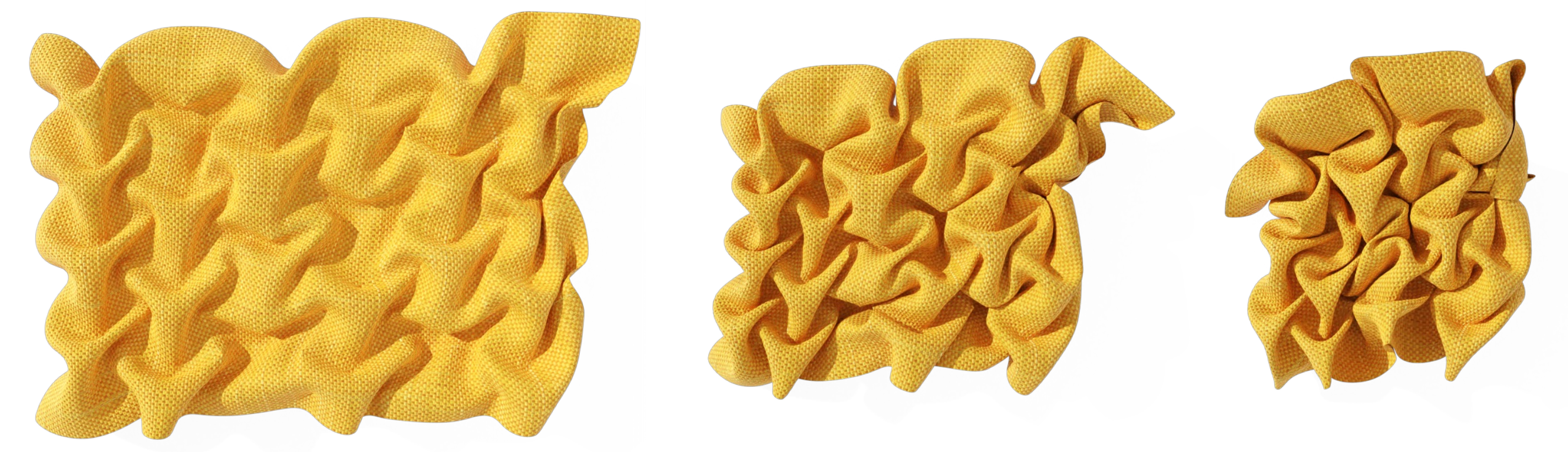}
    \put(12,30){\footnotesize  $\bar{e}_{25}$ = \unit[2.69]{cm}}
    \put(50,30){\footnotesize  $\bar{e}_{50}$ = \unit[1.26]{cm}}
    \put(80,30){\footnotesize  $\bar{e}_{75}$ = \unit[0.97]{cm}}
    \end{overpic}\vspace{-6pt}
    \caption{Simulated smocking design using Blender~\shortcite{blender}. We report the average length in centimeters $\bar{e}_k$ of all sewing lines after $k$ iterations, where $k = 25, 50, 75$ (converged), for a smocking pattern of size \unit[50]{cm} $\times$ \unit[70]{cm}. The stitching lines have initial length of \unit[5.5]{cm} and are expected to reach zero length after stitching.}
    \label{fig:background:blender}
\end{figure}

Computing the smocking design can naturally be formulated as a cloth simulation problem.
We use a popular simulator, {\textsc{cloth}}, implemented in Blender~\shortcite{blender}, which uses the point-based dynamics 
%(PBD) 
of a mass spring system~\cite{bridson2002robust} and incorporates contacts and friction.
%Specifically, the fabric is represented by a set of linear and angular springs that connect the mesh vertices. The cloth dynamics are solved by minimizing the elastic potential energy from the springs with additional constraints that consider collision and friction etc.
To simulate smocking, we add virtual linear springs of rest length $0$, connecting each pair of nodes in each stitching line.
In \figref{fig:background:blender}, we show the simulated result of the pattern in \figref{fig:intro:eg_smocking} over iterations. We observe a similar effect as with \arap: as the sewing lines become shorter, the textile becomes bunched up in an irregular fashion, because the simulator is not aware of the high level regularity of the smocking pattern. 
%
%Standard cloth simulators struggle with smocking because the general cloth parameters do not take the regularity of smocked patterns into consideration. 
%Instead, similar to {\arap}, only the positional constraints on the stitching points are added to the simulation, which is not sufficient to regularize the smocking process.
%
As a comparison, our result shown in \figref{fig:background:arap}(c) achieves regular and realistic smocking with zero-length sewing lines.
See \figref{fig:res:additional_res} for more examples.

\subsection{Observations \& challenges}
\jr{
Through our experiments, we have discovered that \osh{while} it is possible to \osh{find many} smocking designs that meet the criteria outlined in Definition~\ref{def:smocking design}, the definition itself falls short of adequately describing the desired voluminous and regular geometric texture \osh{preferred by} artists.
In practice, a smocking pattern is usually obtained by evenly tiling a unit pattern onto the fabric. As a result, one would anticipate achieving regular pleats with visually repetitive and consistent patterns.
Prior knowledge of regularity is crucial, as the absence of such knowledge causes both state-of-the-art shape deformation methods and cloth simulators to struggle with avoiding visually unpleasant or degenerated local minima.
\osh{At the same time}, formulating regularity in smocking is quite challenging, given that the geometry remains unknown until the fabrication process is completed. Additionally, imposing 3D geometry priors on a 2D input pattern is nontrivial.
}

% \subsubsection{PBD of Mass-Spring System}

% \subsubsection{Extended PBD}

%We draw the sewing lines in black line segments on the gray 2D grid and highlight the unit smocking pattern in red line segments. The red (blue) dots show some example of stitching (pleat) points in the fabric.

%\footnote{https://www.sidefx.com/docs/houdini/nodes/sop/vellumdrape.html}{Houdini}
%\footnote{https://docs.blender.org/manual/en/latest/physics/cloth/index.html}{Blender}

%% file: sections/method.tex
\section{Method}
\label{sec:mtd}
The experiments above reveal that to model smocking, we need to somehow impose a global regular structure on the fabric, because the deformation energies that are based on purely local differential properties have abundant local minima that lack symmetry and yield undesirable results. 
%
%From the previous discussions, we can observe that the main challenge is that only posing constraints to the stitching lines to deform or simulate the fabric is not sufficient enough to solve the smocking design. 
%More specifically, only $\mathcal{O}\left(\left\vert \L\right\vert\right)$ constraints are considered to solve for $\mathcal{O}\left(\left\vert\V\right\vert\right)$ where $\left\vert\L\right\vert \ll \left\vert\V\right\vert$, which leads to a significantly \emph{underdetermined} problem.
%In this case, both the shape deformation techniques and cloth simulators give undesirable results.
%One straightforward fix for this issue would be adding additional regularizers to make the system more constrained and to rule out undesirable solutions or local minima.
%However, it is extremely hard to propose regularizers on the smocking design since it is non-trivial to predict the forms of pleats from the smocking pattern. 
%and different smocking patterns will lead to different final designs in an unpredictable way as well.
%
To tackle this challenge, we solve the smocking design problem in two steps: We consider the input smocking pattern $\P$, defined on a \emph{coarse} representation of the fabric (see \secref{sec:background:formulation}) and optimize its 3D graph embedding. 
We then apply \arap, guided by the computed 3D embedding, on a finer representation of the fabric, $\widetilde{\P}$, to compute the final smocking design.
% We then compute the final smocking design for the same pattern by using a \emph{finer} grid representation of the fabric, $\widetilde{\P}$, by applying {\arap} guided by the coarse embedding of $\P$. 
% \AS{The last sentence is a bit confusing - maybe move the coarser rep. earlier to connect it with the prev. sentence? We then apply ARAP guided by the computed 3D embedding on a finer representation of the grid to compute the final smocking design.}
%
In the following, we explain our method in detail.

%----------------------------------------------------------------------------
%           construct smocked graph
%  .-.
% (o o) boo!
% | O \
%  \   \
%   `~~~'
%----------------------------------------------------------------------------

\subsection{Smocked graph extraction}
We define the \emph{smocked graph} from the input smocking pattern $\P = \left(\G = \left(\V, \E\right), \L = \left\{\ell_i\right\}\right)$ to represent the non-manifold structure of the resulting smocking design.
%
%Note that the smocking pattern $\P = \left(\,\G = \left(\V, \E\right), \L = \left\{\ell_i\right\}\,\right)$ discussed here is a \emph{coarser} level representation of the input fabric, that includes all the vertices in the stitching lines $\L$. We can regard it as the most abstract representation of the pattern and our goal is to extract information from $\P$ to guide the deformation of the final fabric $\widetilde{\P}$.
% 
We first categorize the vertices $v\in\V$ and edges $e\in\E$ as follows:
\begin{defi} 
A vertex $v\in\V$ in a smocking pattern $\P$ is called an \textbf{underlay vertex} if it belongs to a stitching line, i.e., $\exists \ell_i\in\L \ \text{s.t.\ } v\in\ell_i$, and it is called a \textbf{pleat vertex} otherwise.
\end{defi}
\begin{defi} 
An edge $e\in\E$ in a smocking pattern $\P$ is called 
a \textbf{degenerated edge} if its two endpoints belong to the same stitching line,  
an \textbf{underlay edge} if its two endpoints belong to two different stitching lines, and
a \textbf{pleat edge} otherwise.
\end{defi}

\begin{figure}[!t]
\centering
\begin{overpic}[trim=0cm 0cm 0cm -1cm,clip,width=1\linewidth,grid=false]{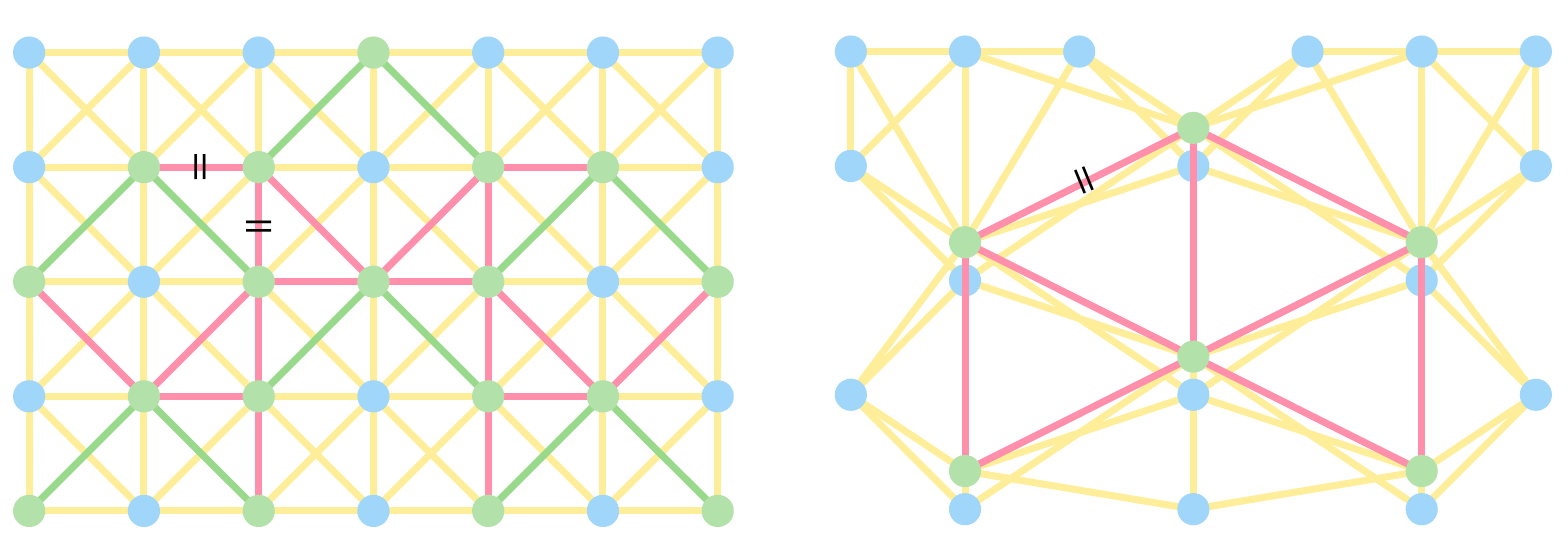}
\put(10,34){\small smocking pattern $\P$}
\put(65,34){\small smocked graph $\S$}
\end{overpic}\vspace{-6pt}
\caption{\emph{Left}: for the smocking pattern $\P$, we color the \textcolor{colvu}{underlay nodes} (resp.\ \textcolor{colvp}{pleat nodes})  in green (resp.\ blue), and the \textcolor{coleu}{underlay edges} (resp.\ \textcolor{colep}{pleat edges}) in pink (resp.\ yellow). The \textcolor{coled}{stitching lines} and the degenerated edges are colored in green. \emph{Right}: we show the corresponding smocked graph $\S$.}\label{fig:mtd:smocked graph}
\end{figure}

For example, in \figref{fig:mtd:smocked graph}
we construct the \emph{smocked graph} $\S = \left(\V_{\S}, \E_{\S}\right)$ from pattern $\P$ by fusing all underlay vertices sharing the same stitching line into one, deleting \emph{degenerated} edges and removing edges that become duplicate as a result of the fusing of underlay vertices. An example of such duplicate edges is marked with `$=$' in \figref{fig:mtd:smocked graph} (left); they correspond to a single edge in the smocked graph.
%Specifically, since multiple vertices belong to the \emph{same} stitching line $\ell_i$ will be stitched together, we only need to keep one arbitrary vertex from this stitching line (denoted as $v_{\ell_i}$) and ignore the rest vertices from $\ell_i$, since they are assumed to have the same location as $v_{\ell_i}$ after sewing. 
%See the smocked graph $\S$ extracted from $\P$ in \figref{fig:mtd:smocked graph}, where we remove (1) the duplicated underlay nodes in the stitching lines, (2) degenerated edges in the stitching lines, and (3) duplicated edges (e.g., see the edges annotated with `$=$' in \figref{fig:mtd:smocked graph}).

The smocked graph $\S$ is a subgraph of $\P$ that encodes the structure of the final smocked design; the vertices and edges of $\S$ inherit the pleat/underlay attributes (the colors in \figref{fig:mtd:smocked graph}) from $\P$.
{We denote the set of underlay (pleat) nodes in $\S$ as $\V_u$ ($\V_p$), and the set of underlay (pleat) edges in $\S$ as $\E_u$ ($\E_p$). We have
\begin{equation}
    \V_{\S} = \V_u \cup \V_p, \quad \E_{\S} = \E_u \cup \E_p.
\end{equation}
Note each vertex in $\V_u$ represents a single stitching line in $\P$, therefore $\vert \V_u \vert = \left\vert \L \right\vert$. 
}
We also define two important subgraphs of $\S$:
% old version of definition 4.3
%The subgraph of the smocked graph $\S$ induced by the underlay edges $\E_u \subset \E_{\S}$  is termed the \textbf{underlay graph}, denoted as $\Su =(\V_u, \E_u)$. It contains all underlay edges and their incident (underlay) vertices. The subgraph of $\S$ induced by the pleat edges $\E_p \subset \E_\S$ is termed the \textbf{pleat graph}, denoted $\Sp = (\V_\S, \E_p)$. It contains all pleat edges and their incident vertices, which can be pleat or underlay vertices.
\begin{defi}
The subgraph of the smocked graph $\S$ induced by the underlay edges is termed the \textbf{underlay graph}, denoted as $\Su$. It contains all underlay edges $\E_{u}$ and their incident underlay vertices $\V_{u}$. The subgraph of $\S$ induced by the pleat edges is termed the \textbf{pleat graph}, denoted $\Sp$. It contains all pleat edges $\E_{p}$ and their incident vertices, including all pleat vertices $\V_p$ and incident underlay vertices.
\end{defi}

\begin{figure}[!t]
\centering
\begin{overpic}[trim=0cm 0cm 0cm -1cm,clip,width=0.98\linewidth,grid=false]{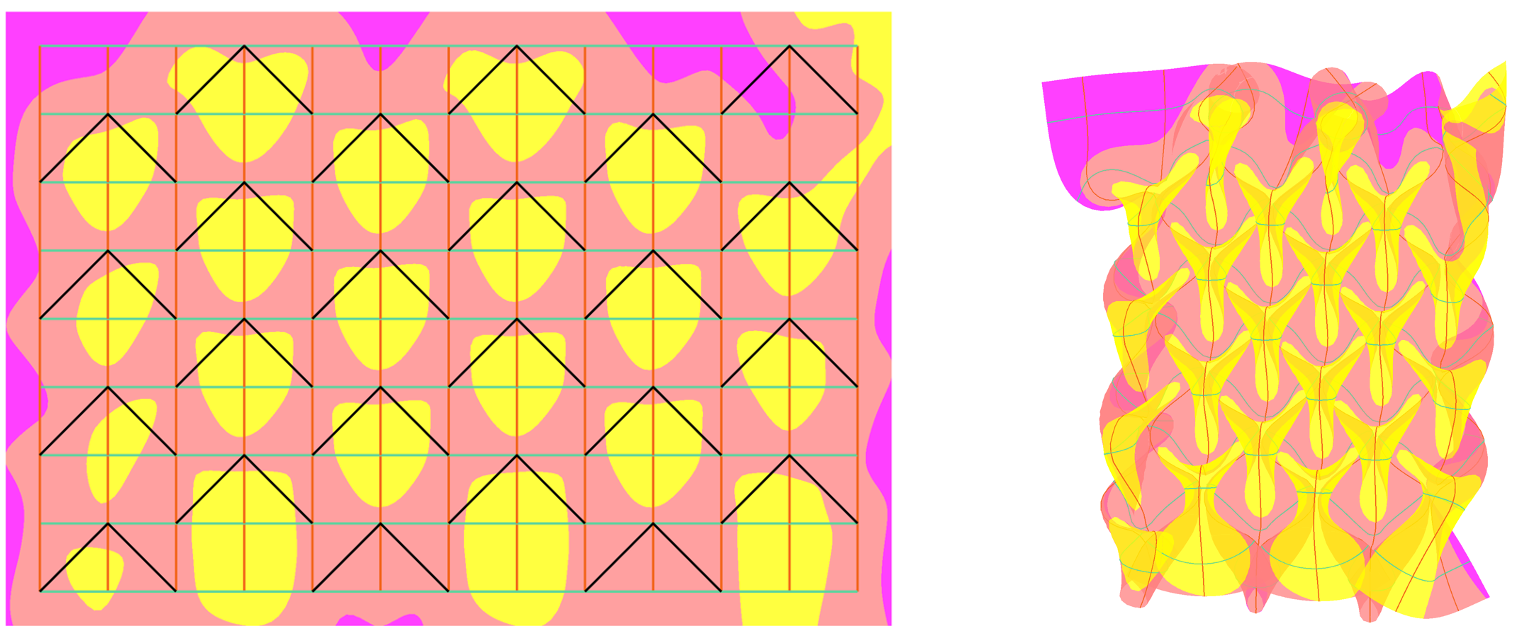}
\put(16,43){\small smocking pattern $\widetilde{\P}$}
\put(71,43){\small smocked design $\widetilde{\M}$}
\end{overpic}\vspace{-6pt}
\caption{Inspired by Lind~\shortcite{lind2019smocked}, we color the smocking pattern w.r.t.\ height after smocking: yellow highlights the regions that form the pleats, while pink highlights the regions that are almost hidden in the smocked result and form the underlay layer that supports the pleats.}
\label{fig:appendix:arrow_height_map}
\end{figure}

We can see that $\S = \Su \cup \Sp$. 
%are chosen from each stitching lines, and we then have $\left\vert \V_u \right\vert = \left\vert \L \right\vert$. Therefore, the number of the pleat nodes satisfies: $\left\vert \V_p \right\vert = \left\vert \V \right\vert - \sum_{\ell_i\in\L}\left\vert \ell_i \right\vert$.
%
\figref{fig:appendix:arrow_height_map} provides an intuition for the smocked graph: we color the smocking pattern by height after smocking, \jr{where yellow corresponds to large height where a pleat pops up (encoded by the pleat graph $\Sp$),  and pink signifies the underlay with low height that forms the base layer of the smocked design (encoded by the underlay graph $\Su$)}.

%----------------------------------------------------------------------------
%           embed smocked graph
%       .-.     
%     _/ ..\    
%    ( \  u/__  
%     \    \__) 
%     /     \   
%  __/       \  
% (   _._.-._/  
%  '-'                          
%----------------------------------------------------------------------------

\subsection{Smocked graph embedding}
The smocked graph is a distilled abstract representation of the smocking pattern, with the stitching constraints already satisfied. Our goal is to find a proper embedding of the smocked graph $\S$, i.e., assign a 3D position for each vertex $v\in\V_{\S}$, such that the embedded smocked graph forms an accurate and realistic 3D structure. We formulate this graph embedding problem as an optimization and design appropriate energies and constraints.

\begin{figure*}[!t]
    \centering
    \includegraphics[width=1\linewidth]{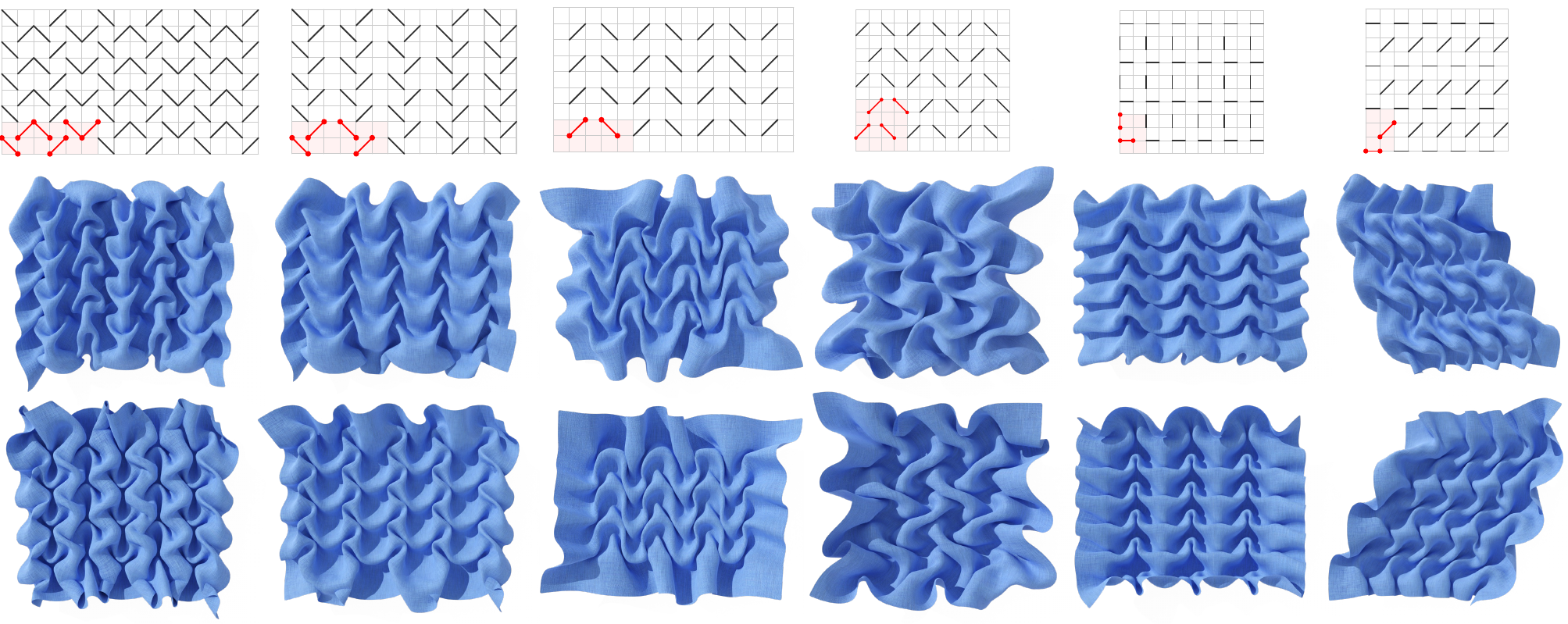}\vspace{-6pt}
    \caption{For different smocking patterns (top row), we show the computed smocking designs' front side ({middle row}) and back side ({bottom row}).}
    \label{fig:res:additional_res}
\end{figure*}

\subsubsection{Embedding distance constraint}\label{sec:mtd:embedding_distance} 
%We first discuss the constraints that the embedding should satisfy. 
We observe that the nodes in the smocked graph $\S$ are constrained by the underlying fabric and cannot move completely freely in space. 
For example, consider two vertices in the underlay graph $v_{\ell_i}, v_{\ell_j} \in \V_u$ (which originated from stitching lines $\ell_i$ and $\ell_j$ in $\P$) with embedded 3D coordinates $\x_{\ell_i}$ and $\x_{\ell_j}$, respectively.
Denote by $d(\cdot\, , \cdot)$ the geodesic distance on the fabric between two vertices, which is approximately equal to their Euclidean distance on the flat fabric.
The Euclidean distance between the embedded underlay vertices is constrained by:
\begin{equation}
    \left\Vert\x_{\ell_i} - \x_{\ell_j}\right\Vert_2  \le \min_{v_p \in \ell_i,\, v_q \in \ell_j} d(v_p, v_q),
\end{equation}
i.e., the \emph{shortest geodesic} distance on the fabric among any pair of stitching vertices on $\ell_i$ and $\ell_j$.
%
%This is a natural constraint, since all the vertices on $\ell_i$ (and on  $\ell_j$) are stitched into a single node. Therefore, the distance in the embedded smocked graph cannot be larger than the smallest geodesic distance attained, say, between $v_p$ and $v_q$, otherwise the fabric would be ripped apart between $v_p$ and $v_q$. 
%
%
For simplicity of exposition, here we assume the fabric cannot stretch. 

\setlength{\columnsep}{5pt}%
\setlength{\intextsep}{0pt}%
\begin{wrapfigure}{r}{0.42\linewidth}%
\centering
\begin{overpic}[trim=0cm 0cm 0cm 0cm,clip,width=1\linewidth,grid=false]{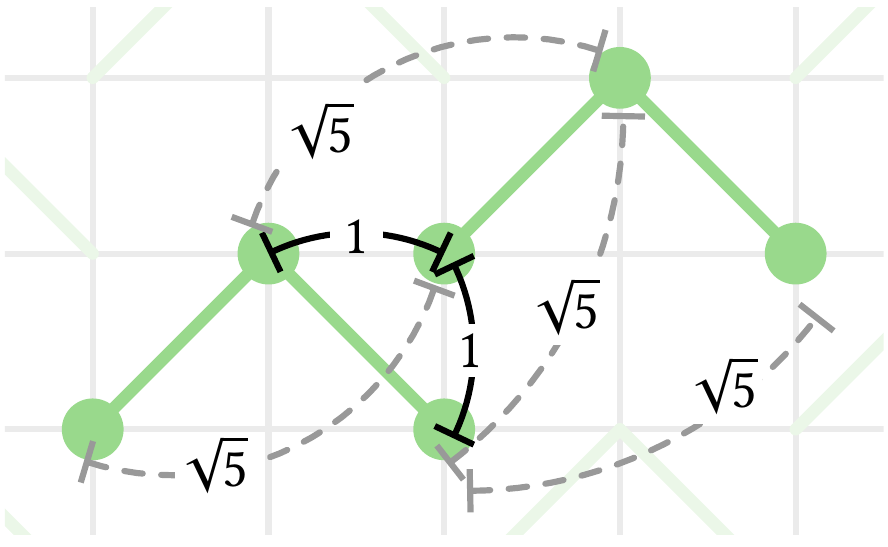}
\put(10,25){\textcolor{colvu}{$\ell_i$}}
\put(85,45){\textcolor{colvu}{$\ell_j$}}
\end{overpic}\vspace{-6pt}%
\caption{The $d_{i,j}$ constraint.}\label{fig:mtd:eg_dij}%
\end{wrapfigure}%
For example, in \figref{fig:mtd:eg_dij} we can see that the constraint for the pair of stitching lines is $d_{i,j} = 1$. If the embedded positions for the two corresponding underlay nodes had a distance larger than $1$, and assuming infinite stiffness, the fabric would tear. 

We can compute such an \emph{embedding distance constraint}, denoted as $\left\Vert\x_i - \x_j\right\Vert_2 \le d_{i,j}$, for any pair of vertices $(v_i, v_j) \in \V_{\S}\times \V_{\S}$. We have $d_{j,i} = d_{i,j}$ and
\begin{equation}\label{eq:mtd:d_ij}
     d_{i,j} = \left\{ \begin{array}{rl}
      \min\limits_{v_r \in \ell_{v_i},\, v_q \in \ell_{v_j}}  d\left(v_r \, , v_q\right), & \text{if} \myspace v_i, v_j\in \V_u, \\
      \min\limits_{v_r \in \ell_{v_i}} d\left(v_r \, , v_j\right), &\text{if} \myspace v_i \in \V_u, \ v_j \in \V_p, \\
      d\left(v_i\, , v_j\right)\mkern3mu , &\text{if} \myspace  v_i, v_j\in \V_p,
\end{array} \right.
\end{equation}
where $\ell_{v_i}$ denotes the stitching line in $\P$ that corresponds to the underlay node $v_i$ in the smocked graph $\V_\S$.
% \begin{cases} 
%       \min_{v_p \in \ell_i,\, v_q \in \ell_j} d(v_p, v_q) & v_i, v_j\in V_u \\
%       \min_{v_p \in \ell_i} d(v_p, v_j) & v_i \in V_u, v_j \in V_p \\
%       d(v_i, v_j) &  v_i, v_j\in V_p 
%    \end{cases}

{
Ideally, we wish to find a {valid} embedding of the smocked graph such that all vertex pairs satisfy the embedding distance constraints. Intuitively, it means that if we physically pin the vertices of the pattern annotated on real fabric to their embedding coordinates, there is no risk of the fabric tearing.
Note that such a valid embedding always exists, since we can simply put all vertices in one point, and all constraints are satisfied in this trivial solution.
}

% \begin{defi}
%     We call an embedding of the smocked graph \textbf{valid}, if we can deform the fabric in real life such that the vertices in the annotated pattern can be fixed to the corresponding embedded position without breaking the fabric. 
% \end{defi}

% \begin{observation}
%     Any embedding with all vertex pairs satisfying the embedding distance constraints discussed above is valid. There are \textbf{infinitely many} valid embeddings of a smocked graph. \OSH{Why is it important that there are infinitely many such embeddings? Maybe we also need to comment that such embeddings exist? put all vertices in one point, then all constraints are satisfied.}
% \end{observation}

\subsubsection{Maximizing embedding energy} 
While the distance constraints determine the \emph{search space} of valid embeddings, we need an objective function to find a desirable embedding and avoid the trivial solution where all nodes get assigned the same location. 
%Otherwise, we can easily get trapped at a trivial solution where all nodes in the smocked graph are cluttered together (i.e. having the same position in the extreme case).
%
%In this case, it is unlikely to display visually pleasant and realistic 3D textures. 
We wish to encourage all nodes to stay as far from each other as possible. Let $\x_i \in \R^3$ be the embedded position of vertex $v_i \in \V_\S$, and $\X$ the stacking of all these positions. We can formulate the following optimization problem for the embedding of $\S$:  
\begin{equation}\begin{split}\label{eq:mtd:prob:trivial}
        \max_{\X \in \R^{|\V_\S|\times 3}} \quad & \sum_{\forall i\neq j} \Vert \x_i - \x_j \Vert_2 \\
        \text{s.t.} \quad & \Vert \x_i - \x_j \Vert_2 \leq d_{i,j} \quad\forall i\neq j.
\end{split}\end{equation}
Despite the simple formulation, this is a difficult, non-convex problem with $\frac{1}{2}{n(n-1)}$ hard non-convex inequality constraints defined on every pair of $n = \left\vert \V_\S\right\vert$ vertices in the smocked graph.

\subsubsection{Simpler formulation as graph embedding}
Our solution is to relax the optimization problem in \eqnref{eq:mtd:prob:trivial} into an easier to solve form, where the inequality constraints are replaced by a significantly smaller set of (possibly soft) {equality} constraints, leading to a classical {graph embedding} problem. 
For simplicity, here we assume the smocking pattern $\P$ is such that the underlay graph $\Su$, as well as the pleat graph $\Sp$,
%of the corresponding smocked graph $\S$ 
is non-empty and has exactly one single connected component. 
%(i.e., has exactly one single connected component).
We discuss other cases in \secref{sec:mtd:non-regular-pattern}.
Moreover, we are particularly interested in \emph{well-constrained} smocking patterns that produce pleasant patterns when fabricated. These patterns have balanced and structured underlay region, such that the pleats are constrained to be regular.
See \secref{sec:usp} for further discussion.

We observe that the pleats that form the geometric textures are constrained by the underlay (see \figref{fig:appendix:arrow_height_map}), 
while the underlay graph encodes the overall structural information and determines the final appearance. 
The distance bounds in \eqnref{eq:mtd:d_ij} hint that the local geometry around the stitching lines gets significantly changed by smocking, since they are pinched together.
%, and their maximum embedding distance is shortened since they are pinched together.
The underlay nodes are heavily constrained by each other and determine the overall smocking structure.
On the other hand, the pleat nodes have more freedom to move in 3D, since they are not stitched to any other points on the fabric and they are expected to form the volumetric 3D textures.
This inspires us to split the embedding problem into two sub-problems: the embedding of the underlay and the pleat graphs in two separate steps, where the embedding of the underlay is employed to constrain the embedding of the pleat nodes.
 
\subsubsection{Embedding the underlay graph}\label{sec:mtd:embed-underlay}
We first try to find the embedding $\X_u$ for the underlay graph $\Su=\left(\V_u \,, \E_u\right)$. 
We observe that for a well-constrained smocking pattern, the underlay graph is \emph{planar} (see the pink subgraph in \figref{fig:mtd:smocked graph} (right)) and therefore can be embedded in 2D.
The maximizing embedding energy encourages large distances between nodes, while the distance constraints bound them by $d_{i,j}$, so we propose to find the 2D embedding of the underlay graph by relaxing \eqnref{eq:mtd:prob:trivial} as:
\begin{equation}\label{eq:mtd:underlay}
\min_{\X\in\R^{|\V_u|\times 2}}  \quad \sum_{(v_i, v_j)\in\E_u} \left(\left\Vert \x_i - \x_j \right\Vert_2 - d_{i,j}\right)^2.
\end{equation}
The relaxation is justified by the fact that in reality the distance constraints are not as stringent, as even the stiffest fabric can stretch a bit.
Instead of considering every pair of underlay nodes, here we only consider the adjacent ones. 
Recall that the distance constraint $d_{i,j}$ is derived from the smocking pattern $\P$, which represents a flat fabric, a connected 2-manifold. Thus it is reasonable to only consider the constraints in local neighborhoods; the constraints beyond 1-ring neighbors should fall in line due to the metric structure and therefore can be ignored. 
See Appendix~\ref{appendix:metric_embedding} for more detailed discussions.

% \begin{figure}[!t]
%     \centering
%     \begin{subfigure}[b]{0.45\linewidth}
%     \includegraphics[width=1\linewidth]{figures/eg_arrow_embedded_graph.pdf}
%     \caption{A 3D embedding of $\S$}\label{fig:mtd:embeded-graph}
%     \end{subfigure}
%     \hfill
%     \begin{subfigure}[b]{0.45\linewidth}
%         \begin{overpic}[trim=0cm 0cm 0cm 6.5cm,clip,width=1\linewidth,grid=false]{figures/eg_arrow_finer.pdf}
%         \end{overpic}\vspace{-5pt}
%         \caption{Smocking design}\label{fig:mtd:design}
%     \end{subfigure}\vspace{-5pt}
    % \caption{Embedding of the coarser (left) and the finer (right) discretization of the smocking pattern shown in \figref{fig:mtd:smocked graph}.}
% \end{figure}

\begin{figure*}[!t]
 \begin{overpic}[trim=0cm 0cm 0cm -2cm,clip,width=1\linewidth,grid=false]{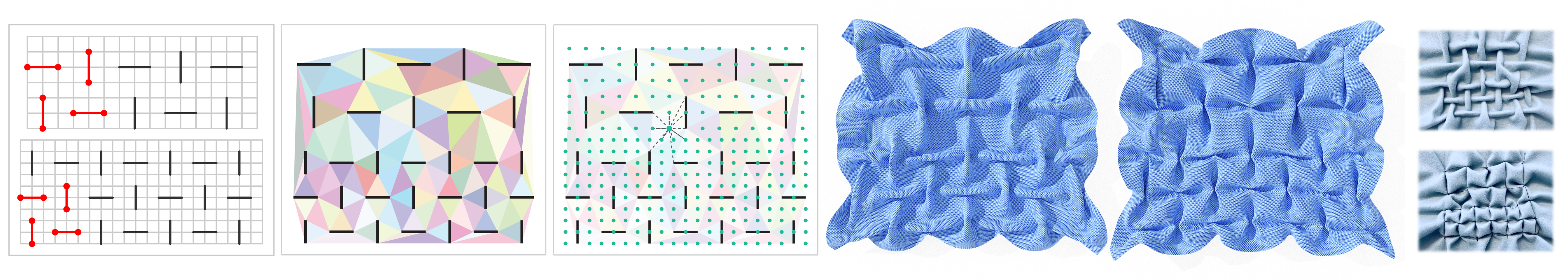}
 \put(1.8,17.5){\footnotesize  (a) input stitching lines}
 \put(18.3,17.5){\footnotesize (b) underlay via Delaunay}
 \put(37,17.5){\footnotesize (c) sample pleat nodes}
 \put(65,17.5){\footnotesize (d) our results}
 \put(60,0.2){\footnotesize\itshape front} 
 \put(79,0){\footnotesize\itshape back} 
 \put(90,17.5){\footnotesize (e) fabrication}
 % \put(97,9){\footnotesize\itshape front} 
 % \put(97,1){\footnotesize\itshape back} 
 \end{overpic}\vspace{-4pt}
 \caption{\textbf{Grid-free smocking design}. (a) We combine two smocking patterns of different scales. It is challenging to design a single regular grid to accommodate all stitching lines simultaneously. (b) Instead of constructing a new grid, we compute a Delaunay triangulation conditioned on the input stitching lines, which gives us the {underlay} graph. (c) We then sample pleat nodes w.r.t.\ the input stitching lines (green dots) and add connectivity between the pleat and the underlay nodes via local Delaunay triangulation (see the dashed lines for some examples), which gives us the {pleat} graph. (d) We can then embed the smocked graph and solve for the smocking design. (e) Physical fabrication of the pattern in (a). }\label{fig:mtd:grid_free}
\end{figure*}
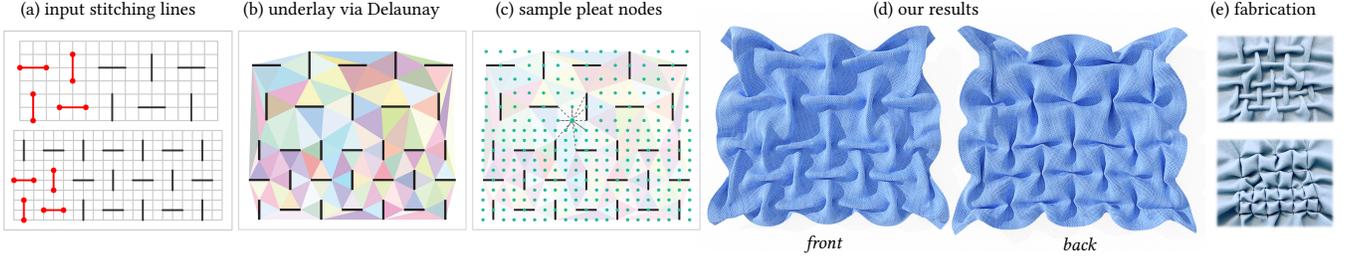

\subsubsection{Embedding the pleat graph}\label{sec:mtd:embed-pleat}
We find the 3D embedding $\X_p$ for the pleat nodes $\V_p$ in the pleat graph $\Sp$ using a similar formulation:
\begin{equation}\label{eq:mtd:pleat}
\min_{\X\in\R^{|\V_p|\times 3}} \sum\limits_{(v_i, v_j)\in\E_p} \left(\left\Vert \x_i - \x_j \right\Vert_2 - d_{i,j}\right)^2,
%- \, w  \sum_{\forall i \ne j} \left\Vert \x_i - \x_j \right\Vert_2
\end{equation}
where we want to stretch each pleat edge to its upper bound $d_{i,j}$ to maximally spread the overall embedding.
Recall that some of the pleat edges in $\E_p$ connect a pleat node and an underlay node. For these edges, the underlay nodes are fixed to the previously solved positions $\X_u$, and only the positions of the pleat nodes are involved in the optimization step in \eqnref{eq:mtd:pleat}.
\jr{The pleat vertices are initialized with the same height to help break the symmetry ambiguity.}
%
% \figref{fig:mtd:embeded-graph} shows the computed 3D embedding of the smocked graph $\S$ illustrated in~\figref{fig:mtd:smocked graph}.
%See Sec.~\ref{sec:res:formulation} for more justifications.

%----------------------------------------------------------------------------
%           ARAP 
%----------------------------------------------------------------------------

\subsection{Smocking design from embedded smocked graph}\label{sec:mtd:arap}

Having solved for the embedding $\X_u \cup \X_p$ of the smocked graph $\S$, we immediately deduce the geometry of the coarse smocking pattern $\P$: all vertices in a stitching line $\ell_i$ have the same location as the embedded position in $\X_u$ of the respective underlay vertex $v_{\ell_i}$, and all remaining (pleat) vertices in $\P$ have their locations in $\X_p$, corresponding to the vertices in the pleat graph.
%
% Now we have solved the embeddings $\X_u$ and $\X_p$ for the underlay nodes and the pleat nodes in the smocked graph $\S$, which is extracted from the coarser smocking pattern $\P$. 
% We can easily find the deformed positions for each vertex $v \in \V$ in the smocking pattern $\P$. Specifically, all the nodes in the stitching line $\ell_i$ have the same location as the embedded position in $\X_u$ for the vertex $v_{\ell_i}$. Each pleat node in $\P$ can find its location in $\X_p$ for the corresponding
% %
% vertex in the pleat graph.
%
To compute the smocking design in finer resolution, we run {\arap} on the high-resolution smocking pattern $\widetilde{\P}$, constraining the positions of the vertices of $\P$ to their embedded locations.
% \figref{fig:mtd:design} shows the final smocking design of the smocking pattern in \figref{fig:mtd:smocked graph} based on the embedded smocked graph shown in \figref{fig:mtd:embeded-graph}. The nodes colored in blue/green in the finer grid are fixed to their computed positions as shown in \figref{fig:mtd:embeded-graph} and \figref{fig:mtd:design}, and the positions of the remaining nodes in the finer grid are solved via {\arap}.
\figref{fig:res:additional_res} shows further results on several interesting smocking patterns.

%----------------------------------------------------------------------------
%           non-regular smocking pattern
%----------------------------------------------------------------------------

% \begin{figure*}[!t]
%     \centering
%     \begin{overpic}[trim=0cm 0cm 14cm -1cm,clip,width=1\linewidth,grid=false]{eg_basket.png}
%     \put(11,17.6){\footnotesize\bfseries (a) smocking pattern \underline{without} pleat nodes}
%     \put(63,17.6){\footnotesize\bfseries (b) add \underline{additional} pleat nodes}
%     \put(25,1){\footnotesize\bfseries\itshape front} 
%     \put(43,1){\footnotesize\bfseries\itshape back} 
%     \put(76.6,1){\footnotesize\bfseries\itshape front} 
%     \put(94.6,1){\footnotesize\bfseries\itshape back} 
%     \end{overpic}\vspace{-12pt}
%     \caption{For a smocking pattern that does not have any pleat nodes (except some free boundary nodes) as shown in (a),  our algorithm can still produce reasonable simulation, but the volumetric textures are less pleasant since no constraints on the pleats are considered during the optimization. An easy fix would be adding additional pleat nodes at the center of each grid cell and adding pleat edges accordingly, as shown by the blue nodes and dashed lines in (b) respectively, which leads to more regular and realistic textures. Note that our algorithm produces similar positions for the underlay nodes in both (a) and (b). \jing{make it single column.}}
%     \label{fig:mtd:eg_basket}
% \end{figure*}

\subsection{Generalizations: Non-regular smocking patterns}\label{sec:mtd:non-regular-pattern}

\subsubsection{Grid-free smocking design}
\label{sec:mtd:grid-free}
We can further generalize our algorithm to more challenging cases where the stitching lines are distributed non-uniformly, making it hard to extract a regular grid to abstract the smocking pattern.
In this case, we can construct a graph from the input stitching lines based on Delaunay triangulation~\cite{delaunay1934sphere,lee1980two} and use it to compute the smocking design as before.
%Specifically, as discussed in 
%\secreflist{sec:mtd:embed-underlay}{sec:mtd:embed-pleat}, the constructed smocked graph from the grid is supposed to encode the local \emph{neighborhood} information which is used to formulate the optimization problems. 
% We first sample some pleat nodes on the fabric.
% We then use Delaunay triangulation to establish the connectivity between the nodes in the stitching lines and the pleat nodes, which leads to a smocked graph. We can then similarly embed the smocked graph and solve for the smocking design.
%
See \figref{fig:mtd:grid_free} for an overview and Algorithm~\ref{alg:mtd:grid_free} in Appendix~\ref{appendix:formulation} for further details.

\setlength{\columnsep}{5pt}%
\setlength{\intextsep}{0pt}%
\begin{wrapfigure}{r}{0.56\linewidth}
\centering
\begin{overpic}[trim=0cm 0cm 0cm 0cm,clip,width=1\linewidth,grid=false]{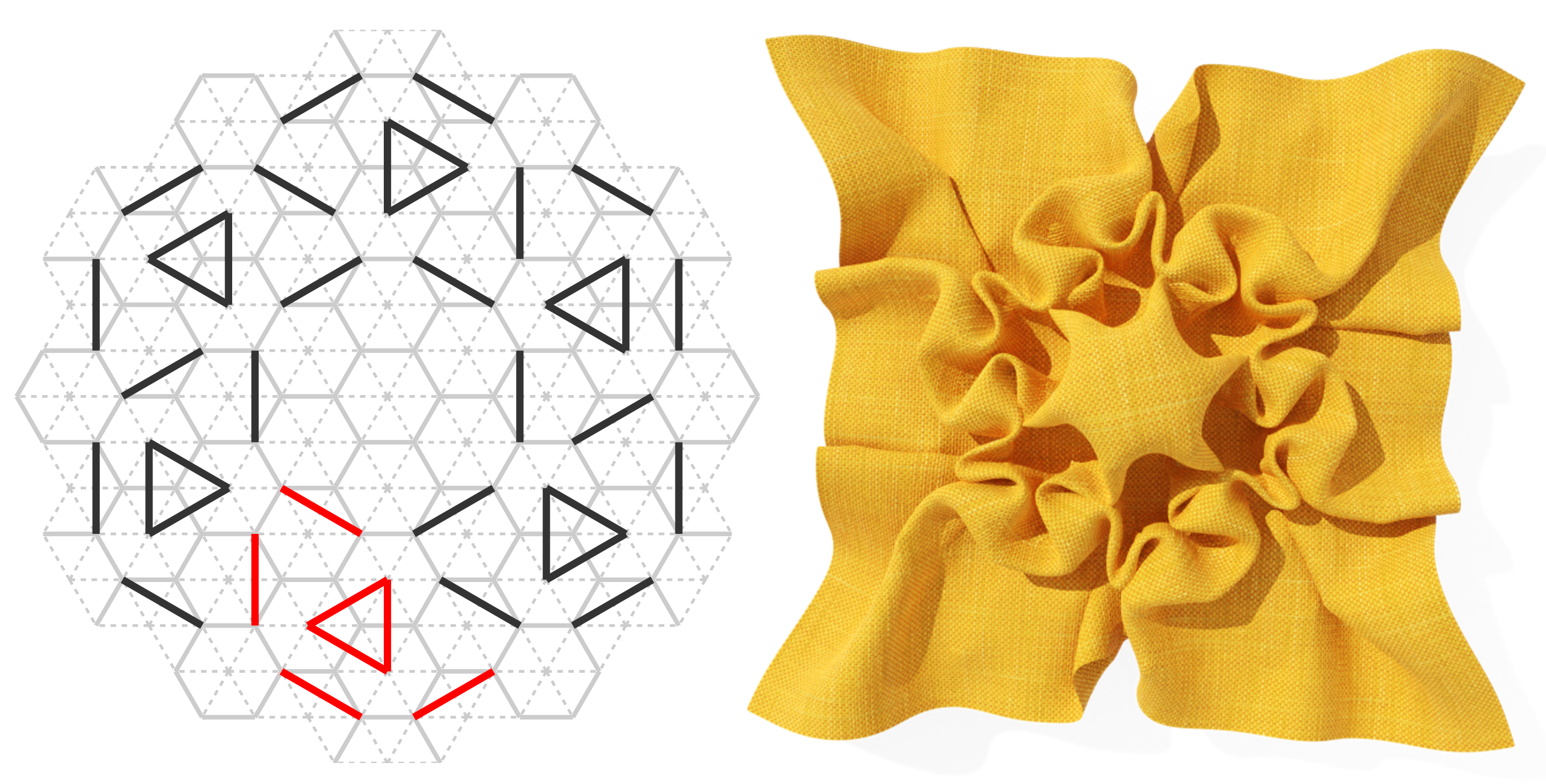}
\end{overpic}\vspace{-9pt}
\caption{{Honeycomb grid.}}\label{fig:mtd:honeycomb}
\end{wrapfigure}
\subsubsection{Honeycomb grid}
\label{sec:mtd:honycomb}
Our formulation does not depend on the exact shape of the grid, we just need to construct the graph $\G $ of the grid, so we can easily apply our algorithm to different types of grids. See \figref{fig:mtd:honeycomb} for an example, where the smocking pattern is defined on a hexagonal grid, and the unit pattern (in red) is tiled in a cyclic fashion.

\begin{figure}[!t]
    \centering
    \begin{overpic}[trim=0cm 0cm 0cm -1cm,clip,width=1\linewidth,grid=false]{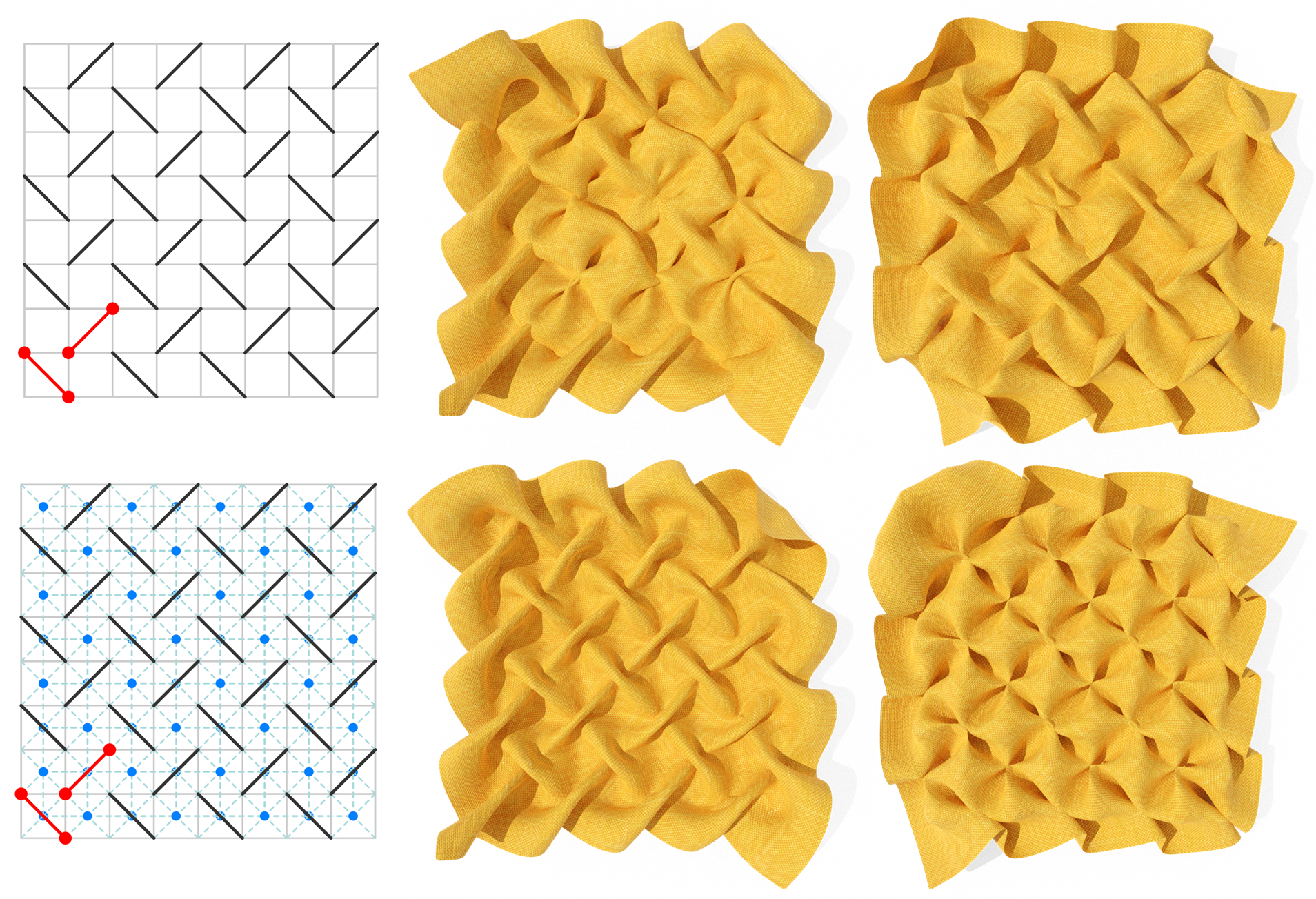}
    \put(1,68){\footnotesize (a) smocking pattern {without} pleat nodes}
    \put(1,34){\footnotesize (b) with {additional} pleat nodes}
    \put(46,0){\footnotesize\itshape front} 
    \put(80,0){\footnotesize\itshape back} 
    \end{overpic}\vspace{-4pt}
    \caption{For a smocking pattern that does not have any pleat nodes (except some free boundary nodes) as shown in (a),  our algorithm can still produce a reasonable result, but the geometric texture features are less pleasing, since no constraints on the pleats are considered during the optimization. An easy fix is to insert additional pleat nodes (colored blue) and pleat edges (dashed lines), leading to a more regular and realistic result.}
    \label{fig:mtd:eg_basket}
\end{figure}
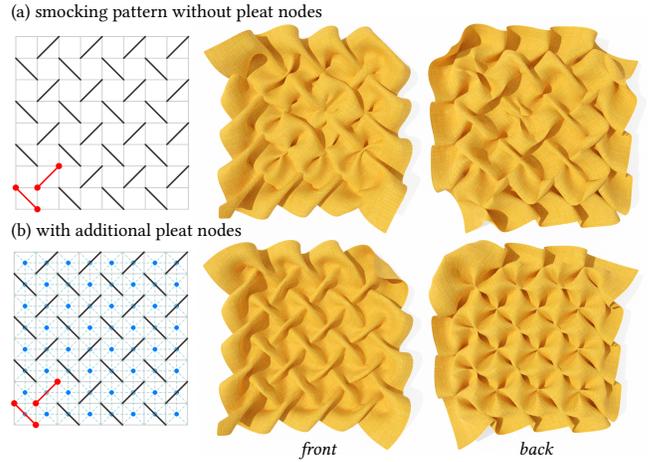
\subsubsection{Empty pleat graph}
It is unlikely to have an empty underlay graph, unless the set of stitching lines $\L$ is empty (then $\V_u = \emptyset$), or the smocking pattern $\P$ is not coarse enough, such that the underlay nodes are not connected to each other ($\E_u = \emptyset$), which can be easily fixed by making $\P$ coarser \jr{(e.g., removing the grid lines that do not contain stitching points or using Delaunay triangulation, as discussed in \secref{sec:mtd:grid-free}, to find $\E_u$)}.
However, it is possible to have stitching lines so densely defined that the pleat node set is empty. In this case, we can insert additional pleat nodes to the smocking pattern and then apply our algorithm. See \figref{fig:mtd:eg_basket} for such an example.

%% file: sections/results.tex
\section{Well-Constrained Smocking Pattern} 
\label{sec:usp}
Most online tutorials discuss how to smock a pre-designed pattern, without providing any heuristics on how to design a good pattern that leads to satisfactory textures in the first place. 
%Our user interface provides a convenient tool for experimenting user specified patterns. 
Here we discuss some observations made during our experiments.
In general, the stitching lines of a good smocking pattern should yield an underlay graph that is well constrained: if the underlay graph is \emph{underconstrained}, it means that the smocked result is ``loose'', its underlay nodes have excess freedom to move in the 2D plane during the embedding, which makes the pleats on top of them less deterministic.
On the other hand, if the underlay graph is \emph{overconstrained}, it means we add too many equality constraints to some underlay nodes, making it impossible to embed the whole underlay graph in 2D. Embedding in 3D would introduce more degrees of freedom and make it harder to obtain regular, visually pleasing textures.

\begin{figure}[!t]
    \centering
    \begin{overpic}[trim=0cm 0cm 0cm -1cm,clip,width=1\linewidth,grid=false]{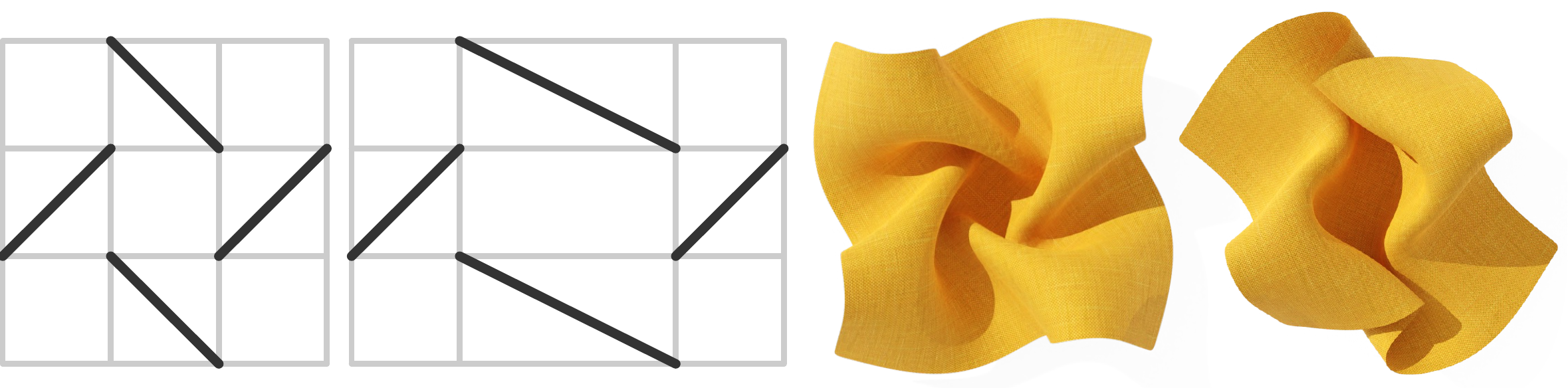}
    \put(10,24){\footnotesize $\P_1$}
    \put(34,24){\footnotesize $\P_2$}
    \put(59,24){\footnotesize $\widetilde{\M}_1$}
    \put(88,24){\footnotesize $\widetilde{\M}_2$}
    \put(4,10){\scriptsize\bfseries $\ell_1$}
    \put(11,20){\scriptsize\bfseries $\ell_2$}
    \put(18,10){\scriptsize\bfseries $\ell_3$}
    \put(27,10){\scriptsize\bfseries $\ell_1$}
    \put(37,20){\scriptsize\bfseries $\ell_2$}
    \put(48,10){\scriptsize\bfseries $\ell_3$}
\end{overpic}\vspace{-6pt}
\caption{We compare $\P_2$, an \textbf{under}-constrained smocking pattern, to $\P_1$, a well-constrained pattern. $\widetilde{M}_i$ visualizes our modeling result of $\P_i$.}\label{fig:usp:underconstrained}
\end{figure}

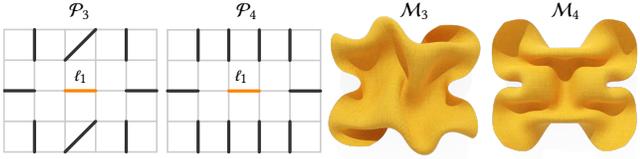
\begin{figure}[!t]
    \centering
    \begin{overpic}[trim=0cm 0cm 0cm 0cm,clip,width=1\linewidth,grid=false]{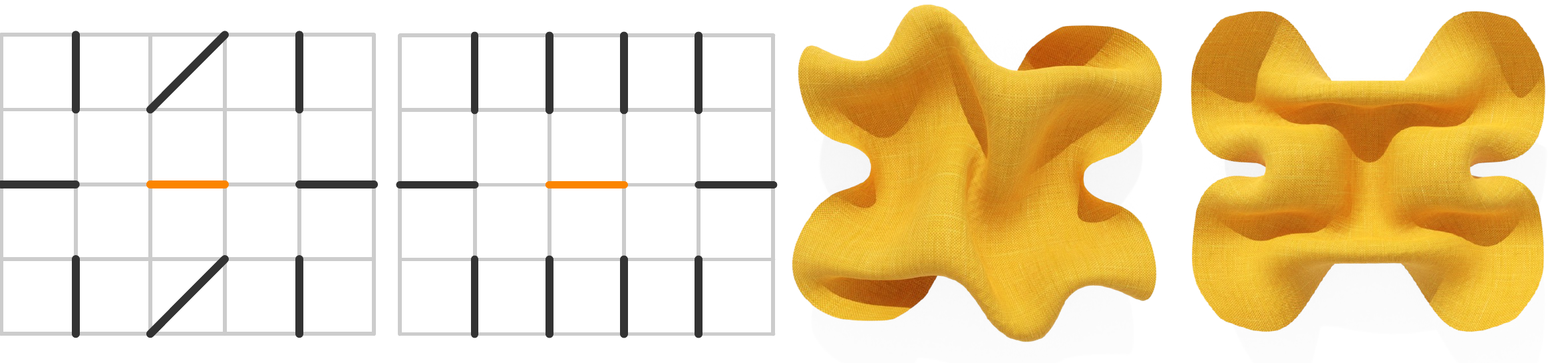}
    \put(10,23){\footnotesize $\P_3$}
    \put(36,23){\footnotesize $\P_4$}
    \put(61.5,23){\footnotesize $\widetilde{\M}_3$}
    \put(85,23){\footnotesize $\widetilde{\M}_4$}
    \put(11,13){\scriptsize $\ell_1$}
    \put(36,13){\scriptsize $\ell_1$}
\end{overpic}\vspace{-6pt}
\caption{We compare $\P_4$, an \textbf{over}-constrained smocking pattern to $\P_3$, a well-constrained pattern. $\widetilde{M}_i$ visualizes our modeling result of $\P_i$.}\label{fig:usp:overconstrained}
\end{figure}

% \begin{table}[!t]
%     \centering
%     \caption{For the smocking patterns $\P_i$ in \figreflist{fig:usp:underconstrained}{fig:usp:overconstrained}, we report the energy in \eqnreflist{eq:mtd:underlay}{eq:mtd:pleat} at convergence. We also report the number of vertex pairs that are stretched (\textit{\# stretched}), i.e. having $\Vert \x_i - \x_j \Vert > d_{i,j}$; and the average stretch ratio (\textit{stretch\%}), i.e. $\nicefrac{\left(\left\Vert \x_i - \x_j \right\Vert -d_{i,j}\right)}{d_{i,j}}$.}\label{tab:energy}
%     \vspace{-3pt}
%     \input{tables/tab_energy.tex}
% \end{table}

\textit{Underconstrained underlay.}
In \figref{fig:usp:underconstrained} we show two patterns $\P_1$ and $\P_2$, where we increase the width of the middle grid cells in $\P_2$ from 1 to 2.
We now consider the embedding distance constraints (defined in Eq.~\eqref{eq:mtd:d_ij}) between the three stitching lines.
For $\P_1$, we have $d_{1,2}=1, d_{2,3}=1, d_{1,3} = \sqrt{2}$. 
We can embed these three underlay nodes in 2D, forming a right triangle, and we call this underlay graph well constrained, since none of the underlay nodes can move locally (only rigid motion of the embedding as a whole is possible).
However, the underlay graph of $\P_2$ is underconstrained. Specifically, we have $d_{1,2}=1, d_{2,3}=1, d_{1,3} = \sqrt{5}$. 
According to triangle inequality,
$\Vert \x_1 - \x_3 \Vert \le d_{1,2} + d_{2,3} = 2 < \sqrt{5} = d_{1,3}$.
%since it is impossible to embed all three nodes such that $\Vert \x_i - \x_j \Vert = d_{i,j}$, due to the violation of the triangle inequality $d_{1,2} + d_{2,3} \le d_{1,3}$.
In other words, $d_{1,3}$ can be removed from \eqnref{eq:mtd:prob:trivial} since this inequality can never be violated.
Therefore, during the embedding of the underlay graph, we would only consider $d_{1,2}$ and $d_{2,3}$, which allows the underlay nodes of $\ell_1, \ell_3$ to move around the node of $\ell_2$.

\textit{Overconstrained underlay.}
$\P_4$ in \figref{fig:usp:overconstrained} shows an example of an overconstrained underlay graph. 
%where it is impossible to embed all the underlay nodes in 2D with their embedding distance constraints satisfied as equality. \OSH{This somehow sounds identical to underconstrained, let's chat about it tomorrow?}
The underlay node that corresponds to $\ell_1$ (colored in orange) is connected to 10 underlay nodes with maximum embedding distance $1$ or $\sqrt{2}$. One can check that, when all the 10 neighboring underlay nodes are coplanar, it is impossible to embed the underlay node of $\ell_1$ on the same plane such that the maximum embedding distance is reached. In this case, the underlay graph is overconstrained, and the embedding of the underlay by our method cannot achieve zero energy as defined in \eqnref{eq:mtd:underlay}. As a comparison, $\P_3$ shows a similar but well constrained pattern.

\begin{figure}[!t]
    \centering
    \includegraphics[width=1\linewidth]{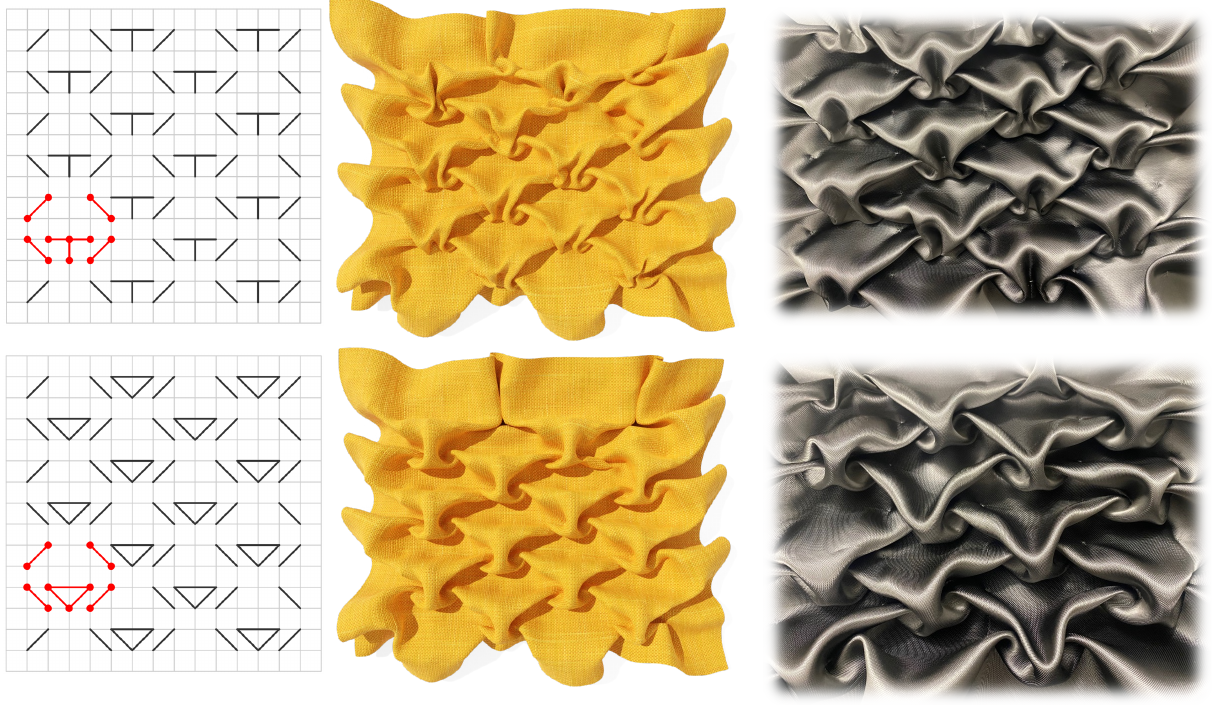}\vspace{-3pt}
    \caption{We compare the smocking designs of two similar smocking patterns, where the top pattern has an extra stitching point on the longest stitching line compared to the bottom pattern.}\label{fig:res:triangle}
\end{figure}

\begin{figure*}[!t]
    \centering
    \begin{overpic}[trim=0cm 0cm 0cm -0.7cm,clip,width=1\linewidth,grid=false]{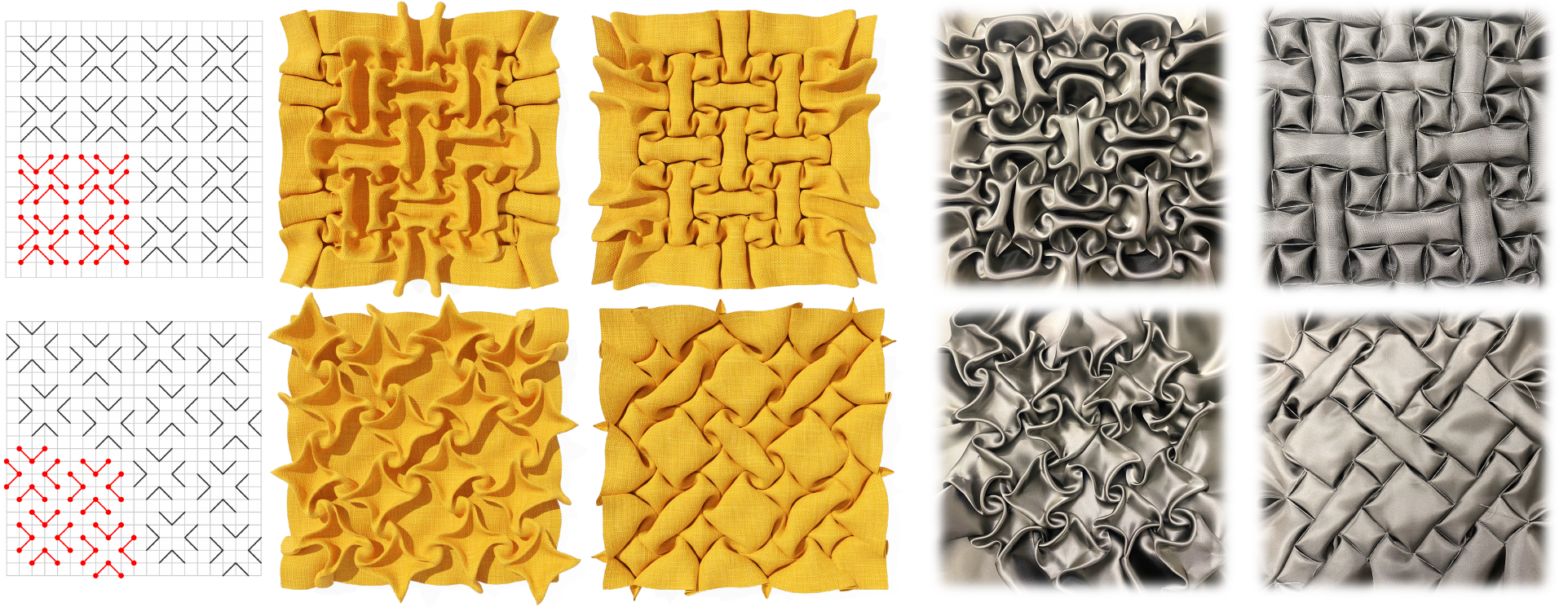}
    \put(3,39){\small smocking pattern}
    \put(33,39){\small our results}
    \put(76,39){\small fabrication}
    \put(25,-1){\footnotesize\itshape front}
    \put(47,-1){\footnotesize\itshape back}
    \put(69,-1){\footnotesize\itshape front}
    \put(88,-1){\footnotesize\itshape back}
    \end{overpic}
    \caption{\textbf{Folded design.} Our method can handle parallel stitching lines, which lead to folded pleats and sharper, less voluminous textures.}
    \label{fig:res:non_volumetric_pattern}
\end{figure*}

Note that our optimization-based formulation works in both cases and produces reasonable smocked results, as shown in \figreflist{fig:usp:underconstrained}{fig:usp:overconstrained}.
We observe that usually the well-constrained smocking patterns can produce more regular and visually pleasant textures.
\jr{Based on these observations, we independently designed the patterns in Fig.~\ref{fig:res:additional_res} (2th, 5th), Fig.~\ref{fig:mtd:honeycomb}, 
Fig.~\ref{fig:res:triangle} (top), Fig.~\ref{fig:res:non_volumetric_pattern} (bottom).}
% but it would converge to some local minimum with {non-zero} energy, which allows us to potentially detect the problematic stitching lines in the user-specified pattern.

%----------------------------------------------------------------------------
%           Results section
%----------------------------------------------------------------------------

\begin{figure*}[!t]
    \begin{overpic}[trim=0cm 0cm 0cm -3cm,clip,width=1\linewidth,grid=false]{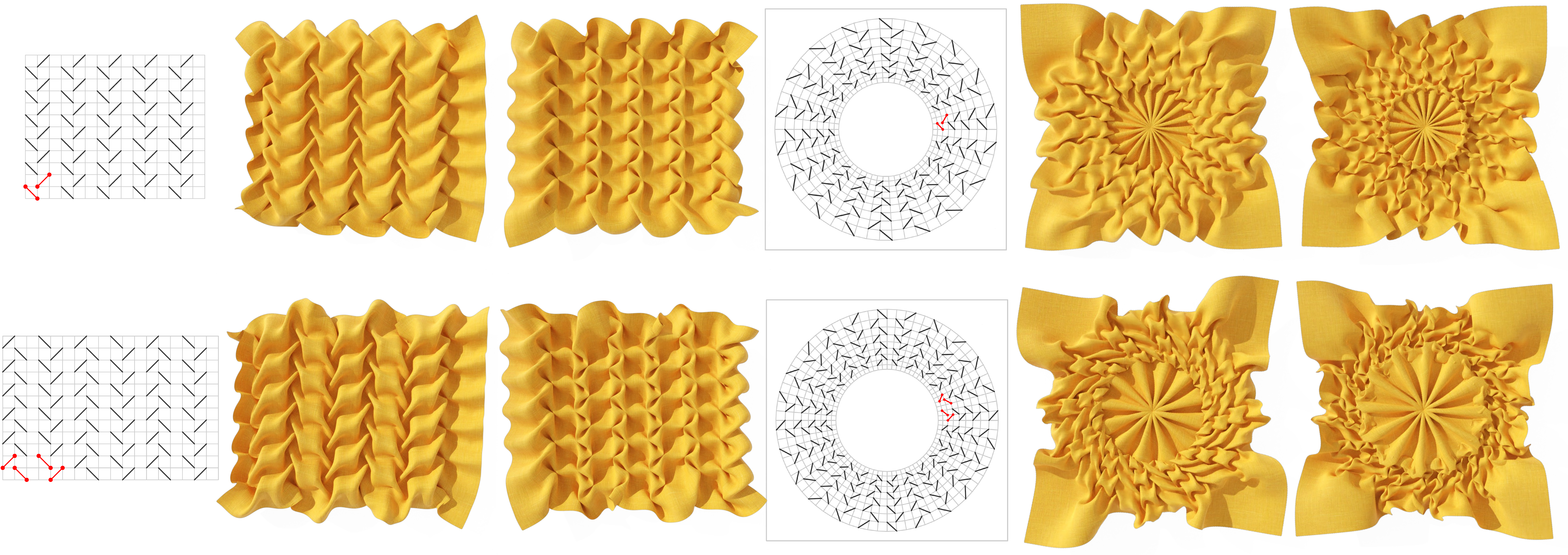}
    \put(2,33){\footnotesize (a) regular \textsc{braid}}
    \put(51,35.5){\footnotesize (b) radial \textsc{braid}}
    \put(2,15){\footnotesize (c) regular \textsc{leaf}}
    \put(51.5,17){\footnotesize (d) radial \textsc{leaf}}
    \put(22,19){\footnotesize\itshape front} 
    \put(39,19){\footnotesize\itshape back} 
    \end{overpic}\vspace{-3pt}
    \caption{\textbf{Radial grids.} We show the smocked shapes from regular (\emph{left}) and radial (\emph{right}) grids for the \jr{\textsc{braid} (\emph{top}) and \textsc{leaf} 
 (\emph{bottom})} patterns.}\label{fig:res:radial_grids}
\end{figure*}

\begin{figure*}[!t]
    \begin{overpic}[trim=0cm 0cm 0cm -1cm,clip,width=0.98\linewidth,grid=false]{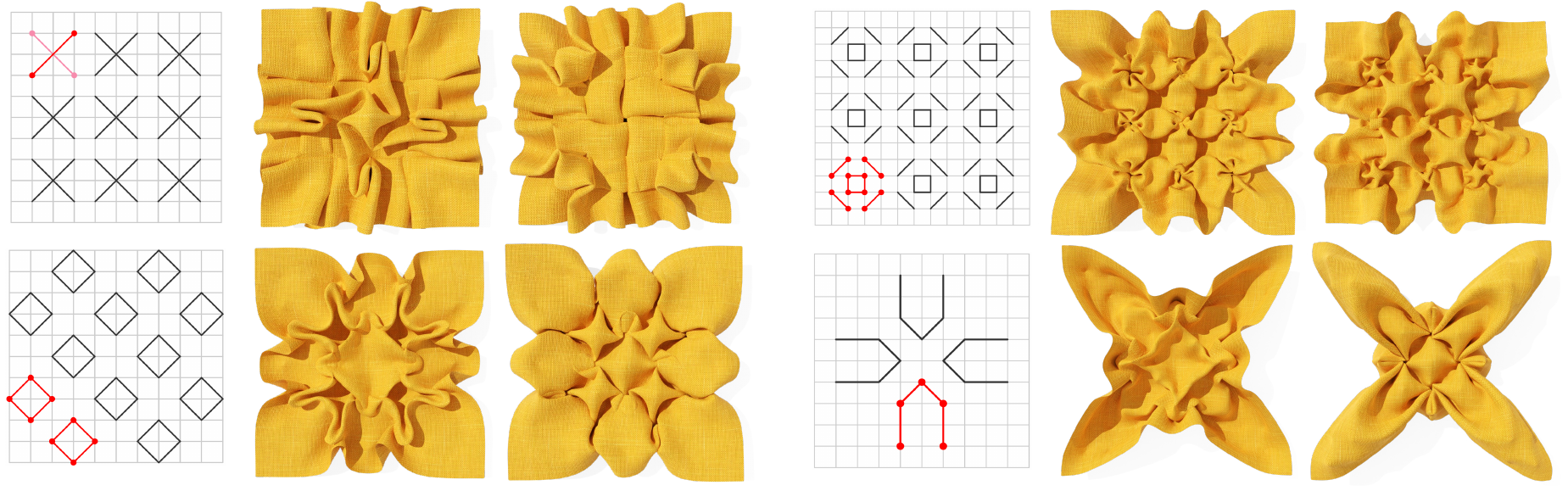}
    \put(22,32){\footnotesize\itshape front}
    \put(37,32){\footnotesize\itshape back}
    \put(-1.5,24){\footnotesize (a)}
    \put(-1.5,8){\footnotesize (b)}
    \put(50,24){\footnotesize (c)}
    \put(50,8){\footnotesize (d)}
    \end{overpic}\vspace{-6pt}
    \caption{\textbf{Long stitching lines}. We show examples of long stitching lines that cross multiple grid cells (a, b, d) and connect many nodes in a small region (c). Note that the unit smocking pattern in (a) contains two separate stitching lines.}\label{fig:res:long_stitching_lines}
\end{figure*}

\begin{figure*}[!t]
    \begin{overpic}[trim=0cm 0cm 0cm 0.76cm,clip,width=1\linewidth,grid=false]{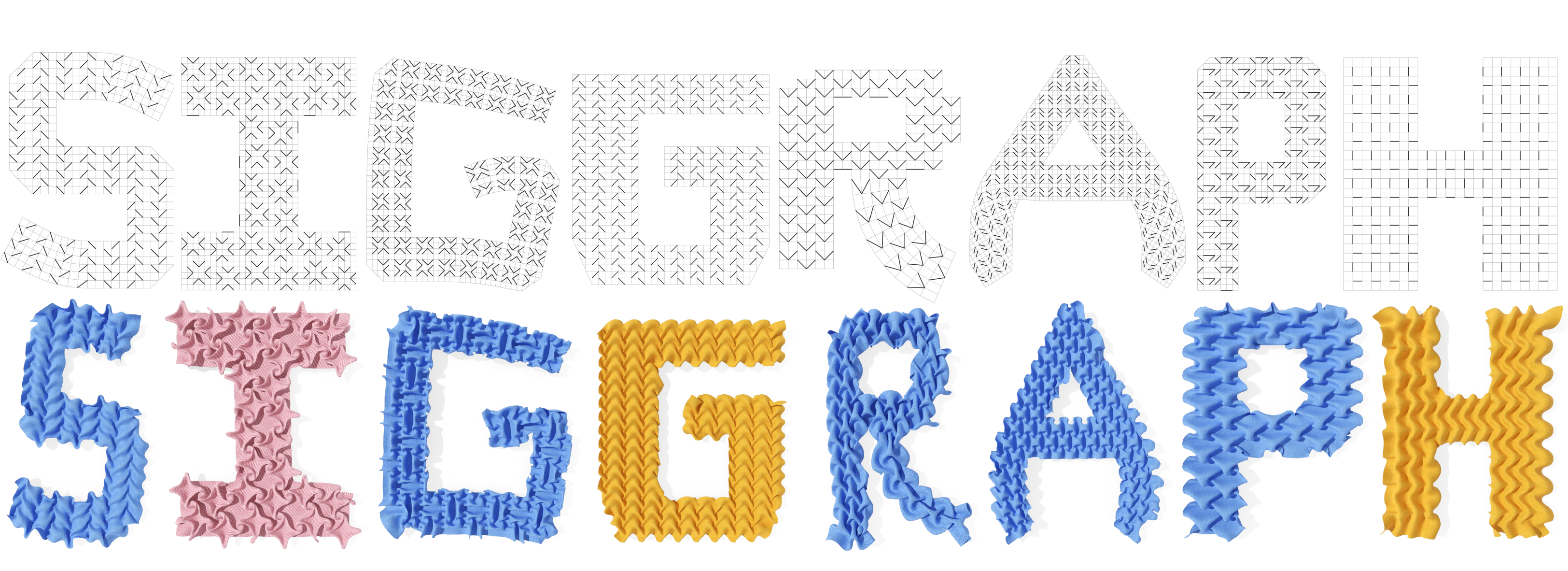}
    \end{overpic}\vspace{-9pt}
    \caption{We use our UI to edit different smocking patterns to decorate letters, including editing operators of cutting and warping grids, removing and adding stitching lines. We show the edited smocking pattern in the top row and the corresponding smocked results in the bottom row. Note that different smocking patterns shrink the fabric in different ratios.}\label{fig:res:sig_letter}\vspace{12pt}
\end{figure*}

\begin{figure*}[!t]
    \begin{overpic}[trim=0cm 0cm 0cm 2.5cm,clip,width=1\linewidth,grid=false]{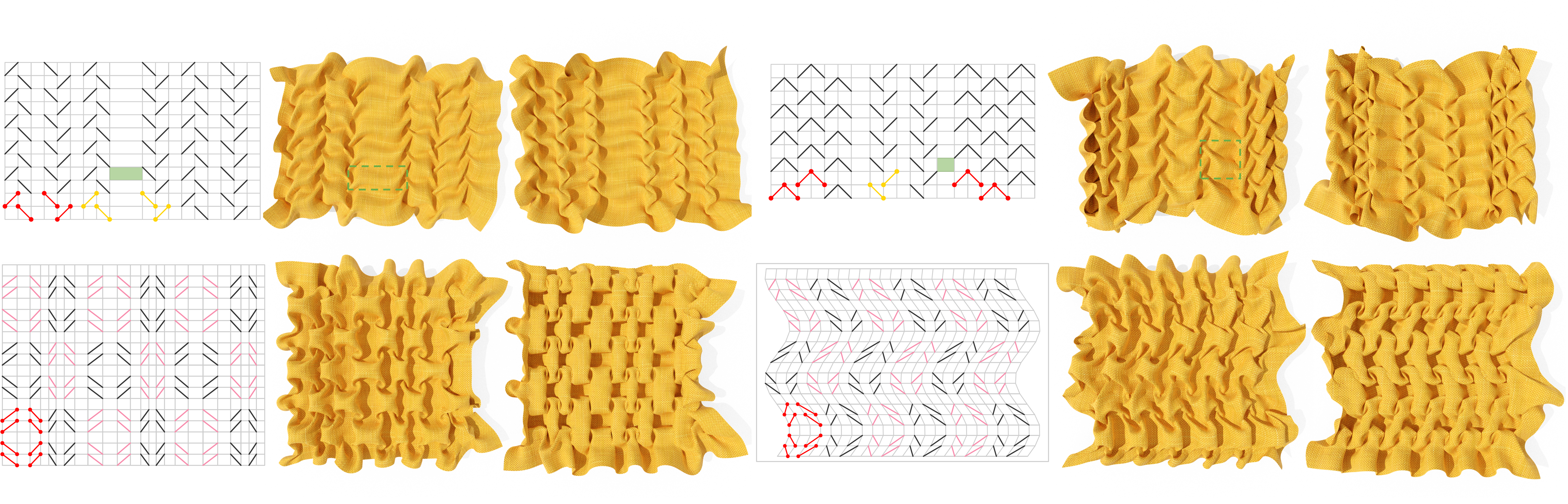}
    \put(2,29){\footnotesize (a) irregular \textsc{leaf}}
    \put(51.2,29){\footnotesize (b) \textsc{arrow} with \textsc{braid}}
    \put(1,16.2){\footnotesize (c) \textsc{box} in different sizes}
    \put(51,16.2){\footnotesize (d) \textsc{box} in curved grids}
    \put(23,16.4){\footnotesize\itshape front} 
    \put(38,16.4){\footnotesize\itshape back} 
    \end{overpic}\vspace{-12pt}
    \caption{\textbf{Irregular grids.} (a) We {increase} the space inside the \textsc{braid} pattern, as highlighted in green. (b) We {mix} the \textsc{arrow} and the \textsc{leaf} patterns, with the gap in-between highlighted in green. (c, d) We {non-linearly} deform the \textsc{box} pattern (adjacent unit patterns are colored in different colors for better visualization).}\label{fig:res:irregular_grids}
\end{figure*}

\section{Results}\label{sec:res}

We demonstrate that our algorithm can produce faithful results that match physical fabrication for different types of smocking patterns, as can be seen in the figures throughout the paper and in the accompanying video.
We also provide an interactive UI for smocking pattern exploration.
We will release our implementation. 
%please see the accompanying video.

%----------------------------------------------------------------------------
%           smocking design results
%----------------------------------------------------------------------------

\subsection{Smocking design}
%Here we focus on various types of patterns that can demonstrate the capabilities of our algorithm.

\paragraph{Folded smocking design.} 
During the experiments, we observe that there are roughly two different styles of designing stitching lines. The first is \emph{conflicting} stitching lines, where if extended, pairs of stitching lines would intersect with each other; such stitching lines create {concave} features after stitching (see, e.g.,  \figref{fig:res:radial_grids}). The second kind is \emph{parallel} stitching lines, where after stitching, the in-between fabric is folded flatly, leading to {less voluminous} textures (see \figref{fig:res:non_volumetric_pattern}). Our method can handle both cases.

\paragraph{Local modification.} 
Our method is intuitive and predictable with respect to local changes of the \emph{unit} smocking pattern. 
As shown in \figref{fig:res:triangle}, when we modify a stitching line, the final smocking design does not change drastically. Instead, the final results differ locally, as intuitively expected.

\paragraph{Irregular grids.} 
Our method is not limited to uniform square grids. We can handle hexagonal grids (\figref{fig:mtd:honeycomb}), radial grids (\figref{fig:res:radial_grids}), combinations of grids (\figref{fig:mtd:grid_free}) and other irregular grids (see  \secref{sec:res:ui}).

%
%Specifically, in \figref{fig:res:radial_grids}, we show the smocking designs of three different patterns tiled in a \emph{radial} grid with a comparison to the regular grid.

\paragraph{Long stitching lines.}
Computing smocking designs with long stitching lines can be quite challenging.
Stitching lines that stretch across multiple grid cells, as in \figref{fig:res:long_stitching_lines}(a,b,d), can potentially create large protruding features and allow the pleat nodes to have more freedom to move during optimization. 
Stitching lines that connect multiple nodes in a single grid cell, as in \figref{fig:res:long_stitching_lines}(c) and \figref{fig:res:triangle}, can lead to complicated texture in a small region, which is difficult to model in general. Our method can produce reasonable and visually pleasing results in both cases.

%----------------------------------------------------------------------------
%           Interactive UI
%----------------------------------------------------------------------------

\subsection{Interactive UI}\label{sec:res:ui}
We integrate our method into Blender as an add-on with an interactive interface that allows users to design and modify smocking patterns, as well as visualize the computed smocking designs. 
Efrat \etal~\shortcite{efrat2016hybrid} also provide a UI for smocking pattern design that allows users to tile 5 known patterns with different spacing or rotation. The tiled smocking pattern needs to be printed out for fabrication to see the resulting smocking design.
In comparison, our UI is more flexible, it supports mesh-level modifications (see Fig.~\ref{fig:res:sig_letter}) and allows the user to design stylish patterns by drawing stitching lines freely. 
Our method can also be used to explore variations of existing patterns.
For example, in Fig.~\ref{fig:res:irregular_grids} we show smocking designs with modified grids using our UI.
Since the smocking design computation and visualization are integrated into the UI,
it becomes much easier and more efficient for casual users to explore different patterns. 
To stress this point: it can take a few \emph{hours} to smock a piece of physical fabric, including drawing grids, annotating stitching lines, and sewing every single stitching line with pleating and knotting of the threads. In contrast, our algorithm  demonstrates the computed smocking design in \emph{seconds}.
%The UI can be used to efficiently check whether a user-specified unit pattern is satisfactory or not, as discussed in \secref{sec:usp}. 
%
As a proof of concept, we prototype smocked sleeve designs, as shown in \figref{fig:teaser:sleeves}, by computing the smocking designs with extra margins, which leads to natural folds on the boundary. We then deform the smocked shape w.r.t.\ a hand model using the ``\textsf{bend}'' function in Blender. These preliminary results suggest that our algorithm can be potentially used for digital garment design.
See Appendix~\ref{appendix:ui} and the supplementary video for a more detailed exposition of the functionalities of our UI.

% %----------------------------------------------------------------------------
% %           Digital Garments
% %----------------------------------------------------------------------------

% \subsection{Application: Smocking for Digital Design} 
% We also show an application in digital garments of our method.
% Fig.~\ref{fig:teaser:sleeves}

%----------------------------------------------------------------------------
%          baselines
%----------------------------------------------------------------------------

\subsection{Comparison to baselines}
In \figref{fig:background:blender} and the supplementary material, we show comparisons to the cloth simulator of the open-source software Blender~\cite{blender}. 
In this section, we provide additional comparisons to the state-of-the-art cloth simulators, {\arcsim}~\cite{narain2012adaptive} and {\cipc}~\cite{Li2021CIPC}, and the commercial software~\cite{MD}, which is closed-source. For all the comparisons to baselines, we use the fabric in the same resolution as ours. In particular, for {\arcsim} and {\cipc} we additionally provide the \emph{non-planar} initial configuration for the fabric, where all stitching points are \emph{offset} \jr{out of the fabric plane in the same direction} (see \figref{fig:res:cipc:all}~(e)). If starting from a planar configuration, {\arcsim} gets stuck in the first iteration, and {\cipc} produces a cluttered result with irregular pleats.

\begin{figure}[!t]
    \centering
    \begin{overpic}[trim=0cm 0cm 0cm -1.35cm,clip,width=1\linewidth,grid=false]{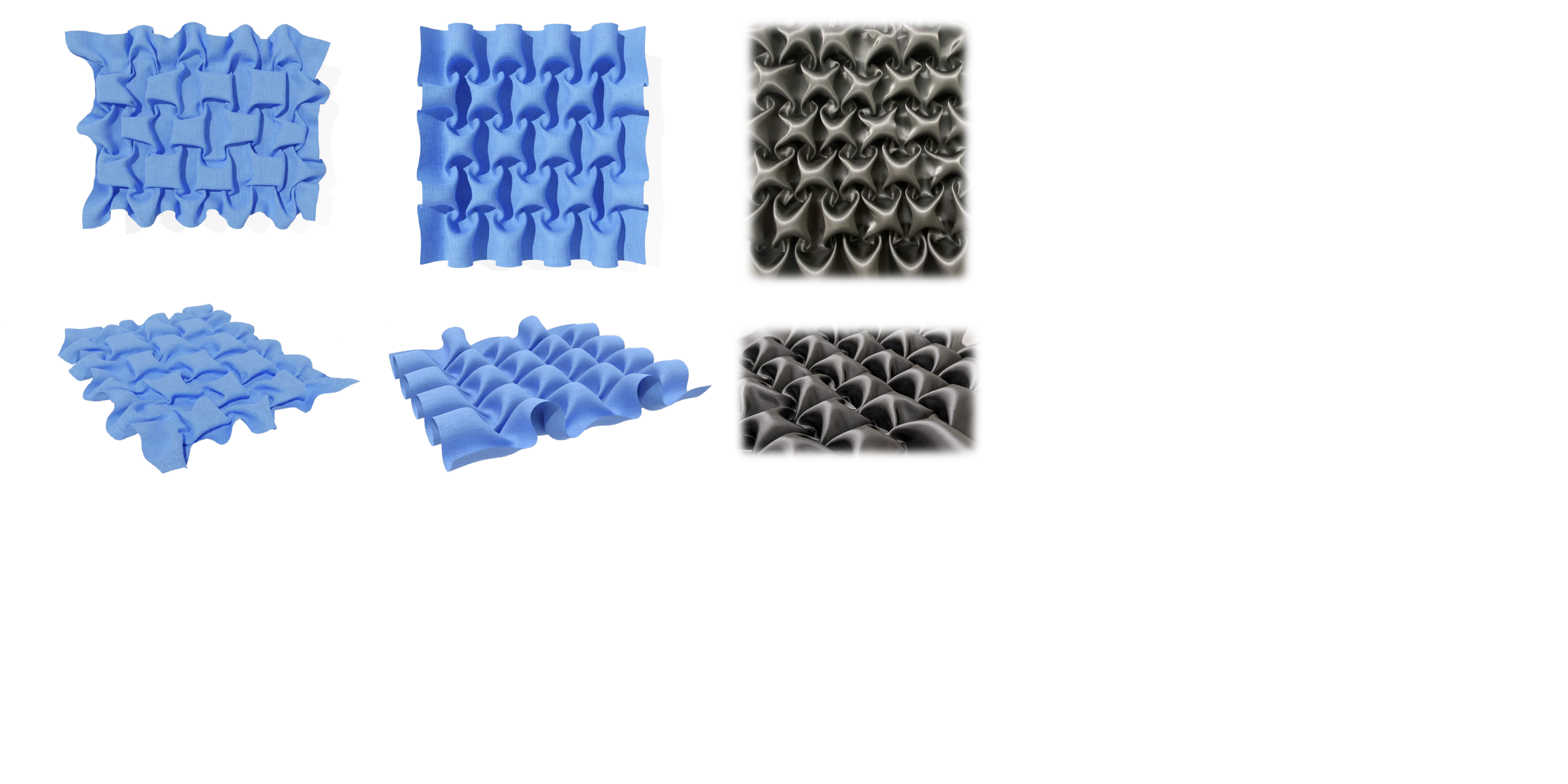}
    \put(5,51){\footnotesize Marvelous Designer}
    \put(50,51){\footnotesize ours}
    \put(80,51){\footnotesize fabrication}
    \put(-1.5,32){\scriptsize\itshape view 1}
    \put(-1.5,5){\scriptsize\itshape view 2}
    \end{overpic}\vspace{-3pt}
    \caption{Comparison to the commercial software~\cite{MD}.}
    \label{fig:res:md}
\end{figure}

\begin{figure}[!t]
    \centering
    \begin{overpic}[trim=0cm 0cm 0cm -0.5cm,clip,width=1\linewidth,grid=false]{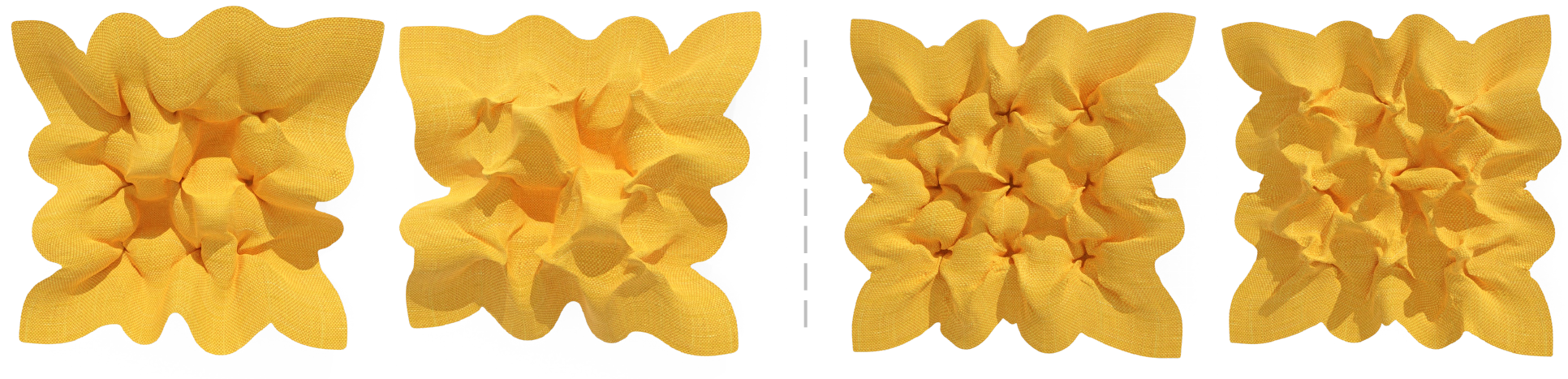}
    \put(8,-1){\footnotesize\itshape front}
    \put(34,-1){\footnotesize\itshape back}
    \put(63,-1){\footnotesize\itshape front}
    \put(87,-1){\footnotesize\itshape back}
    \put(18,25){\footnotesize $d_p = $ \unit[20]{mm}}
    \put(70,25){\footnotesize $d_p = $ \unit[8]{mm}}
    \end{overpic}
    \caption{
    Using Marvelous Designer to simulate the pattern in Fig.~\ref{fig:res:long_stitching_lines} (c), with particle distance $d_p$ set to \unit[20]{mm} (left) and \unit[8]{mm} (right).}
    \label{fig:res:md2}
\end{figure}
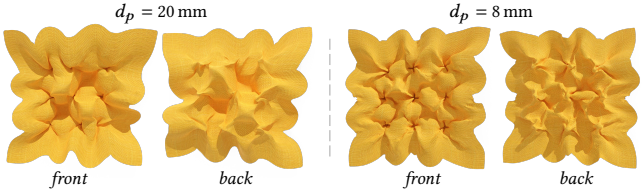

\paragraph{Comparison to Marvelous Designer (MD)~\shortcite{MD}.}
\jr{To prepare the input for the commercial software Marvelous Designer, each stitching line in the fabric needs to be specified using the ``tack'' tool, which adds extra complexity for smocking simulation in MD.}
In \figref{fig:res:md}, we report the best result obtained from this software, \jr{where we experimented with different parameters such as stiffness, damping, pressure, sewing distance and so on}.
Please see the supplementary materials for full simulations with different parameter settings. 
We observe that the ``solidify'' function, which is designed to maintain the desired draping state per pattern unit, is the key factor in helping  Marvelous Designer achieve the expected box-like geometric features.
However, the simulated smocking details in Marvelous Designer are less regular and do not match the physically fabricated result as well as our approach does.
In \figref{fig:res:md2} we show the results of Marvelous Designer on a much more complicated pattern, depicted in \figref{fig:res:long_stitching_lines} (c), containing long stitching lines. With the ``solidify'' function and high enough resolution, Marvelous Designer struggles to produce meaningful results, while our method produces a faithful preview of the fine details of the smocking results.

\begin{figure}[!t]
    \centering
    \begin{overpic}[trim=0cm 0cm 0cm -2cm,clip,width=1\linewidth,grid=false]{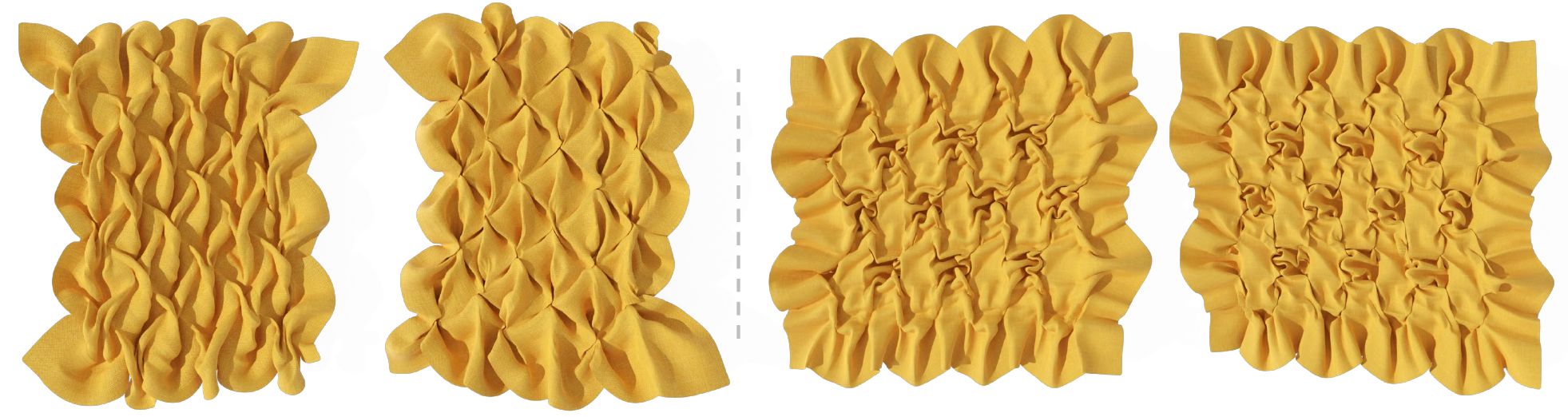}
    \put(8,28){\footnotesize (a) \textsc{arrow} pattern (Fig.~\ref{fig:intro:eg_smocking})}
    \put(60,28){\footnotesize (b) \textsc{box} pattern (Fig.~\ref{fig:res:md})}
    \put(8,-2.5){\footnotesize\itshape front}
    \put(32,-2.5){\footnotesize\itshape back}
    \put(59,-2.5){\footnotesize\itshape front}
    \put(85,-2.5){\footnotesize\itshape back}
    \end{overpic}
    \caption{\textbf{Results of {\arcsim}}. We provide two examples of using \cite{narain2012adaptive}, starting from initial configurations with offsets to break the symmetry. It takes \unit[3]{min} and \unit[5]{min} to obtain the results for the \textsc{arrow} and \textsc{box} pattern, respectively. Without a geometric prior, the simulated pleats are not voluminous and do not realistically reflect the physical fabrications.}
    \label{fig:res:arcsim}
\end{figure}

\paragraph{Comparison to {\arcsim}}
{\arcsim}~\cite{narain2012adaptive} is a powerful method for simulating fine features, such as wrinkles and creases for cloth deformations. 
We adapt the more advanced implementation\footnote{https://git.ista.ac.at/gsperl/ARCSim-HYLC} of {\arcsim}~\cite{sperl2020homogenized} for smocking, where the to-be-stitched vertex pairs are specified using the ``glue'' constraints.
In \figref{fig:res:arcsim} we show the best results we attained in consultation with the authors of the method. 
We ran the simulation from the initial configuration where all stitching points are offset. 
We experimented with the parameters of repulsion thickness, collision stiffness and different fabric materials.  We also disabled the ``remeshing'' option to preserve the glue constraints.
We can see that the shrinking ratio of the smocked fabric is more accurate than Blender and Marvelous Designer. 
However, the lack of volume and realism in the simulated pleats suggests that this method may not be suitable for use as a direct preview tool for artists designing smocking patterns.

\begin{figure*}[!t]
     \begin{overpic}[trim=0cm 0cm 0cm -1.5cm,clip,width=1\linewidth,grid=false]{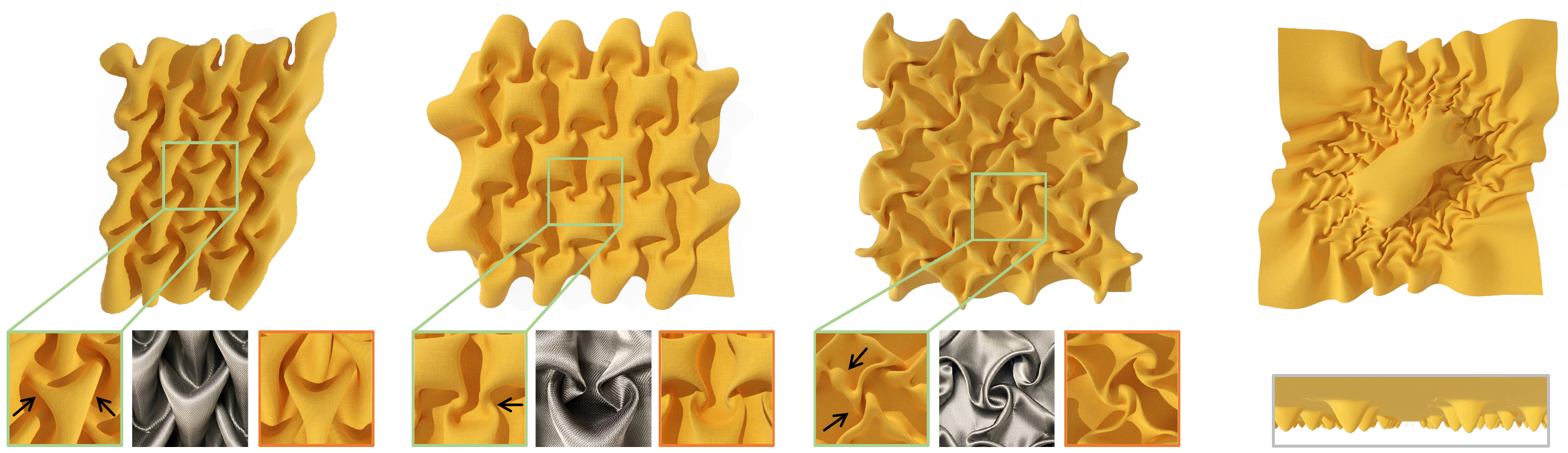}
    \put(5,31){\footnotesize (a) \textsc{arrow} pattern (Fig.~\ref{fig:intro:eg_smocking})}
    \put(30,31){\footnotesize (b) \textsc{box} pattern (Fig.~\ref{fig:res:md})}
    \put(55,31){\footnotesize (c) \textsc{diamond} pattern (Fig.~\ref{fig:res:non_volumetric_pattern})}
    \put(83,31){\footnotesize (d) \textsc{braid} pattern (Fig.~\ref{fig:res:radial_grids})}
    \put(82,6){\footnotesize (e) add offset to initial fabric}
    % \put(2,20){\footnotesize {\cipc}}
    \put(3,8){\scriptsize {\cipc}}
    \put(9.5,8){\scriptsize fabrication}
    \put(19,8){\scriptsize \textbf{ours}}
    
    \put(29,8){\scriptsize {\cipc}}
    \put(35.4,8){\scriptsize fabrication}
    \put(44.5,8){\scriptsize \textbf{ours}}
    
    \put(54,8){\scriptsize {\cipc}}
    \put(61,8){\scriptsize fabrication}
    \put(70.5,8){\scriptsize \textbf{ours}}

    \put(5,29){\footnotesize $t_{\text{cipc}} = $ \unit[1]{min}, \quad $ t_{\text{ours}} = $ \unit[2]{sec}}
    \put(30,29){\footnotesize $t_{\text{cipc}} = $ \unit[3]{min}, \quad $ t_{\text{ours}} = $ \unit[2]{sec}}
    \put(56.5,29){\footnotesize $t_{\text{cipc}} = $ \unit[6]{min}, \quad $ t_{\text{ours}} = $ \unit[4]{sec}}
    \put(82,29){\footnotesize $t_{\text{cipc}} = $ \unit[21]{min}, \quad $ t_{\text{ours}} = $ \unit[22]{sec}}
    \end{overpic}\vspace{-3pt}
    \caption{\textbf{Comparison to {\cipc}}. We show four smocked results (a-d) using {\cipc}~\cite{Li2021CIPC}, where all stitching points in the initial configuration are offset along the $z$-axis, as shown in (e), to obtain more regular results. In comparison, our method provides a more realistic preview of the smocked fabric in a much shorter time, which allows the artists to iterate the smocking design \emph{interactively}. For example, for the pattern of (d), it takes our method 22 seconds, while {\cipc} takes  21 minutes.}
    \label{fig:res:cipc:all}
\end{figure*}
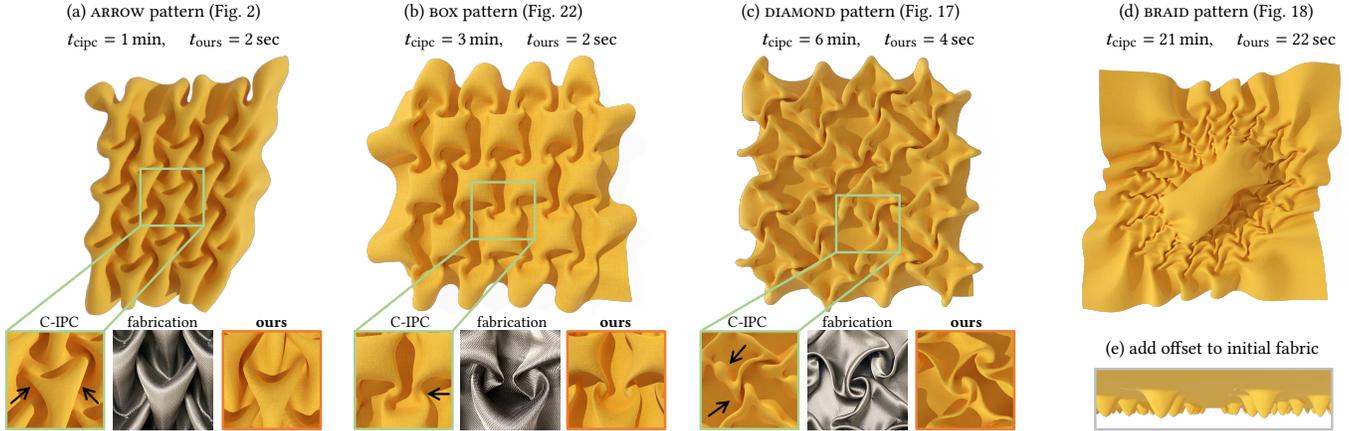

\paragraph{Comparison to {\cipc}}
Co-dimensional incremental potential contact ({\cipc})~\cite{Li2021CIPC} is a current state-of-the-art method for cloth simulation that can model thickness and handle collision and frictional contact.
To run {\cipc}\footnote{https://github.com/ipc-sim/Codim-IPC}, we rescale the smocking pattern to its intended dimensions in centimeters and offset all stitching points to guide the simulation. 
In \figref{fig:res:cipc:all} we show the best attained results on four examples in consultation with the authors of the method.
We experimented with various values of bending, stretching, stitching force, time step size, offset value, etc.
We also tried using both the static solver and the dynamic solver with various time steps.
When using the dynamic solver, we found that using a large time step (e.g., $dt =\ $\unit[10]{sec.}) can achieve much less wrinkled and more realistic results than the default time step ($dt =\ $\unit[0.01]{sec.}).
\jr{The dynamic solver without collision handling is much more efficient than the static solver. However, finding a suitable equilibrium state for the dynamic solver is challenging. For example,} running the dynamic solver until convergence, where the change of the vertex positions is smaller than a threshold while setting the vertex velocity to zero at each iteration, leads to a cluttered configuration. 
The best intermediate results are similar to the ones shown in \figref{fig:res:cipc:all}, where the static solver is used.
Overall, {\cipc} achieves better and more realistic results than the other baselines. However, the whole fabric gets sheared, and the geometric shape of the pleats is not as accurate as in our method. For example, as highlighted in \figref{fig:res:cipc:all}, the transition regions between the box shapes are wrong in example (b), and the bumps along the edges of the diamond shapes are unnatural in example (c). The method takes minutes to execute.
Moreover, expertise in cloth simulation is needed to tune the parameters in order to obtain reasonable results, as the default values did not work out of the box.

\paragraph{Summary}
\begin{table}[!t]
    \caption{Comparing different solutions to (pre-)visualize a smocking pattern. MD stands for Marvelous Designer, and ``fabric.'' stands for manual physical fabrication.}
    \label{tab:baseline}
    \input{tables/tab_baseline}
\end{table}
\jr{In comparison, our method is much simpler to use, requiring no domain knowledge, and it is more efficient for previewing purposes, taking only a few seconds. This enables interactive design iterations for artists. See \tabref{tab:baseline} for a comparison summary.}

\subsection{Ablation study \& justifications}

\begin{figure}[!t]
    \centering
    \includegraphics[width=1\linewidth]{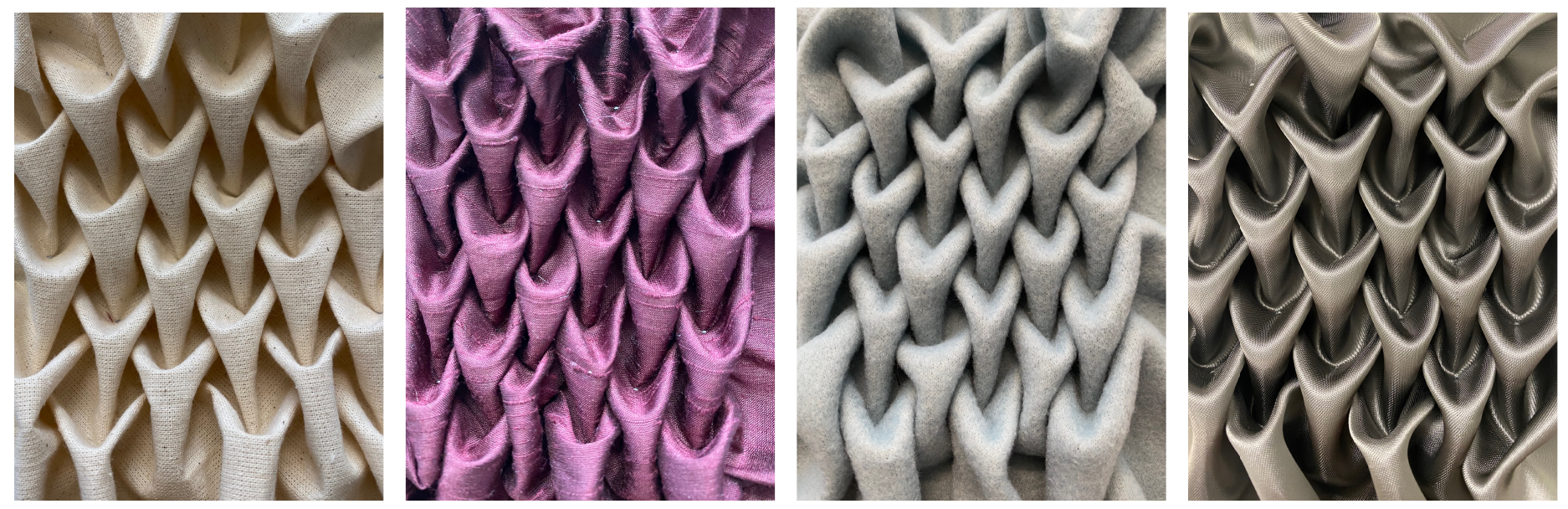}\vspace{-3pt}
    \caption{Fabricating the \textsc{ARROW} pattern using different fabric materials including canvas, polyester (crisp, thin), polyester (soft, thick), and satin, from left to right respectively.}
    \label{fig:res:arrow_diff_mat}
\end{figure}

\paragraph{Geometric appearance}
In this work, we aim to preview the shape of a smocked pattern solely based on its geometric features, disregarding the impact of different fabric materials. Indeed, the final outcome of smocking can be influenced by the type of fabric used, which may possess varying levels of stretchiness. 
However, as evidenced by the multitude of examples available online, the \emph{geometric appearance} of a smocked pattern remains \osh{very similar} regardless of the fabric used, including our experiments with canvas, satin, and polyester (see \figref{fig:res:arrow_diff_mat}), as well as numerous examples found on YouTube and Pinterest featuring silk, leather, wool, cotton, lace, denim fabrics and so on. It is, in fact, the stitch structure, not so much the specific material, that ultimately determines the geometric \osh{structure} of the pattern.
% \OSH{I propose to drop this part of the sentence.}, which is formed by bending deformation of the fabric, not by stretching (for example, the canvas fabric is almost non-stretchable). 
Therefore, it is reasonable to model the geometric appearance for preview purposes \jr{and delegate the material-dependent characteristics, such as bending stiffness, to cloth simulators.}

\paragraph{Smocked graph.} 
The key component of our method is the formulation of the \emph{smocked graph}, extracted from the smocking pattern, 
\setlength{\columnsep}{3pt}%
\setlength{\intextsep}{0pt}%
\begin{wrapfigure}{r}{0.5\linewidth}
\centering
\begin{overpic}[trim=0cm 0cm 0cm 0cm,clip,width=1\linewidth,grid=false]{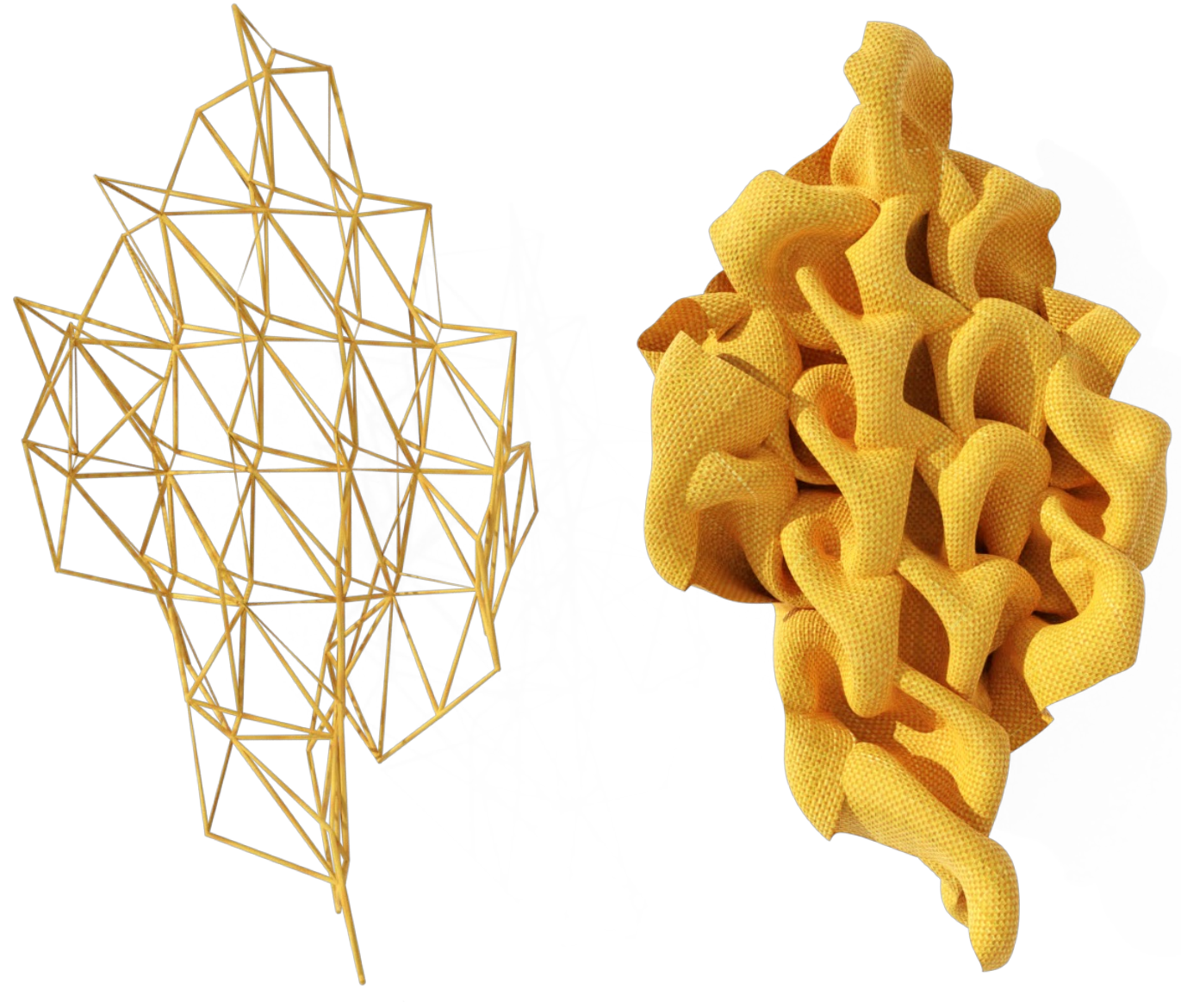}
\end{overpic}\vspace{-3pt}
\caption{Coarse-to-fine {\arap}.}
\label{fig:coarse-to-fine-arap}
\end{wrapfigure}
\jr{which explicitly encodes the modified geometry after stitching (as discussed in \secref{sec:mtd:embedding_distance}). To check that our embedded smocked graph}
% As discussed in \secref{sec:mtd}, the smocked graph is decomposed into two subgraphs, the \emph{underlay} and the \emph{pleat} graph.
is indeed critical for the successful computation of the smocking design, we try applying {\arap} on the coarse smocking pattern $\P$ instead of utilizing the smocked graph.
We optimize \eqnref{eq:background:arap} on the coarse smocking pattern $\P$ and use the result to deform the finer fabric discretization $\widetilde{\P}$,  as discussed in \secref{sec:mtd:arap}. \figref{fig:coarse-to-fine-arap} shows the resulting $\P$ and $\widetilde{\P}$.
%Running {\arap} on the coarser grid $\P$ is directly comparable to our embedded smocked graph.
We can see that the result is more regular than applying {\arap} to the fine grid $\widetilde{\P}$ directly (cf.\ \figref{fig:background:arap}).
However, the overall geometric texture is still not as well structured as ours. The reason is that the pleat vertices have too many degrees of freedom and are not sufficiently regularized in this approach, whereas our smocked graph encodes the global structural information and firmly sets the relationship between the underlay and the pleat nodes, yielding more regular results.

\paragraph{Pleat graph embedding.}
Our method embeds both the underlay graph and the pleat graph to guide the fabric deformation. To justify that the pleat graph embedding is indeed helpful, we try
\setlength{\columnsep}{3pt}%
\setlength{\intextsep}{1pt}%
\begin{wrapfigure}{r}{0.55\linewidth}
\centering
\begin{overpic}[trim=0cm 0cm 0cm 0cm,clip,width=1\linewidth,grid=false]{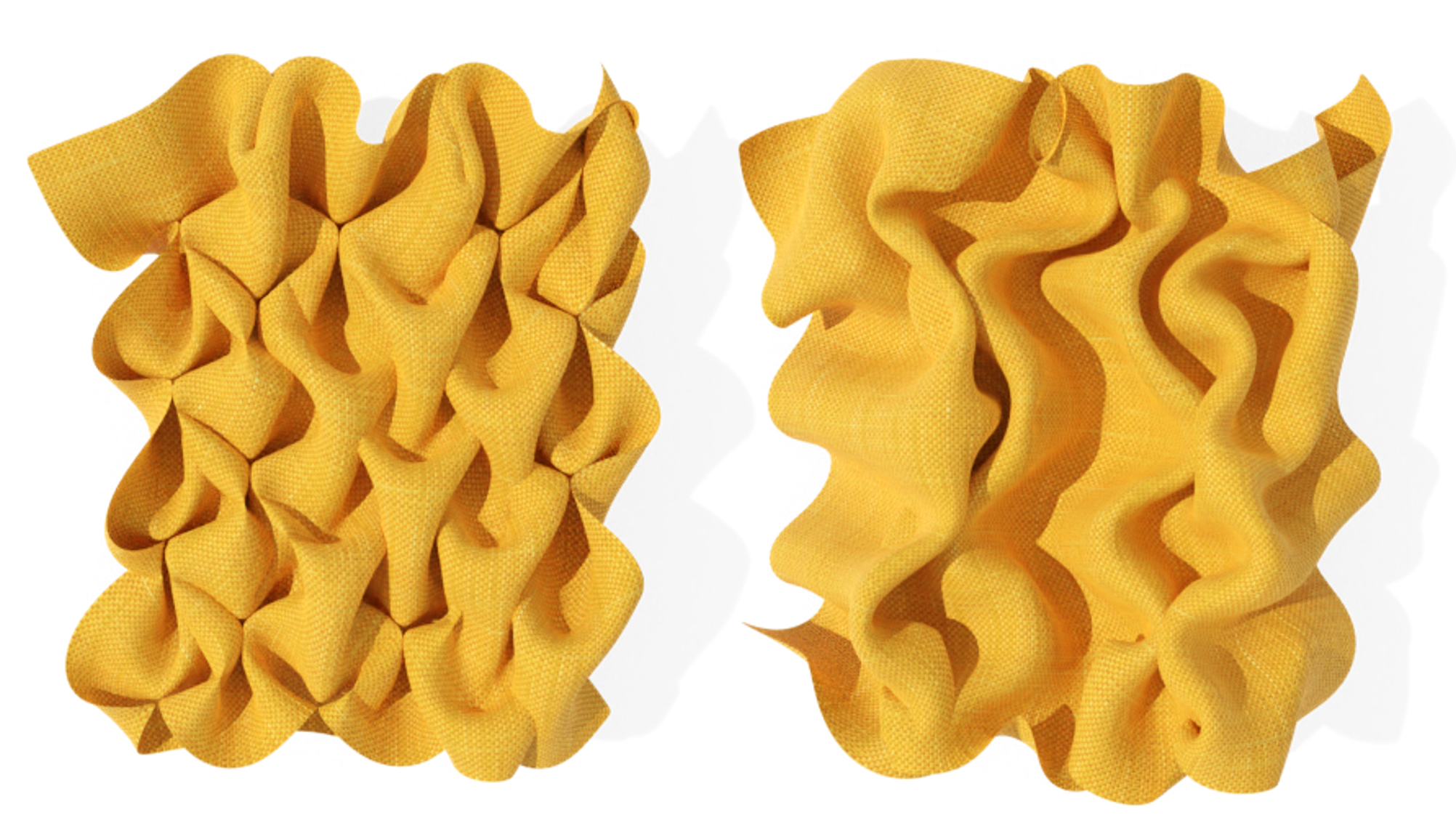}
\put(7,-4){\footnotesize\itshape only fix underlay}
\put(60,-4){\footnotesize\itshape only fix pleat}
\end{overpic}
\caption{Ablation on pleat graph.}
\label{fig:res:ablation_pleat}
\end{wrapfigure}
using only the optimized embedding of the \emph{underlay} graph to guide the {\arap} deformation, see the left part of \figref{fig:res:ablation_pleat}. For completeness, we also show the result of only using the optimized embedding of the \emph{pleat} graph to guide the deformation on the right of \figref{fig:res:ablation_pleat}.
We can see that the optimized embedding of the underlay graph can help to guide the deformation to achieve a less cluttered result compared to the other {\arap}-baselines. However, without guidance from the pleat graph to reduce the search space, the pleats exhibit inconsistent orientations and irregular shapes, resulting in an unpleasant (but still feasible) preview. This ablation justifies that both the underlay and the pleat graphs contribute to form regular and faithful appearance of the geometric texture.

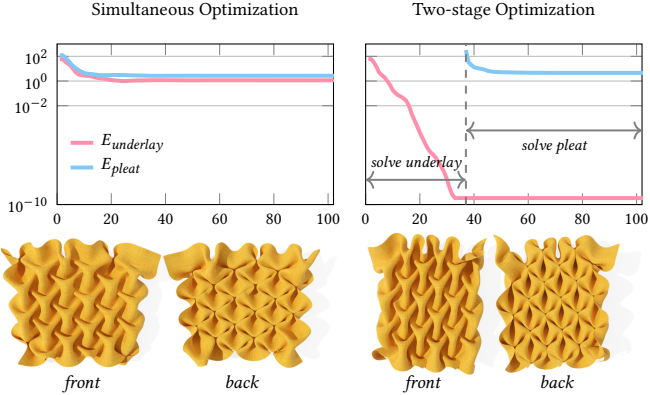
\begin{figure}[!t]
    \centering
    \input{figures/tmp/ablation_optimization}
    \begin{overpic}[trim=0cm 0cm 0cm 0cm,clip,width=1\linewidth,grid=false]{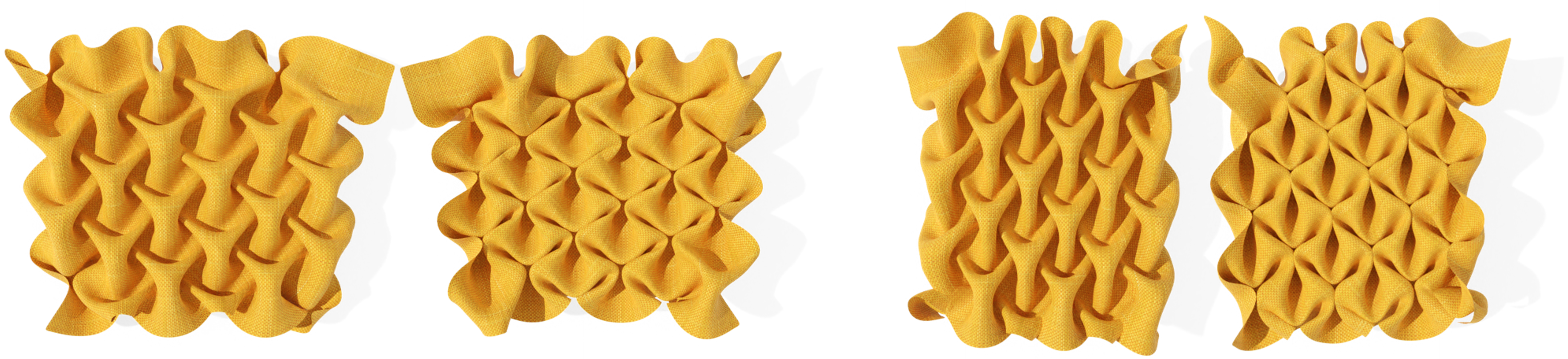}
    \put(10,-2){\footnotesize\itshape front}
    \put(35,-2){\footnotesize\itshape back}
    \put(63,-2){\footnotesize\itshape front}
    \put(84,-2){\footnotesize\itshape back}
    \end{overpic}
    \caption{\textbf{Ablation on solver.} We compare our two-stage optimization scheme to a simultaneous setting, where the underlay and the pleat graph are embedded at the same time. We show the energies over iterations on the top and the final results on the bottom. \osh{The spatial energy distributions are visualized in \figref{fig:err_distribution}.}}
    \label{fig:res:ablation_solver}
\end{figure}

\paragraph{Two-stage optimization.}
It is natural to have a two-stage optimization of embedding the underlay and pleat graph separately, since the pleats are induced by the fixed underlay graph.
To further justify it, we compare to the setting where the underlay and pleat graphs are solved simultaneously by minimizing the sum of energies in \eqnref{eq:mtd:underlay} and \eqnref{eq:mtd:pleat}. We show the corresponding energy over iterations in \figref{fig:res:ablation_solver}.
We can see that the simultaneous optimization still produces reasonable results, since the proposed distance constraints $d_{i,j}$ properly encode the modified local structure after smocking. However, our two-stage optimization leads to a more faithful result w.r.t.\ the real fabrications shown in \figref{fig:background:arap} (d), in 3 times shorter computation time due to the smaller number of variables to optimize in each stage.
More specifically, in our two-stage optimization process, we first focus on embedding the underlay graph accurately, then use it to constrain and embed the pleat graph. Solving the two embeddings simultaneously is more likely to land in an undesirable local minimum.

%----------------------------------------------------------------------------
%           Implementation
%----------------------------------------------------------------------------
\subsection{Implementation}

\begin{table}[!t]
\caption{\textbf{Smocking pattern complexity and modeling runtime}. We report the topology of the smocked graph, including the number of underlay/pleat vertices and edges, and the resolution of the fine grid $\vert \V_{\widetilde{\P}}\vert$. The runtimes of embedding the underlay graph $\S_u$, the pleat graph $\S_p$, and solving for the full smocking design $\widetilde{\P}$ are reported in {seconds}.}\label{tab:appendix:runtime}
%\vspace{-3pt}
\input{tables/tab_runtime.tex}

\end{table}

\paragraph{Implementation and runtime.}
We implement our algorithm in Python, and design the GUI as an addon in Blender.
\jr{The full implementation can be found at [INSERT LINK].}
The Python implementation uses the projected Newton solver for optimization.
Recall that our algorithm has three main steps: (1) embedding the underlay  graph $\S_u$, (2) embedding the pleat graph $\S_p$ with fixed underlay graph, (3) solving for the smocking design on a finer grid $\widetilde{\P}$ based on the embedded smocked graph.
In Table~\ref{tab:appendix:runtime} we report the runtime of each step of our method on multiple smocking patterns with different complexities.  
Note that the number of stitching lines equals to the number of underlay nodes,  $\vert \L \vert = \vert \V_u \vert$, so we do not report it separately in the table. 
Our method takes a few seconds on the medium-sized smocking patterns, and up to half a minute on the large ones. 
As a comparison, it usually takes up to a few \emph{hours} to smock a pattern, including drawing the grid on the fabric, annotating all the stitching lines and sewing them. 
Sewing and making knots for all the stitching points are the most time-consuming parts of the process.
Usually it takes about 2 to 3 minutes to finish a \emph{single} stitching line for an experienced maker. 
We can see that our method is more efficient, convenient, and error-tolerant.

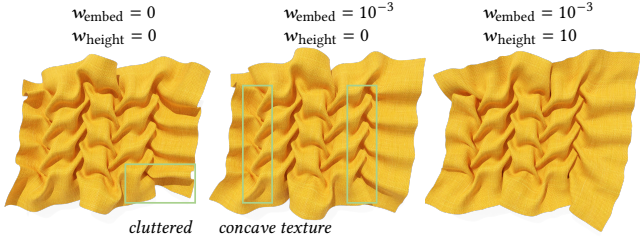
\begin{figure}[!t]
    \centering
    \begin{overpic}[trim=0cm 0cm 0cm -3cm,clip,width=1\linewidth,grid=false]{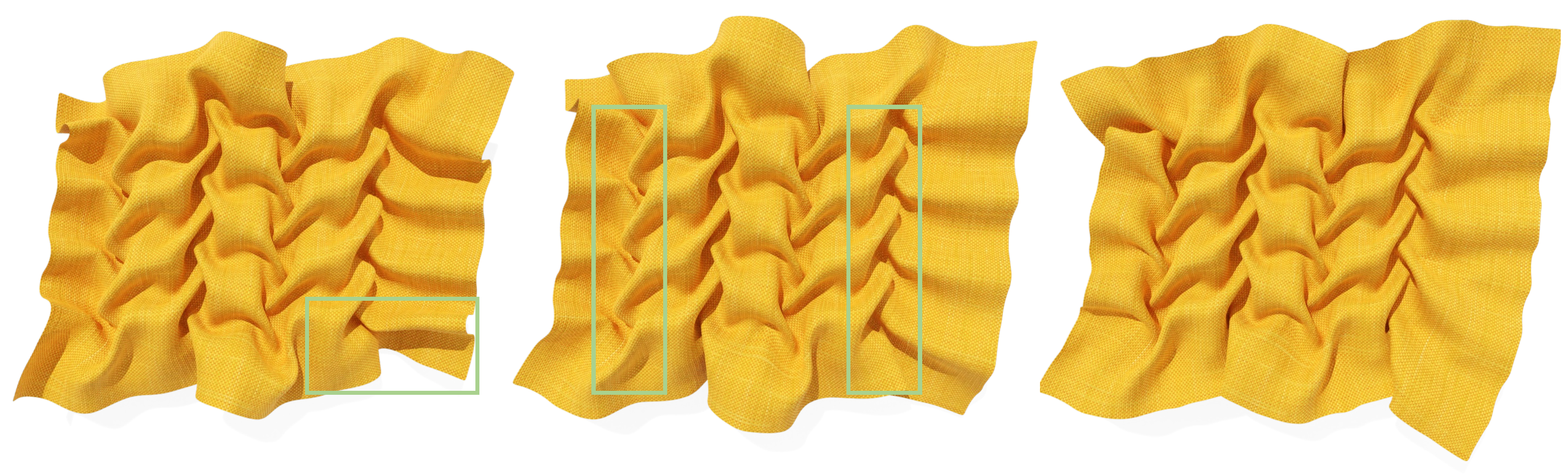}
    \put(11,34){\footnotesize $w_{\text{embed}} = 0$}
    \put(11,30.5){\footnotesize $w_{\,\text{height}} = 0$}
    \put(44,34){\footnotesize $w_{\text{embed}} = 10^{-3}$}
    \put(44,30.5){\footnotesize $w_{\,\text{height}} = 0$}
    \put(75,34){\footnotesize $w_{\text{embed}} = 10^{-3}$}
    \put(75,30.5){\footnotesize $w_{\,\text{height}} = 10$}
    \put(20,0.5){\footnotesize\itshape cluttered}
    \put(34,0.5){\footnotesize\itshape concave texture}
    \end{overpic}\vspace{-3pt}
    \caption{The \textsc{LEAF} pattern with different parameters in Eq.~\eqref{eq:appendix:pleat_height}.}
    \label{fig:appendix:leaf_param}
\end{figure}

\paragraph{Regularizers.} When optimizing for the embedding of the pleat graph as discussed in \secref{sec:mtd:embed-pleat}, we can add extra regularizers to make the pleats more regular:
\begin{equation}\begin{split}\label{eq:appendix:pleat_height}
\min_{\X\in\R^{\vert \V_p \vert \times 3}}  \quad & \sum_{(v_i, v_j)\in\E_p} \left(\left\Vert \x_i - \x_j \right\Vert_2 - d_{i,j}\right)^2 \\
& - \, w_{\text{\scriptsize{embed}}}  \sum_{\forall i \ne j} \left\Vert \x_i - \x_j \right\Vert_2  + 
\, w_{\text{\scriptsize{height}}} \, \mathrm{Var}[h],
\end{split}\end{equation}
where $\mathrm{Var}[h]$ is the variance of the heights (the $z$-coordinates) of the pleat nodes.
Here we add the maximizing embedding energy with a negative sign to make the {underconstrained} pleat nodes (e.g., boundary pleats) stay away from each other. 
We also encourage the geometric texture to keep a uniform height distribution by penalizing its variance.
In \figref{fig:appendix:leaf_param} we show a simple ablation study. We can see that adding the maximizing embedding term leads to a less cluttered boundary. Meanwhile, adding the pleat height regularizer can push the concave pleats (with negative height) upwards to form a more regular pattern. We observe that even without these regularizers our method produces good results away from the fabric boundary, and we use very small weights $w_{\text{\scriptsize{embed}}} = w_{\text{\scriptsize{height}}} = 10^{-3}$ to make the boundary pleats more attractive.

\paragraph{Initialization and parameters.} We initialize each underlay node $v_{\ell_i}$ by the average position on the flat fabric of all the stitching points in $\ell_i$, with zero height.
We initialize each pleat node by its original position on the fabric, with the initial height set to $1$.
The weights $w_{\text{\scriptsize{embed}}}$ and  $w_{\text{\scriptsize{height}}}$ in Eq.~\eqref{eq:appendix:pleat_height} are both set to $10^{-3}$ for all the experiments.
We run the embedding optimization until convergence. The underlay embedding energy at convergence is smaller than $10^{-8}$ for a well-constrained pattern.

%% file: tables/tab_baseline.tex
\definecolor{tabeasy}{HTML}{99d98c}
\definecolor{tabhard}{HTML}{f08080}
\definecolor{tabmed}{HTML}{ffd000}
\newcommand{\easy}{\cellcolor{tabeasy!50}\cmark}
\newcommand{\medium}{\cellcolor{tabmed!50}\checked}
\newcommand{\hard}{\cellcolor{tabhard!50}\xmark}
\footnotesize
{\def\arraystretch{1.2}
\begin{tabular}{lcccccc}
\toprule[1pt]
 \multicolumn{1}{c}{\textit{Properties} $\backslash$ Solutions}
 & \makebox[0.55cm][c]{Fabric.} & \makebox[0.55cm][c]{Blender} & \makebox[0.55cm][c]{MD} & \makebox[0.55cm][c]{{\arcsim}} & \makebox[0.55cm][c]{{\cipc}}& \makebox[0.55cm][c]{\textbf{Ours}}\\ \midrule[1pt]
\textit{\textbf{Easy} to prepare input?} &
  \hard & \medium  & \hard & \medium  & \medium  & \easy \\
\textit{\textbf{Easy} to use (fabricate)}?&
  \hard & \medium & \medium  &\hard & \hard & \easy \\
\textit{Ere the pleats \textbf{\jr{accurate}?}} &
  \easy & \hard & \hard & \hard & \medium & \easy \\
\textit{Fabric shrinks \textbf{realistically}?} &
  \easy & \hard & \hard & \medium & \medium & \easy \\  
\textit{\textbf{Efficient} for preview?} &
  \hard & \medium & \medium & \medium & \medium & \easy \\ \bottomrule[1pt]
\end{tabular}
}

%% file: figures/tmp/ablation_optimization.tex
\definecolor{colvu}{HTML}{99d98c}
\definecolor{colvp}{HTML}{7fc8f8}
\definecolor{coleu}{HTML}{ff8fab}
\definecolor{colep}{HTML}{ffee99}
\definecolor{coled}{HTML}{99d98c}
\definecolor{mycolor1}{HTML}{ff8fab}%
\definecolor{mycolor2}{HTML}{7fc8f8}%
\definecolor{mycolor3}{HTML}{7fc8f8}%
\definecolor{mycolor4}{HTML}{99d98c}%
\begin{tikzpicture}

\begin{axis}[%
width=0.43\linewidth,
height=0.25\linewidth,
at={(0,0)},
scale only axis,
xmin=0,
xmax=102,
ymode=log,
ymin=1e-10,
ymax=1000,
yminorticks=true,
ymajorgrids,
ytick={1e-10,1e-2,1,1e2},
axis background/.style={fill=white},
tick label style={font=\scriptsize},
title style={font=\footnotesize},
title={Simultaneous Optimization}
]
\draw[dashed,dash pattern=on 0.3cm off 5cm,mycolor1,line width=1.5pt] (5,1e-5) -- (20,1e-5) node[midway, right] {\scriptsize\itshape \textcolor{black}{$E_{\text{underlay}}$}};
\draw[dashed,dash pattern=on 0.3cm off 5cm,mycolor2,line width=1.5pt] (5,1e-7) -- (20,1e-7) node[midway, right] {\scriptsize\itshape \textcolor{black}{$E_{\text{pleat}}$}};
\addplot [color=mycolor1, line width=1.5pt, forget plot]
  table[row sep=crcr]{%
1	59.6518650841519\\
2	59.6518650841519\\
3	36.4675089149489\\
4	25.8322627478523\\
5	16.9971878864285\\
6	9.01004962094357\\
7	4.85096690703511\\
8	3.38693245004259\\
9	2.95624945184466\\
10	2.78105876025243\\
11	2.68007047030701\\
12	2.56222238651858\\
13	2.27095834672262\\
14	2.00139499362693\\
15	1.75134018697766\\
16	1.66155723691639\\
17	1.53463166651817\\
18	1.44564653644818\\
19	1.31722059202916\\
20	1.21135985271851\\
21	1.13877452656822\\
22	1.08580458584621\\
23	1.0446981482875\\
24	1.01677122156759\\
25	1.00349982449724\\
26	1.0188169590863\\
27	1.04751867913345\\
28	1.08576121914196\\
29	1.10113318954312\\
30	1.12968436011081\\
31	1.12209465330352\\
32	1.12911042961046\\
33	1.14142810863977\\
34	1.15046904189158\\
35	1.16346181291975\\
36	1.16839257778471\\
37	1.17166810810353\\
38	1.16826401307991\\
39	1.1664889354692\\
40	1.16257257303337\\
41	1.16117114770903\\
42	1.15897043396891\\
43	1.15680198139841\\
44	1.15394665491533\\
45	1.1506828971289\\
46	1.14729569736177\\
47	1.1450303857702\\
48	1.14428040609246\\
49	1.1446628072251\\
50	1.14565783882465\\
51	1.14691256701027\\
52	1.14810841135857\\
53	1.14917391247228\\
54	1.15011412327003\\
55	1.15093621321627\\
56	1.15159968306142\\
57	1.15210166005542\\
58	1.15237393694808\\
59	1.15233614253223\\
60	1.15203998432591\\
61	1.15167930613961\\
62	1.15143130560197\\
63	1.15141074257653\\
64	1.15157915009715\\
65	1.15182638907683\\
66	1.15202391916512\\
67	1.15202542006998\\
68	1.15173610607202\\
69	1.15122059774119\\
70	1.15061276755682\\
71	1.15005762822962\\
72	1.14969872384195\\
73	1.14958307043869\\
74	1.14960920355161\\
75	1.14964745299123\\
76	1.14961981659826\\
77	1.14951571561523\\
78	1.14936957916326\\
79	1.14924411725603\\
80	1.14919607772577\\
81	1.14923561416501\\
82	1.14932108241401\\
83	1.14941306148007\\
84	1.14946992996654\\
85	1.14947588840655\\
86	1.14944597244181\\
87	1.14940345408659\\
88	1.14937226634698\\
89	1.14935831438721\\
90	1.14935288905832\\
91	1.14934798100255\\
92	1.14934038328304\\
93	1.14933286620585\\
94	1.14932644827698\\
95	1.14932339049191\\
96	1.14932327804884\\
97	1.14932252246831\\
98	1.14932019181529\\
99	1.1493155422217\\
100	1.14931450286756\\
101	1.14931684943967\\
102	1.14932847932289\\
103	1.14934428112798\\
104	1.14936447980282\\
105	1.1493799122124\\
106	1.14938935027135\\
107	1.14939105597795\\
108	1.14938896713517\\
109	1.14938878384902\\
110	1.14939104973393\\
111	1.14939615032257\\
112	1.14940184384538\\
113	1.1494051141328\\
114	1.14940624536552\\
115	1.14940349740847\\
116	1.14939995965966\\
117	1.14939845605221\\
118	1.1493974070208\\
119	1.14939979230353\\
120	1.14940467075805\\
121	1.14940848115022\\
122	1.14941214269928\\
123	1.14941262014753\\
124	1.14941389147176\\
125	1.14941407141746\\
126	1.14941500792645\\
127	1.14941736656886\\
128	1.14942140781545\\
129	1.1494235438825\\
130	1.1494235438825\\
};
% \addlegendentry{test1}

\addplot [color=mycolor2, line width=1.5pt, forget plot]
  table[row sep=crcr]{%
1	124.130631666023\\
2	124.130631666023\\
3	78.0434061172656\\
4	48.8765292707864\\
5	25.6910014447962\\
6	15.3550321627265\\
7	10.6962286671207\\
8	7.44602701881951\\
9	5.63301877383328\\
10	4.4972188995709\\
11	4.02473276604843\\
12	3.78067775592302\\
13	3.58202537016806\\
14	3.38271233399269\\
15	3.13743155887351\\
16	2.99519618518701\\
17	2.94998543132317\\
18	2.95318387683768\\
19	3.01082696495924\\
20	3.06187975373657\\
21	3.08433693992695\\
22	3.08272561688888\\
23	3.07065389156088\\
24	3.0462045050909\\
25	3.01780266012346\\
26	2.97513590784156\\
27	2.93487773868698\\
28	2.88965811345996\\
29	2.86132332693936\\
30	2.80834326255861\\
31	2.78410689535458\\
32	2.74191389146626\\
33	2.70451707680983\\
34	2.68417347185041\\
35	2.6676148754036\\
36	2.66211767137057\\
37	2.65767649412477\\
38	2.66018267070732\\
39	2.66129257910341\\
40	2.66587696794914\\
41	2.6683892056674\\
42	2.67170992701191\\
43	2.67418760311855\\
44	2.67646432749761\\
45	2.67858077575896\\
46	2.68080665537994\\
47	2.68265904784092\\
48	2.68397268506107\\
49	2.68478699162495\\
50	2.68526624931274\\
51	2.68527556863287\\
52	2.68475761781546\\
53	2.683728661311\\
54	2.68245140631994\\
55	2.68128358498212\\
56	2.68046006815274\\
57	2.67992251755249\\
58	2.67959267805534\\
59	2.67942929764983\\
60	2.67937008726396\\
61	2.67933954314682\\
62	2.67929261440775\\
63	2.67922729733895\\
64	2.67917425875223\\
65	2.67916701321405\\
66	2.67924657587714\\
67	2.67944291386512\\
68	2.67973411124358\\
69	2.68006067270954\\
70	2.68037458656565\\
71	2.68064829833027\\
72	2.68084702441355\\
73	2.6809573498766\\
74	2.68100438422625\\
75	2.68101890399286\\
76	2.68101530601772\\
77	2.6809896794028\\
78	2.68093768022379\\
79	2.68086268359415\\
80	2.68078589215026\\
81	2.6807349957314\\
82	2.68071897368265\\
83	2.68073253092538\\
84	2.68076072193647\\
85	2.68078202673291\\
86	2.68078284217941\\
87	2.68076497317143\\
88	2.6807442652591\\
89	2.68073847412385\\
90	2.68075261164876\\
91	2.68078329264447\\
92	2.68082087977036\\
93	2.68085216805095\\
94	2.68087146172154\\
95	2.68087838163819\\
96	2.68087532161334\\
97	2.68086894468944\\
98	2.68086587351635\\
99	2.6808684084421\\
100	2.68087380841228\\
101	2.68087805390736\\
102	2.68087399318903\\
103	2.68086278783442\\
104	2.68084293160103\\
105	2.68082396716105\\
106	2.68080989999058\\
107	2.6808053786923\\
108	2.68080974703841\\
109	2.68081647629827\\
110	2.68082293126726\\
111	2.68082567104592\\
112	2.68082589355617\\
113	2.6808234836308\\
114	2.68082170734368\\
115	2.68082157485735\\
116	2.6808214552998\\
117	2.68082442350852\\
118	2.68082597064278\\
119	2.68082637250659\\
120	2.68082564668238\\
121	2.68082428607998\\
122	2.68082175644056\\
123	2.68082021428333\\
124	2.68081843144024\\
125	2.68081686912203\\
126	2.68081521552038\\
127	2.68081344394644\\
128	2.68081233243389\\
129	2.68081111664611\\
130	2.68081111664611\\
};
\end{axis}

\begin{axis}[%
width=0.43\linewidth,
height=0.25\linewidth,
at={(0.48\linewidth,0)},
scale only axis,
xmin=0,
xmax=102,
ymode=log,
ymin=1e-10,
ymax=1000,
yminorticks=true,
ymajorgrids,
ytick={1e-10,1e-2,1,1e2},
yticklabels={},
axis background/.style={fill=white},
tick label style={font=\scriptsize},
title style={font=\footnotesize},
title={Two-stage Optimization}
]
\addplot [color=mycolor1, line width=1.5pt, forget plot]
  table[row sep=crcr]{%
1	59.6518650841519\\
2	59.6518650841519\\
3	33.2427299989499\\
4	14.0520523681332\\
5	4.83599113823487\\
6	2.950192569726\\
7	1.96536204082345\\
8	0.972541706557014\\
9	0.478421670343124\\
10	0.185529422363798\\
11	0.105417429935132\\
12	0.0722821526655873\\
13	0.0545737475444297\\
14	0.0397116103667762\\
15	0.0248796715110922\\
16	0.00974159514972225\\
17	0.00218192326153756\\
18	0.000679641764353892\\
19	0.000227696656747045\\
20	6.82713448992361e-05\\
21	2.3544645760978e-05\\
22	8.56644377952096e-06\\
23	3.28549278272077e-06\\
24	1.67966133026416e-06\\
25	1.03591982487771e-06\\
26	5.7200770369014e-07\\
27	2.54232961852473e-07\\
28	1.10878453160697e-07\\
29	4.12730519063801e-08\\
30	1.10335307706386e-08\\
31	2.12355261583555e-09\\
32	6.22573804353285e-10\\
33	3.32243890557089e-10\\
34	3.32243890557089e-10\\
35	3.32243890557089e-10\\
36	3.32243890557089e-10\\
37	3.32243890557089e-10\\
38	3.32243890557089e-10\\
39	3.32243890557089e-10\\
40	3.32243890557089e-10\\
41	3.32243890557089e-10\\
42	3.32243890557089e-10\\
43	3.32243890557089e-10\\
44	3.32243890557089e-10\\
45	3.32243890557089e-10\\
46	3.32243890557089e-10\\
47	3.32243890557089e-10\\
48	3.32243890557089e-10\\
49	3.32243890557089e-10\\
50	3.32243890557089e-10\\
51	3.32243890557089e-10\\
52	3.32243890557089e-10\\
53	3.32243890557089e-10\\
54	3.32243890557089e-10\\
55	3.32243890557089e-10\\
56	3.32243890557089e-10\\
57	3.32243890557089e-10\\
58	3.32243890557089e-10\\
59	3.32243890557089e-10\\
60	3.32243890557089e-10\\
61	3.32243890557089e-10\\
62	3.32243890557089e-10\\
63	3.32243890557089e-10\\
64	3.32243890557089e-10\\
65	3.32243890557089e-10\\
66	3.32243890557089e-10\\
67	3.32243890557089e-10\\
68	3.32243890557089e-10\\
69	3.32243890557089e-10\\
70	3.32243890557089e-10\\
71	3.32243890557089e-10\\
72	3.32243890557089e-10\\
73	3.32243890557089e-10\\
74	3.32243890557089e-10\\
75	3.32243890557089e-10\\
76	3.32243890557089e-10\\
77	3.32243890557089e-10\\
78	3.32243890557089e-10\\
79	3.32243890557089e-10\\
80	3.32243890557089e-10\\
81	3.32243890557089e-10\\
82	3.32243890557089e-10\\
83	3.32243890557089e-10\\
84	3.32243890557089e-10\\
85	3.32243890557089e-10\\
86	3.32243890557089e-10\\
87	3.32243890557089e-10\\
88	3.32243890557089e-10\\
89	3.32243890557089e-10\\
90	3.32243890557089e-10\\
91	3.32243890557089e-10\\
92	3.32243890557089e-10\\
93	3.32243890557089e-10\\
94	3.32243890557089e-10\\
95	3.32243890557089e-10\\
96	3.32243890557089e-10\\
97	3.32243890557089e-10\\
98	3.32243890557089e-10\\
99	3.32243890557089e-10\\
100	3.32243890557089e-10\\
101	3.32243890557089e-10\\
102	3.32243890557089e-10\\
103	3.32243890557089e-10\\
104	3.32243890557089e-10\\
105	3.32243890557089e-10\\
106	3.32243890557089e-10\\
107	3.32243890557089e-10\\
108	3.32243890557089e-10\\
109	3.32243890557089e-10\\
110	3.32243890557089e-10\\
111	3.32243890557089e-10\\
112	3.32243890557089e-10\\
113	3.32243890557089e-10\\
114	3.32243890557089e-10\\
115	3.32243890557089e-10\\
116	3.32243890557089e-10\\
117	3.32243890557089e-10\\
118	3.32243890557089e-10\\
119	3.32243890557089e-10\\
120	3.32243890557089e-10\\
121	3.32243890557089e-10\\
122	3.32243890557089e-10\\
123	3.32243890557089e-10\\
124	3.32243890557089e-10\\
125	3.32243890557089e-10\\
126	3.32243890557089e-10\\
127	3.32243890557089e-10\\
128	3.32243890557089e-10\\
129	3.32243890557089e-10\\
130	3.32243890557089e-10\\
131	3.32243890557089e-10\\
132	3.32243890557089e-10\\
133	3.32243890557089e-10\\
134	3.32243890557089e-10\\
135	3.32243890557089e-10\\
136	3.32243890557089e-10\\
137	3.32243890557089e-10\\
138	3.32243890557089e-10\\
139	3.32243890557089e-10\\
140	3.32243890557089e-10\\
141	3.32243890557089e-10\\
142	3.32243890557089e-10\\
143	3.32243890557089e-10\\
144	3.32243890557089e-10\\
145	3.32243890557089e-10\\
146	3.32243890557089e-10\\
147	3.32243890557089e-10\\
148	3.32243890557089e-10\\
149	3.32243890557089e-10\\
150	3.32243890557089e-10\\
151	3.32243890557089e-10\\
152	3.32243890557089e-10\\
153	3.32243890557089e-10\\
154	3.32243890557089e-10\\
155	3.32243890557089e-10\\
156	3.32243890557089e-10\\
157	3.32243890557089e-10\\
158	3.32243890557089e-10\\
159	3.32243890557089e-10\\
160	3.32243890557089e-10\\
161	3.32243890557089e-10\\
162	3.32243890557089e-10\\
163	3.32243890557089e-10\\
164	3.32243890557089e-10\\
165	3.32243890557089e-10\\
166	3.32243890557089e-10\\
167	3.32243890557089e-10\\
168	3.32243890557089e-10\\
169	3.32243890557089e-10\\
170	3.32243890557089e-10\\
171	3.32243890557089e-10\\
172	3.32243890557089e-10\\
173	3.32243890557089e-10\\
174	3.32243890557089e-10\\
175	3.32243890557089e-10\\
176	3.32243890557089e-10\\
177	3.32243890557089e-10\\
178	3.32243890557089e-10\\
179	3.32243890557089e-10\\
180	3.32243890557089e-10\\
181	3.32243890557089e-10\\
182	3.32243890557089e-10\\
183	3.32243890557089e-10\\
184	3.32243890557089e-10\\
185	3.32243890557089e-10\\
};
\addplot [color=mycolor2,line width=1.5pt, forget plot]
  table[row sep=crcr]{%
37	323.203804135825\\
38	32.8074189723239\\
39	19.9014987511225\\
40	13.1294020238727\\
41	11.4552189113388\\
42	10.272718780302\\
43	9.27997071667305\\
44	8.20223124085519\\
45	6.70762101624698\\
46	6.10675151262529\\
47	5.62752465599546\\
48	5.42227089039119\\
49	5.32417987718949\\
50	5.21147097791365\\
51	5.13997267467135\\
52	5.06664142223233\\
53	5.00316775540773\\
54	4.9472173893148\\
55	4.89489366320105\\
56	4.85331529158756\\
57	4.81718601229491\\
58	4.7791219134539\\
59	4.73792783589036\\
60	4.70590537253165\\
61	4.67043325981096\\
62	4.65769133579663\\
63	4.63527009747443\\
64	4.62219585370503\\
65	4.61704835514092\\
66	4.61448210496311\\
67	4.61106185040942\\
68	4.60580749524219\\
69	4.59877458572026\\
70	4.59246246706704\\
71	4.58725633248076\\
72	4.58520364425375\\
73	4.58414103735178\\
74	4.5838033605339\\
75	4.58341351147924\\
76	4.58265834863773\\
77	4.58152366503702\\
78	4.5801960508653\\
79	4.57888309049841\\
80	4.57755287183485\\
81	4.57597117284193\\
82	4.5740451545021\\
83	4.57188631686771\\
84	4.56999853848654\\
85	4.56765453687449\\
86	4.56616199684207\\
87	4.56528891940259\\
88	4.5647674791582\\
89	4.56521345829159\\
90	4.56660687671683\\
91	4.56821363747958\\
92	4.56937166918632\\
93	4.57130705298639\\
94	4.57074312190505\\
95	4.56877565642381\\
96	4.56738911483399\\
97	4.56709538695802\\
98	4.56656157990562\\
99	4.56733550502332\\
100	4.5668665874124\\
101	4.56668636646532\\
102	4.56642849589511\\
103	4.56485056118134\\
104	4.56470273896784\\
105	4.56664094598668\\
106	4.56584460424819\\
107	4.56576865203413\\
108	4.56703826620551\\
109	4.56668898699538\\
110	4.56592212655444\\
111	4.5648050262641\\
112	4.56401087666895\\
113	4.56344579029876\\
114	4.56478902444529\\
115	4.56667301362112\\
116	4.56895565534761\\
117	4.5702860852894\\
118	4.57081081509834\\
119	4.57025841225649\\
120	4.56904759168988\\
121	4.56803258816489\\
122	4.56755802334589\\
123	4.56748823318679\\
124	4.56764111877553\\
125	4.56786310684882\\
126	4.56803706599054\\
127	4.56816671608682\\
128	4.56832186222339\\
129	4.56854079092793\\
130	4.56879429980774\\
131	4.56900573877245\\
132	4.56912317840736\\
133	4.56911573252462\\
134	4.5690114032721\\
135	4.56888883244032\\
136	4.56884891179544\\
137	4.56900042512084\\
138	4.56943154303095\\
139	4.56997797669118\\
140	4.57041601575269\\
141	4.57061254175874\\
142	4.57062596835322\\
143	4.57031458229278\\
144	4.57004287714743\\
145	4.56976974492878\\
146	4.56968634345422\\
147	4.56964783832487\\
148	4.5696391773336\\
149	4.56963015765938\\
150	4.56961348840553\\
151	4.56959043661048\\
152	4.56956959508566\\
153	4.56955628438991\\
154	4.56954828871526\\
155	4.56954468557457\\
156	4.56954404642749\\
157	4.56954319824557\\
158	4.56954136297334\\
159	4.56953883992803\\
160	4.56953578874024\\
161	4.56953441998182\\
162	4.56953556054595\\
163	4.56953928897472\\
164	4.56954412865288\\
165	4.56954963694137\\
166	4.56955565506821\\
167	4.56956183313831\\
168	4.56956793266923\\
169	4.56957246883557\\
170	4.56957471952749\\
171	4.56957515119836\\
172	4.56957376272941\\
173	4.56956925488569\\
174	4.5695610256805\\
175	4.56955078044712\\
176	4.5695424258082\\
177	4.56953722904462\\
178	4.56953411241643\\
179	4.5695322857847\\
180	4.56953117695409\\
181	4.56952898630406\\
182	4.56952486824817\\
183	4.56951924879224\\
184	4.56951258335005\\
185	4.56951258335005\\
};
\draw[dashed,gray,line width=0.75pt] (axis cs:37,1e-10) -- (axis cs:37,1e3);
\draw[<->,gray,line width=0.75pt] (0,1e-8) -- (37,1e-8) node[midway, above] {\scriptsize\itshape \textcolor{black}{solve underlay}};
\draw[<->,gray,line width=0.75pt] (37,1e-4) -- (102,1e-4) node[midway, below] {\scriptsize\itshape \textcolor{black}{solve pleat}};

\end{axis}
\end{tikzpicture}%

%% file: tables/tab_runtime.tex
\footnotesize
{\def\arraystretch{1.3}\tabcolsep=0.74em 
\begin{tabular}{lrrrrrccc}
\toprule[1pt]
\multirow{2}{*}{\begin{tabular}[c]{@{}c@{}}\itshape smocking\\ 
\itshape pattern\end{tabular}} & \multicolumn{4}{c}{\itshape smocked graph complexity} & \multirow{2}{*}{\begin{tabular}[c]{@{}c@{}} $\vert \V_{\widetilde{\P}}\vert$ \end{tabular}} & \multicolumn{3}{c}{\itshape optimization (sec.)} \\ \cmidrule[0.8pt](l){2-5} \cmidrule[0.8pt](l){7-9}
  & $\vert \V_u \vert$ &  $\vert \E_u \vert$ &   $\vert \V_p \vert$ &  $\vert \E_p \vert$ &  & $\; \S_u$ & $\; \S_p$ & $\widetilde{\P}$ \\ \midrule[1pt]
 Fig.~\ref{fig:intro:eg_smocking} & 24 & 53 & 45 &  186 & 5074 &  0.0015 & 0.130 &  1.920   \\
\rowcolor{tabyellow!15}
Fig.~\ref{fig:mtd:honeycomb} &  30  & 66 & 121 & 360 & 8613 & 0.0012 & 1.161  & 3.370  \\
Fig.~\ref{fig:res:non_volumetric_pattern} (a)  & 64 & 210  & 97 & 382 & 12769 & 0.0017 & 0.714 & 3.768  \\ 
\rowcolor{tabyellow!15}
Fig.~\ref{fig:res:non_volumetric_pattern} (b)  & 64 & 98 & 249 & 1038 & 19321 & 0.0014 & 1.972 &  4.088 \\
Fig.~\ref{fig:res:triangle} (a)  & 49  &  106  &  130 &  537 &  14994 &  0.0016 & 0.852 & 3.269  \\ 
\rowcolor{tabyellow!15}
Fig.~\ref{fig:res:triangle} (b) & 49 &  106  &  144  & 621 & 11236  & 0.0014 & 0.836 &  3.215 \\
Fig.~\ref{fig:res:radial_grids} (a)  & 60 & 149 & 88 & 418 & 9116 & 0.0015 & 0.402 & 3.044  \\
\rowcolor{tabyellow!15}
Fig.~\ref{fig:res:radial_grids} (b)  & 144 & 353 & 262 &  1222 & 67600 & 0.0023  & 2.705 & 21.25  \\
Fig.~\ref{fig:res:radial_grids} (c)  & 72 & 153 & 103 & 525 & 10836 & 0.0007 & 0.364 & 2.882  \\ 
\rowcolor{tabyellow!15}
Fig.~\ref{fig:res:radial_grids} (d)  & 192  & 392  & 346 & 1705 & 81796  & 0.0076 & 7.527 &  25.89  \\
\bottomrule[1pt]
\end{tabular}
}

%% file: sections/conclusion.tex
\section{Conclusion, Limitations and Future Work}
\label{sec:conclusion}
In this paper, we discuss how to mathematically formulate Canadian smocking, a decorative and practically beneficial surface embroidery technique.
We introduce a simple yet effective method that solves for the smocking design with 3D geometric textures based on an input smocking pattern, which is represented by a set of stitching lines drawn on top of the fabric.
We first extract the smocked graph from the input pattern, where the points in the same stitching lines are merged into a single vertex, and the degenerated or redundant edges are removed. This smocked graph encodes the geometric features of the final smocking design. 
To obtain the smocking design, we first embed the smocked graph in 3D, where we embed the underlay graph and the pleat graph in two steps. We then use the embedded smocked graph to guide the deformation of the fabric represented via a finer grid using {\arap}.
Our method is efficient and accurate, and our computed smocking designs are very similar to real fabrications for a large set of patterns, which allows us to design a user interface for smocking design exploration.

In this work, we formulate smocking as a pure shape modeling problem without considering cloth dynamics and collision response. 
%Therefore, self-intersections are not considered or handled. For example, we can see some self-intersections in the Fig.~\ref{fig:res:long_stitching_lines}. 
Though the self-intersections do not significantly affect the visual appearance of the digital smocking design, it would be interesting to take them into consideration during modeling. 
\jr{We also wish to investigate smocking from the perspective of cloth simulation, as we can see that the state-of-the-art simulators cannot tackle this problem directly. 
Our preview results can effectively offer a trustworthy initial guess for cloth simulators, enabling the incorporation of additional material-dependent parameters, such as bending stiffness, to generate more realistic results at fine scales.}
Another limitation of our current approach is that we do not fully explore all possible smocking designs from an input pattern. For some complicated smocking patterns (such as \figref{fig:res:long_stitching_lines}~(d)), multiple visually appealing local minima (i.e., multiple final smocking designs) are possible. These results can guide the user or designer to iron or steam the smocked fabric into different shapes. We leave this shape space exploration as future work.
Another interesting direction is to investigate the inverse problem of smocking, i.e., finding the arrangement of stitching lines such that the final result is close to an expected 3D texture or shape.
Since smocking is a popular embroidery technique used by high-end fashion designers, we wish to explore smocking design directly on surfaces in 3D, so that it could be integrated with garment design.
In our experiments, we notice that the smocked shapes can serve as meta-materials, since the pleats on top of the rigid underlay create extra thickness and elastic cushioning. In future work, it would be interesting to design smocking patterns with particular physical properties.

%% file: appendix/appendix_metric.tex
%\vspace{1cm}
\section{Smocked Graph Embedding}\label{appendix:metric_embedding}
In this section we discuss the intuition behind our relaxation of the problem defined in \eqnref{eq:mtd:prob:trivial} for the embedding of the smocked graph, presented in \secreflist{sec:mtd:embed-underlay}{sec:mtd:embed-pleat}.

\textit{Embedding the underlay.} 
We embed the underlay graph in 2D by fully stretching the underlay edges to their upper bound $d_{i,j}$ as defined in \eqnref{eq:mtd:d_ij}.
Intuitively, gathering the underlay nodes in a stitching line is equivalent to translating all these nodes in the $xy$-plane (the initial fabric plane) to the same position. After smocking (translation), the underlay nodes still stay on the $xy$-plane, i.e., they remain co-planar.
This co-planarity property allows us to solve the embedding of the underlay in 2D, which significantly reduces the search space.
Indeed, we observe that the underlay regions of the fabricated results are co-planar, which validates our 2D search space.
%Another physical explanation of the co-planarity of the underlay nodes is that: usually a smocking pattern is obtained by tiling the same unit pattern in a grid. In this case, all the stitching lines (or underlay nodes in the smocked graph) except those at the boundary are \emph{isotropic} to each other and therefore stay co-planar.

%The valid embedding of the underlay that maximizes the embedding energy (defined in \eqnref{eq:mtd:prob:trivial}) is when all underlay nodes are co-planar.
%
%Specifically, we can regard the maximizing embedding energy as the projected area of the embedded graph onto $xy$-plane. %: we prefer to have a fully stretched smocking design instead of a clustered one, which means the projected area should be as large as possible. 
%The largest projected area is achieved when all the underlay nodes are co-planar.
%Otherwise, we can project the non-planar underlay node to $xy$-plane and further stretch the embedding to achieve larger projected area. Therefore, we can assume the height ($z$-axis value) of the underlay nodes to be zero during the optimization and only solve for the $2$D embeddings.

%for some underlay node pairs $(v_i, v_j)\in\V_u \times \V_u$, we have the \emph{equality} $\left\Vert \x_i - \x_j \right\Vert_2 = d_{i,j}$ satisfied. We can ignore the other inequality constraints. 

%The key idea is to find out which set of inequality constraints become equality constraints. Actually, the edges $\E_u$ in the underlay graph suggest the vertex pairs where the maximum embedding distance will be reached. 
%

Our reformulation is based on the fact that the optimum to \eqnref{eq:mtd:prob:trivial} is at the \emph{exact} boundary of some inequality constraints.
We can picture the inequality constraints in \eqnref{eq:mtd:prob:trivial} as follows: imagine we have a set of tiny balls or beads placed on the $xy$-plane, and each bead represents an underlay node. 
For each pair of beads, e.g., the beads that represent the $i,j$-th underlay node, we use a string with length $d_{i,j}$ to connect the two beads (see the inset figure). We then move the beads on the plane such that they are far from each other and none of the strings are broken. We can imagine that at the optimal situation, some of the strings become \emph{tight}, which means if we move the attached beads any further, these strings will break. At
\setlength{\columnsep}{5pt}%
\setlength{\intextsep}{0pt}%
\begin{wrapfigure}{r}{0.35\linewidth}%
\centering
\begin{overpic}[trim=0cm 0cm 0cm 0cm,clip,width=1\linewidth,grid=false]{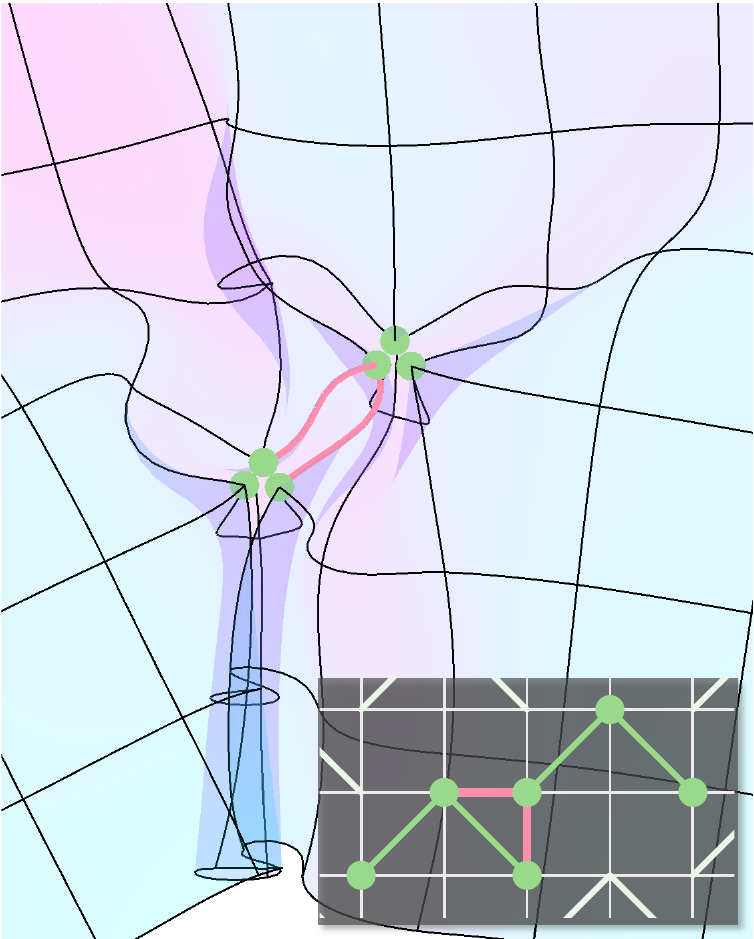}
\put(14,50){\textcolor{black}{$\x_{\ell_i}$}}
\put(40,70){\textcolor{black}{$\x_{\ell_j}$}}
\put(35,48){$d_{i,j}$}
\put(38,11){\textcolor{colvu}{$\ell_i$}}
\put(56,22){\textcolor{colvu}{$\ell_j$}}
\end{overpic}
%\caption{\textbf{Smocking Pattern}}\label{fig:mtd:dij}
\end{wrapfigure}%
the same time, there are other strings that stay \emph{loose}, which means their extreme can never be reached and they are useless in constraining the beads.
In other words, if we simply remove the loose strings, and do the above experiment again, we will end up with the same configuration (up to translation and rotation). 
Moreover, the short strings are more likely to be tight at the optimum compared to the long strings. 
Since the underlay graph is planar and the $d_{i,j}$ values are derived from the fabric, which is a connected and presumably inextensible 2-manifold, we can conclude that the underlay edges (e.g., pink edges/strings highlighted in the inset figure) become tight at the optimum. 
Therefore, in \eqnref{eq:mtd:underlay} we only consider the underlay nodes that are connected to each other.

\textit{Embedding the pleats.} We embed the pleat nodes in 3D on top of the solved underlay graph. 
Our reformulation in \eqnref{eq:mtd:pleat} is based on the observation that the 3D embedding of a pleat node is only constrained by the embeddings of the \emph{neighboring} nodes. 
Specifically, the embedding distance constraint between a pleat node and a faraway node can be ignored. 
The intuition behind this is that the distance constraint between a pleat node $v_1$ and a faraway node $v_2$ is satisfied automatically according to the triangle inequality if the distance constraint between $v_2$ and a neighboring node $v_3$, and the distance constraint between $v_3$ and $v_1$ are satisfied.
Thus, we only need to consider the pleat nodes and their neighboring nodes, i.e., the node pairs in the pleat edge set $\E_p$. 
%
% Moreover, since the pleat nodes have more freedom to move in 3D, we also know the maximum embedding is achieved at the exact boundary of the inequality constraints.

%% file: appendix/appendix_ui.tex
%\newpage
\section{Interactive User Interface}\label{appendix:ui}

\begin{figure}[!t]
    \centering
    \includegraphics[width=0.9\linewidth]{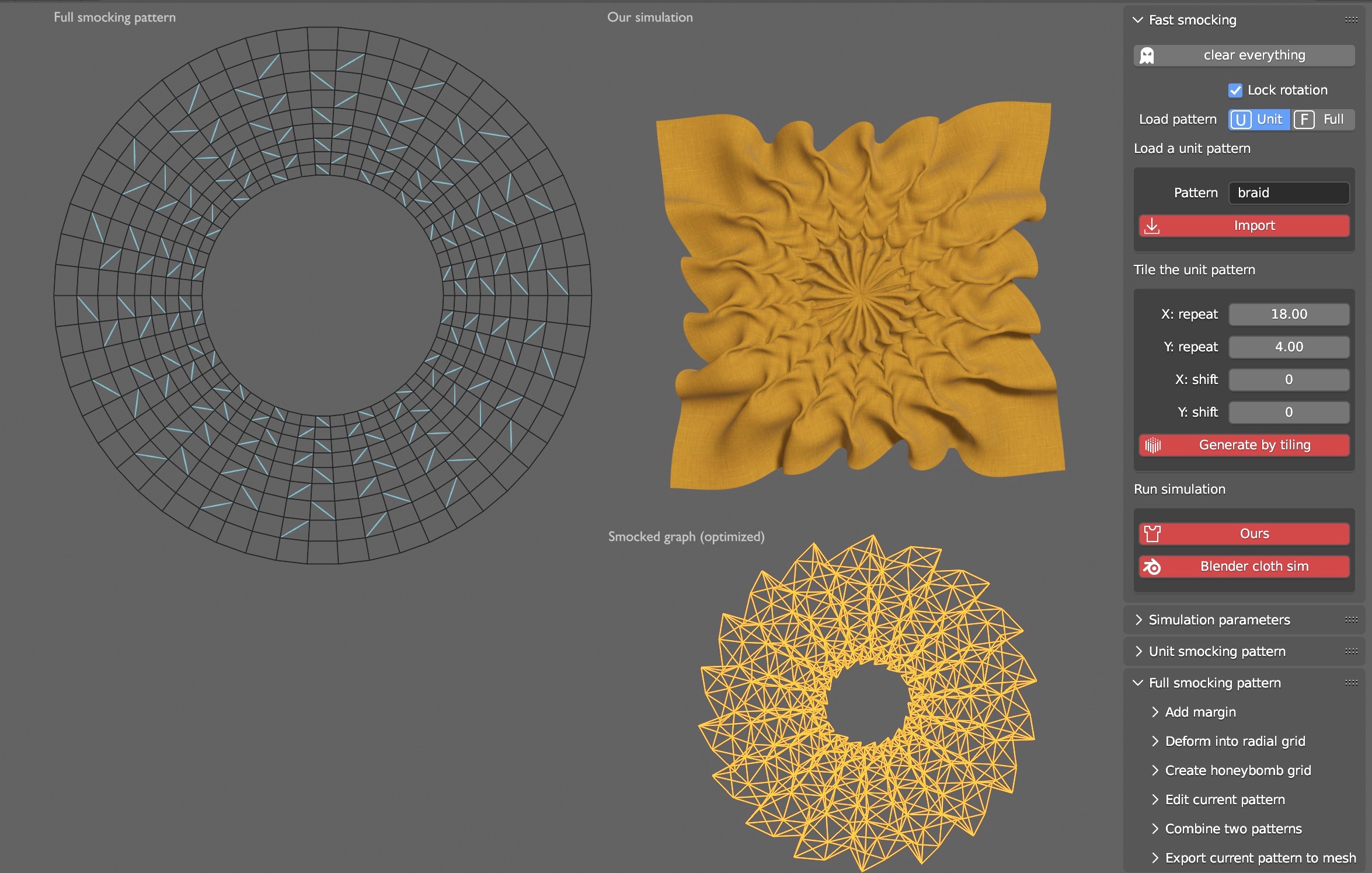}\vspace{-6pt}
    \caption{We introduce an interactive user interface for smocking design, implemented in Blender as an add-on.}\label{fig:appendix:ui}\vspace{-6pt}
\end{figure}

We implement an interactive user interface in Blender as an add-on (see \figref{fig:appendix:ui}) including the following functionalities:
\begin{itemize}[leftmargin=*]
    \item \emph{define a unit smocking pattern} by creating a 2D grid and drawing stitching lines
    \item \emph{define a full smocking pattern} by
    \begin{itemize}
        \item tiling the loaded unit smocking pattern (with user-defined repetition and shift of the unit pattern)
        \item drawing stitching lines directly on a square or hexagonal grid to define the full pattern
    \end{itemize}
    \item \emph{modify a full smocking pattern} by
    \begin{itemize}
        \item deforming the square grid into a \emph{radial} grid (with user-defined radius)
        \item adding margins to the pattern
        \item combining it with another smocking pattern (along user-specified axis and space)
        \item deleting/adding stitching lines from/to the pattern
    \end{itemize}
    \item \emph{simulate the smocked pattern} with intermediate steps including
    \begin{itemize}
        \item extracting the smocked graph from the pattern
        \item embedding the underlay and pleat subgraphs of the smocked graph
        \item applying {\arap} to compute the smocking design
    \end{itemize}
    \item \emph{render the smocking design}
    \item \emph{run cloth simulator} implemented in Blender on the fine-resolution smocking pattern
    % \item \emph{prototype for digital garment design} by wrapping the simulated smocked pattern to garments based on parameterization.
\end{itemize}
Please see the supplementary videos for the real-time demonstrations of our UI. 
\jr{The Blender add-on can be found at [INSERT LINK]}.

%% file: appendix/appendix_formulation.tex
%\vspace{1.1cm}

\section{Algorithmic Details}\label{appendix:formulation}
Algorithm~\ref{alg:mtd:grid_free} shows the pseudo-code for grid-free smocking design discussed in \secref{sec:mtd:grid-free}, which gives instructions on how to construct a graph from input stitching lines without given grids. 
\jr{
Specifically, given a set of stitching lines as input, we first extract the underlay nodes $\V$ from the endpoints of the stitching lines.
We can then construct the underlay graph by computing a Delaunay triangulation~\cite{delaunay1934sphere,lee1980two} conditioned on the input stitching lines.
We then  sample a set of pleat nodes and construct the pleat graph.
As discussed in \secref{sec:mtd} and demonstrated by \figref{fig:appendix:arrow_height_map}, the pleat region pops up from the base layer (underlay region) to form the texture. We therefore focus on connecting the sampled pleat nodes to the underlay graph to construct the pleat graph.
For each sampled pleat node $v$, we compute the Delaunay triangulation again on the underlay nodes and this pleat node, i.e., $\V  \bigcup \{v\}$, from which we can extract the pleat edges between $v$ and the neighboring underlay vertices.
We can additionally connect the pleat nodes to the neighboring pleat nodes by Delaunay triangulating all the pleat nodes only. 
In this way, we can construct the underlay graph and \osh{the} pleat graph for smocking design computation from stitching lines alone. 
One important observation is that the pleat nodes need to be evenly sampled w.r.t. the input stitching lines. For example, one can take the middle points of the stitching lines as the pleat nodes, as shown in \figref{fig:mtd:eg_basket}. In this way, the regularity encoded in the input stitching lines is kept during the pleat graph construction, and therefore leads to desirable simulated results. 
% \figref{fig:appendix:random_pleat} shows an example of randomly sampled pleat nodes, which produces a less regular result.
}

\begin{algorithm}[!t]
\DontPrintSemicolon
\SetKwData{Left}{left}\SetKwData{This}{this}\SetKwData{Up}{up}
\SetKwFunction{Union}{Union}\SetKwFunction{FindCompress}{FindCompress}
\SetKwInOut{Input}{Input}\SetKwInOut{Output}{Output}
\Input{ A set of stitching lines $\L = \left\{\, \ell_i \,\right\}$}
\Output{A graph $\G = (\V, \E)$ to complete the smocking pattern}
$\V \gets \left.\left\{v\in\ell_i \right\vert \forall \ell_i\in \L\right\}$ \;
\textcolor{gray}{\footnotesize // create underlay edges} \;
$\E \gets \text{DelaunayTriangulation}\left(\V\right)$ \;
regularly sample a set of pleat nodes $\V'$ inside the bounding box of $\L$ \;
\ForEach{$v\in \V'$}{
\textcolor{gray}{\footnotesize // create pleat edges for $v$} \;
$\E' \gets \text{DelaunayTriangulation}\left(\V \bigcup \left\{v\right\}\right)$\;
$\E \gets \E \bigcup \left.\left\{ e\in \E' \right\vert v \in e\right\}$\;
}
$\E' \gets \text{DelaunayTriangulation}\left(\V'\right)$\;
$\E \gets \E \bigcup \E'$\;
$\V \gets \V \bigcup \V'$\;
\caption[caption]{Smocking Pattern from Stitching Lines}
\label{alg:mtd:grid_free}
\end{algorithm}

% \begin{figure}
%     \centering
%     \begin{overpic}[trim=0cm 0cm 0cm 0cm,clip,width=0.9\linewidth,grid=false]{figures/eg_braid_rand_pleat.pdf}
%     \end{overpic}\vspace{-3pt}
%     \caption{
%     \jr{
%     We sample \emph{random} pleat nodes to the pattern shown in \figref{fig:mtd:eg_basket}. \emph{Left}:~the randomly sampled pleat nodes are colored in green. \emph{Right}: we show the corresponding computed smocking design.
%     }
%     }
%     \label{fig:appendix:random_pleat}
% \end{figure}

% \subsection{Solver for Embedding the Underlay Graph}
% Recall that we would like to find the 2D embedding $x_p \in \R^{2}$ for the underlay nodes $v_p$ such that:
% %
% \begin{equation}
%     \begin{split}\label{eq:mtd:prob:reform}
%         \max_{x_{p}} \quad & \sum_{\forall p\neq q} \Vert x_p - x_q \Vert_2 \\
%         \text{s.t.} \quad & \Vert x_p - x_q \Vert_2 = d_{p,q} \quad\forall (p, q)\in \E_{\,=}\\
%         & \Vert x_p - x_q \Vert_2 \leq d_{p,q} \quad\forall (p, q)\in \E_{\le}
%     \end{split}
% \end{equation}
% %
% For simplicity of notation, we denote the set of vertex pairs where their maximum distance is reached as $\E_{\,=}$, and the set of rest vertex pairs as $\E_{\le}$.

%% file: appendix/appendix_additional_res.tex
\begin{figure}[!t]
    \centering
    \begin{subfigure}[b]{0.45\linewidth}
    \includegraphics[width=1\linewidth]{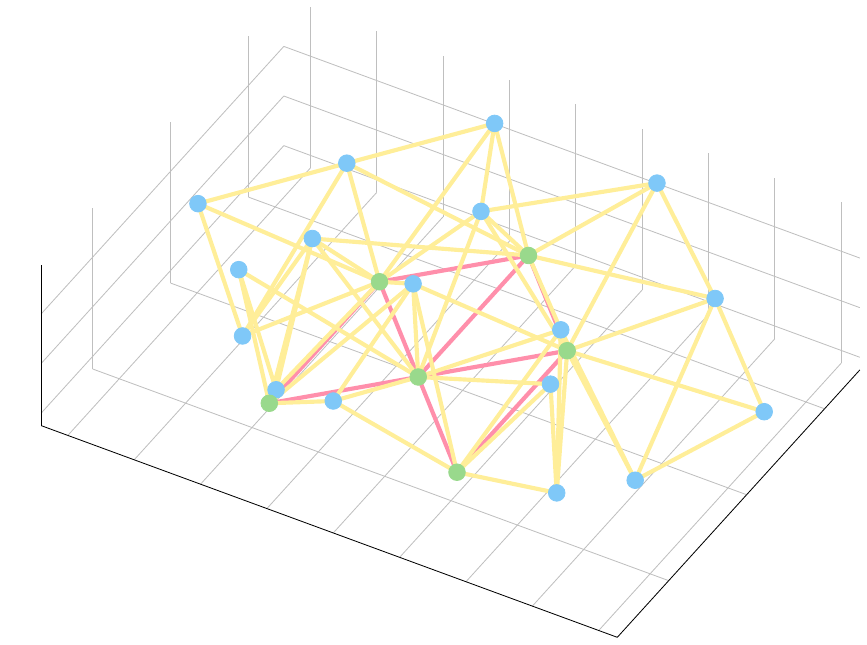}
    \caption{A 3D embedding of $\S$}\label{fig:mtd:embeded-graph}
    \end{subfigure}
    \hfill
    \begin{subfigure}[b]{0.45\linewidth}
        \begin{overpic}[trim=0cm 0cm 0cm 6.5cm,clip,width=1\linewidth,grid=false]{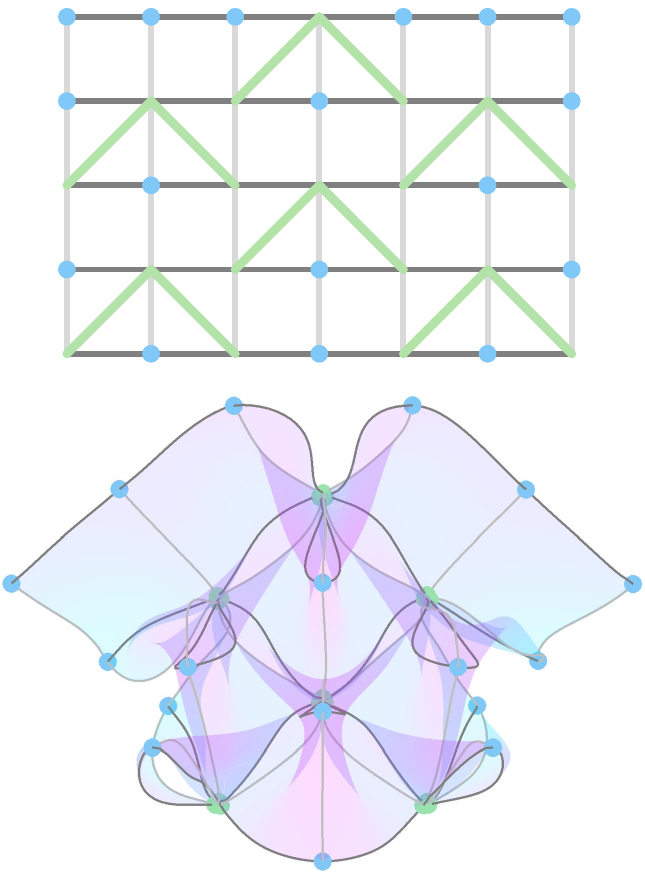}
        \end{overpic}\vspace{-5pt}
        \caption{Smocking design}\label{fig:mtd:design}
    \end{subfigure}\vspace{-5pt}
   \caption{\jr{Embedding of the coarser (left) and the finer (right) discretization of the smocking pattern shown in \figref{fig:mtd:smocked graph}.}}
\end{figure}

\section{Additional Results}\label{appendix:step_arap}

\paragraph{C-IPC without collision}
\begin{figure}[t]
    \centering
    \begin{overpic}[trim=0.0cm 0cm 0cm -0.5cm,clip,width=1\linewidth,grid=false]{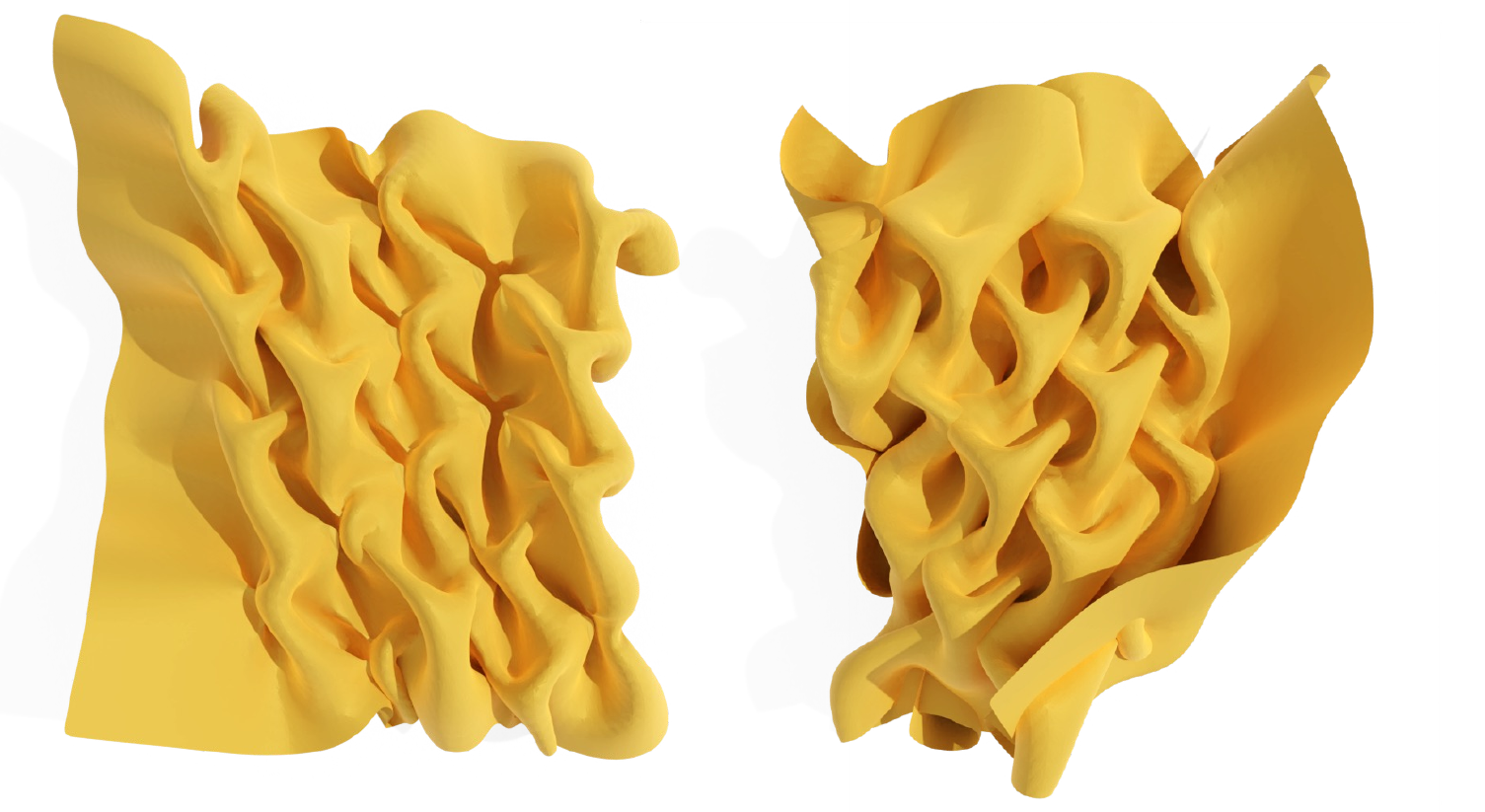}
    \put(12,53){\footnotesize static solver, $t=\unit[1]{min}$}
    \put(55,53){\footnotesize dynamic solver, $t=\unit[18]{sec}$}
    \end{overpic}\vspace{-3pt}
    \caption{\jr{C-IPC on \textsc{arrow} pattern \emph{without} collision handling.}}
    \label{fig:res:cipc_no_collision}
\end{figure}
\jr{
\figref{fig:res:cipc_no_collision} shows the results of CIPC~\cite{Li2021CIPC} on the \textsc{arrow} pattern without collision handling. 
Disabling self-collision handling for the \osh{static} solver does not improve computational efficiency; instead, it leads to worse results compared to \figref{fig:res:cipc:all}. 
Using the dynamic solver without self-collision handling results in faster iterations. However, running the dynamic solver until convergence leads to a cluttered configuration. Here, we select an intermediate iteration where the pleats are sufficiently formed and the overall shape starts to show signs of \osh{clumping}.
}

\begin{figure}[!t]
    \centering
    \begin{overpic}[trim=0.7cm 0cm 0.5cm 0cm,clip,width=1\linewidth,grid=false]{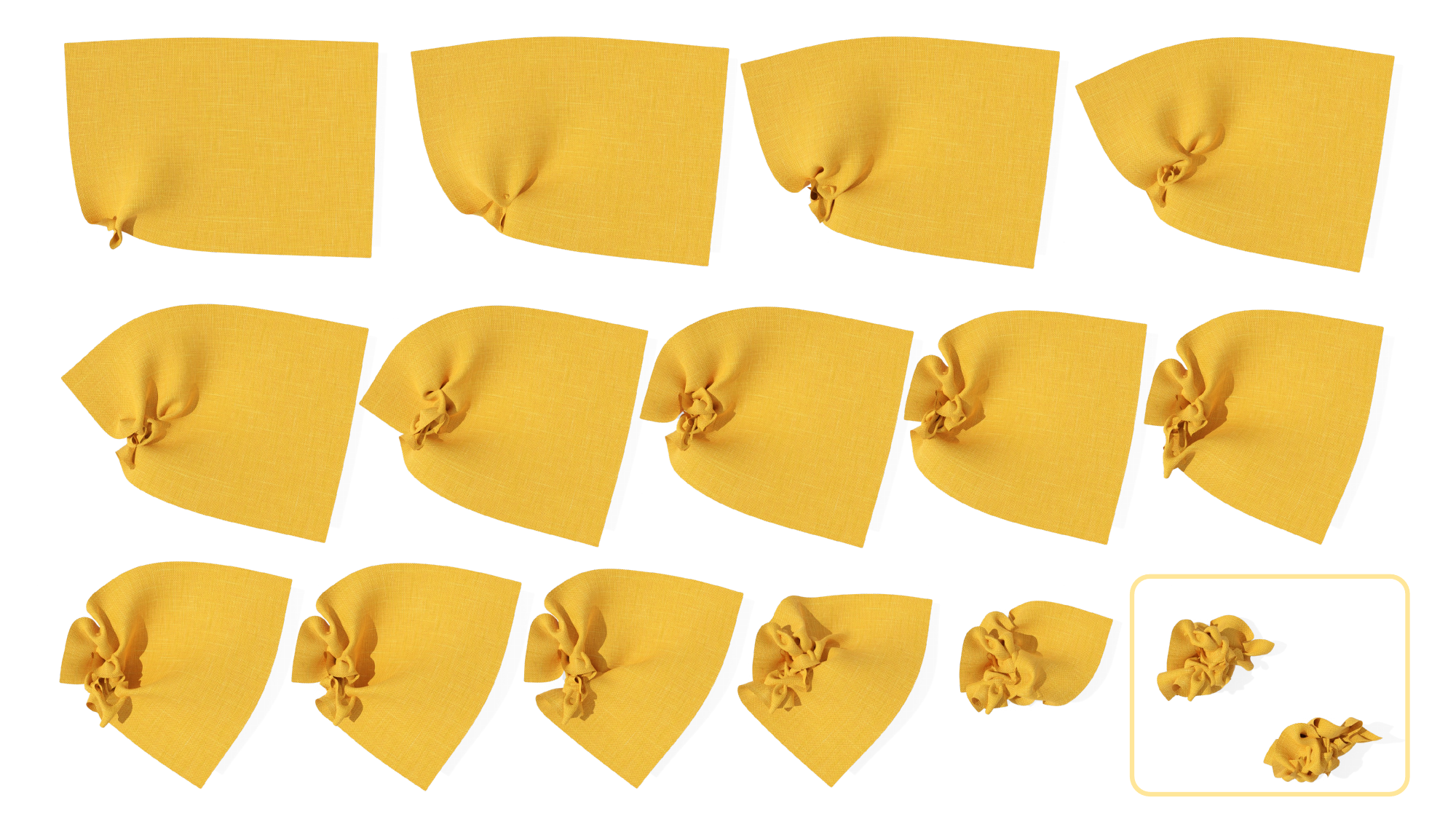}
    \put(2,40){\scriptsize $\ell_1$}
    \put(28,40){\scriptsize $\ell_2$}
    \put(52,40){\scriptsize $\ell_3$}
    \put(78,40){\scriptsize $\ell_4$}
    \put(2,21){\scriptsize $\ell_5$}
    \put(25,21){\scriptsize $\ell_6$}
    \put(45,21){\scriptsize $\ell_7$}
    \put(62,21){\scriptsize $\ell_8$}
    \put(82,21){\scriptsize $\ell_9$}
    \put(2,3){\scriptsize $\ell_{10}$}
    \put(18,3){\scriptsize $\ell_{11}$}
    \put(36,3){\scriptsize $\ell_{12}$}
    \put(51,3){\scriptsize $\ell_{15}$}
    \put(67,3){\scriptsize $\ell_{20}$}
    \put(80,3){\scriptsize $\ell_{24}$}
    \put(80,15.5){\scriptsize\itshape top view}
    \put(88,8){\scriptsize\itshape side view}
    \end{overpic}\vspace{-3pt}
    \caption{
    \textbf{Step-by-step {\arap}.} We run {\arap} to sew each stitching line $\ell_i,~i=1, \ldots, 24$ in the smocking pattern shown in \figref{fig:intro:eg_smocking}. Note that all visualizations are in a consistent scale.}\label{fig:res:arap_step}\vspace{-6pt}
\end{figure}
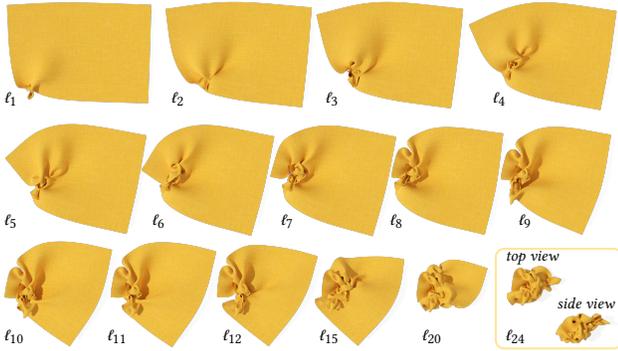

\paragraph{Step-by-step {\arap}}
Another straightforward solution to model smocking is by applying {\arap} to each stitching line separately to mimic the manufacturing process. 
In \figref{fig:res:arap_step}, we adopt this strategy to model the smocking pattern depicted in \figref{fig:intro:eg_smocking}, following a left-to-right and bottom-to-top sequence of stitching. 
However, we observe that this approach causes the fabric to bend inward instead of shrinking towards the center as it happens during actual fabrication (illustrated in \figref{fig:background:arap}(d)).
As a result, the final outcome exhibits a messy appearance with extremely cluttered pleats with many self-intersections, as shown in \figref{fig:res:arap_step} of the final result when viewed from the side. 
As a comparison, our progressive {\arap} simultaneously stitches all the stitching lines, ensuring more uniform ``stitching forces'' to promote fabric shrinkage in more accurate directions. This results in a superior baseline compared to the step-by-step {\arap} approach.

% \begin{table}[!t]
%     \centering
%     \caption{
%     \jr{
%     For the smocking patterns $\P_i$ in \figreflist{fig:usp:underconstrained}{fig:usp:overconstrained}, we report the energy in \eqnreflist{eq:mtd:underlay}{eq:mtd:pleat} at convergence. We also report the number of vertex pairs that are stretched (\textit{\# stretched}), i.e. having $\Vert \x_i - \x_j \Vert > d_{i,j}$; and the average stretch ratio (\textit{stretch\%}), i.e. $\nicefrac{\left(\left\Vert \x_i - \x_j \right\Vert -d_{i,j}\right)}{d_{i,j}}$.
%     }
%     }\label{tab:energy}
%     \vspace{-3pt}
%     \input{tables/tab_energy.tex}
% \end{table}

\begin{figure}[t]
    \centering
    \begin{overpic}[trim=2cm 9.8cm 1cm 9cm,clip,width=0.9\linewidth,grid=false]{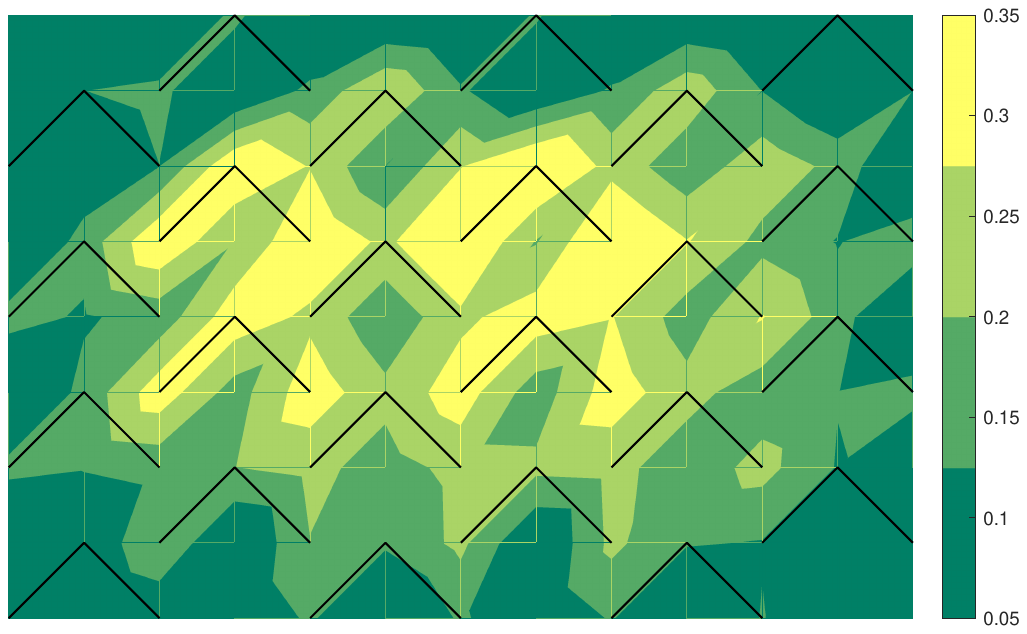}
    \end{overpic}
    \begin{overpic}[trim=2cm 9.8cm 1cm 9cm,clip,width=0.9\linewidth,grid=false]{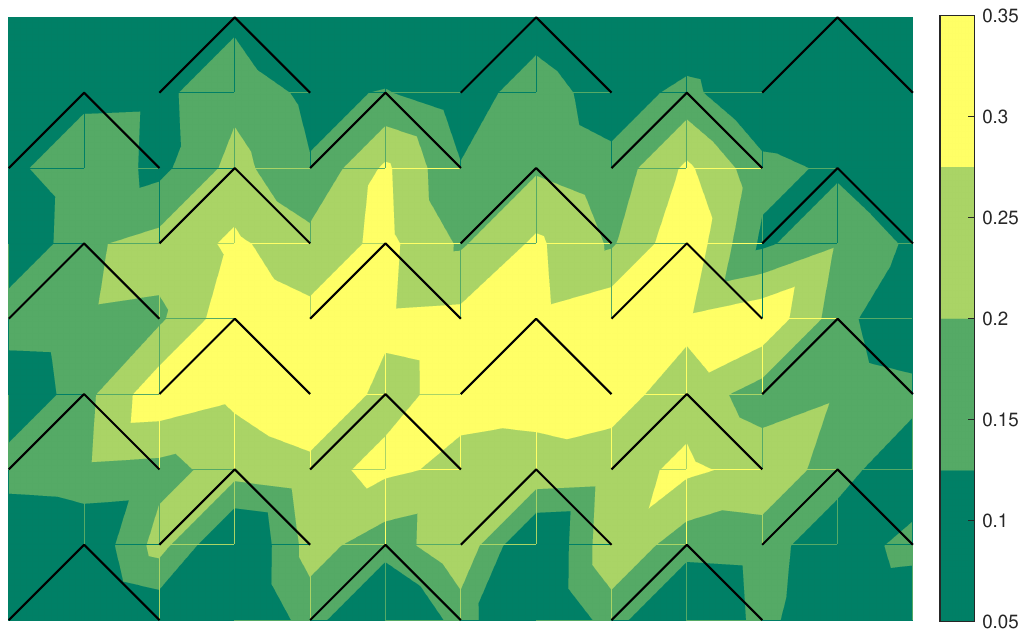}
    \end{overpic}\vspace{-3pt}
    \caption{
    \jr{
    Optimized energy distribution after simultaneous optimization (\emph{top}) and our two-stage optimization (\emph{bottom}), as discussed in \figref{fig:res:ablation_solver}. We also draw the stitching lines in black.
    }
    }
    \label{fig:err_distribution}
\end{figure}

\paragraph{Optimized energy distributions}
\jr{
% In \tabref{tab:energy}, we report the energy and the number of vertex pairs that are stretched at convergence for the smocking patterns shown in \figreflist{fig:usp:underconstrained}{fig:usp:overconstrained}.
%
In \figref{fig:err_distribution} we visualize the energy (the sum of \eqnref{eq:mtd:underlay} and \eqnref{eq:mtd:pleat}) of the computed smocked design shown in \figref{fig:res:ablation_solver} after optimization using simultaneous solver and our two-stage solver. For easier visual comparison, we visualize the per-vertex errors on the original smocking pattern.
We can see that the result from simultaneous optimization shows more prominent error in the underlay region (edges that connect two different stitching lines), while the result of the two-stage optimization has a \osh{smoother error distribution} in the pleat region.
}